 \documentclass[numberedappendix,iop,revtex4]{emulateapj}

\usepackage{color}

\usepackage{placeins}

\newcommand{\MACSeleven}{MACS J1149.5+2223}
\newcommand{\MACSzeroseven}{MACS J0717.5+3745}
\newcommand{\MACSzerofour}{MACS J0416.1-2403}

\newcommand{\hst}{\textit{HST}}
\newcommand{\ds}{D_\mathrm{S}}
\newcommand{\dls}{D_\mathrm{LS}}
\newcommand{\dl}{D_\mathrm{L}}
\newcommand{\zphot}{z_\mathrm{phot}}
\newcommand{\zl}{z_\mathrm{l}}
\newcommand{\zs}{z_\mathrm{s}}
\newcommand{\rcut}{r_\mathrm{cut}}
\newcommand{\rcore}{r_\mathrm{core}}
\newcommand{\Lstar}{L^{\star}}

\shorttitle{Lens models of the HFF clusters}
\slugcomment{Accepted to ApJ: draft date \today}
\shortauthors{Johnson et al.}

\begin{document}

\title{Lens models and magnification maps of the six Hubble Frontier Fields clusters$^*$}

\author{
Traci L. Johnson\altaffilmark{1}, Keren Sharon\altaffilmark{1}, Matthew B. Bayliss\altaffilmark{2,3}, Michael D. Gladders\altaffilmark{4,5}, Dan Coe\altaffilmark{6}, Harald Ebeling\altaffilmark{7}}
\altaffiltext{*}{Based on observations made with the 6.5 meter Magellan Telescopes located at Las Campanas Observatory, Chile and the NASA/ESA Hubble Space Telescope, obtained through the Mikulski Archive for Space Telescopes (MAST) at the Space Telescope Science Institute, which is operated by the Association of Universities for Research in Astronomy, Inc., under NASA contract NAS 5-26555. These observations are associated with program \#13495.}
\email{tljohn@umich.edu}
\altaffiltext{1}{University of Michigan, Department of Astronomy, 1085 South University Avenue, Ann Arbor, MI 48109, USA}
\altaffiltext{2}{Department of Physics, Harvard University, 17 Oxford Street, Cambridge, MA 02138, USA}
\altaffiltext{3}{Harvard-Smithsonian Center for Astrophysics, 60 Garden Street, Cambridge, MA 02138, USA}
\altaffiltext{4}{Kavli Institute for Cosmological Physics at the University of Chicago}
\altaffiltext{5}{Department of Astronomy \& Astrophysics, The University of Chicago, 5640 South Ellis Avenue, Chicago, IL 60637, USA}
\altaffiltext{6}{Space Telescope Science Institute, Baltimore, MD, USA}
\altaffiltext{7}{Institute for Astronomy, University of Hawaii, 2680 Woodlawn Drive, Honolulu, HI 96822, USA}

\begin{abstract}

We present strong-lensing models, as well as mass and magnification maps, for the cores of the six \hst\ Frontier Fields galaxy clusters. Our parametric lens models are constrained by the locations and redshifts of multiple image systems of lensed background galaxies. We use a combination of photometric redshifts and spectroscopic redshifts of the lensed background sources obtained by us (for Abell 2744 and Abell S1063), collected from the literature, or kindly provided by the lensing community. Using our results, we (1) compare the derived mass distribution of each cluster to its light distribution, (2) quantify the cumulative magnification power of the HFF clusters, (3) describe how our models can be used to estimate the magnification and image multiplicity of lensed background sources at all redshifts and at any position within the cluster cores, and (4) discuss systematic effects and caveats resulting from our modeling methods. We specifically investigate the effect of the use of spectroscopic and photometric redshift constraints on the uncertainties of the resulting models. We find that the photometric redshift estimates of lensed galaxies are generally in excellent agreement with spectroscopic redshifts, where available. However, the flexibility associated with relaxed redshift priors may cause the complexity of large-scale structure that is needed to account for the lensing signal to be underestimated. Our findings thus underline the importance of spectroscopic arc redshifts, or tight photometric redshift constraints, for high precision lens models.

All products from our best-fit lens models (magnification, convergence, shear, deflection field) and model simulations for estimating errors are made available via the Mikulski Archive for Space Telescopes. 

\end{abstract}


\section{Introduction}

Our knowledge of galaxy formation, galaxy evolution, and universal star formation history from shortly after the Big Bang to the present era has greatly increased through deep imaging surveys with the \emph{Hubble Space Telescope} (\hst). These surveys include the Hubble Deep Field \citep{Williams:1996vn}, its successive iterations \citep{Beckwith:2006rt, Bouwens:2011ys, Ellis:2013fr, Illingworth:2013yq}, and the Cosmic Assembly Near-IR Deep Extragalactic Legacy Survey \citep[CANDELS;][]{Grogin:2011ly,Koekemoer:2011fk}. One of the ``final frontiers" for the \hst\ is the detection of the first galaxies to have formed after the Big Bang at $z>8$, the factories that formed the first stars responsible for reionizing the universe \citep{Finkelstein:2012bh}. These deep field surveys have so far discovered a handful of the \emph{brightest} galaxies at these redshifts \citep{Ellis:2013fr,Oesch:2013lq}. Nevertheless, even with the exquisite resolution and depth of \hst, current studies are reaching the limit for finding the \emph{typical} luminosity galaxy at these redshifts; their detectability is affected by cosmic dimming of surface brightness, and most of the redshifted spectral energy distribution we can detect falls in the reddest photometric bands of optical and infrared telescopes. By combining the power of \hst\ resolution with the magnification boost from gravitational lensing, we can overcome some of those observational barriers and attempt to reach survey depths comparable to those of the \emph{James Webb Space Telescope} and 30-meter telescopes, several years before they come online.

Using director's discretionary time as part of the Hubble Deep Field Initiative, the \hst\ Frontier Fields program (HFF; PI: J. Lotz, \hst\ PID 13498) will image four (and up to six) strong lensing galaxy clusters and nearby parallel fields and use them as cosmic telescopes to observe the high redshift universe. Over the next three years, each of these clusters will be imaged in seven photometric bands with ACS and WFC3/IR for a total of 140 orbits per cluster, achieving a depth of 28-29 magnitudes across all bands, much deeper than any cluster observed by \hst\ to date. Typical lensing magnifications for background galaxies in these fields may be $\mu\sim\times2-5$, which increases the depth by an additional $2.5\log_{10}\mu\simeq0.8-1.7$ magnitudes. Therefore, much of the fields can exceed an equivalent depth of 30 magnitudes with the aid of lensing. The HFF will uncover the faint end of the luminosity function at $z\sim8$ and detect the most distant galaxies at $z>10$, the primordial predecessors to today's $L^\star$ galaxies.

The added magnification factor from the lensing clusters can influence the derived physical properties of galaxies (i.e., source luminosity, size, stellar mass, star formation rate) and can thus have an effect on the number counts and resulting luminosity functions.\footnote{For more information, see the gravitational lensing primer for HFF users (\url{http://archive.stsci.edu/prepds/frontier/lensmodels/webtool/ hlsp\_frontier\_model\_lensing\_primer.pdf}).} Proper analysis of the galaxies in these fields will require a deep understanding of the intervening lensing system, thus adding a layer of complexity compared to the analysis of ``blank" (i.e., not lensed) fields. To enable the analysis of the lensed sources as soon as the HFF observations become available, five independent teams were chosen to create preliminary lens models (this work included) for each cluster based on existing archival \hst\ imaging and ground-based data. These models of the lensing potential, derived from a variety of lens modeling techniques, are publicly available to the community and can be used to compute the magnification factor at any location in the HFF field-of-view (FOV), for any source redshift.

In this paper, we present our parametric, strong-lensing models of the HFF clusters and new spectroscopic measurements of lensed galaxies. In \S \ref{sec:HSTimaging}, we discuss the archival \hst\ imaging used to identify galaxy cluster members and systems of multiple images for the modeling process. We describe in \S \ref{sec:magellan} our spectroscopic redshift measurements of lensed galaxies in two of the HFF clusters. In \S \ref{sec:models}, we describe our lens modeling methods, the publicly available lens model products (i.e. convergence, shear, deflection, magnification maps), and mass measurements computed from the lens models. We present the detailed results of the lens models for each cluster in \S \ref{sec:results} -- the image constraints and their redshifts, the selection of cluster member galaxies, and choice of halos and parameters to optimize in each model, and we compare the mass measurements to those in the literature. In \S \ref{sec:discussion}, we discuss the use of the models, modeling precision, the importance of including spectroscopic redshifts in lens modeling, and our plan for future model revisions based on the Frontier Fields data. The appendix shows the lensed arc spectra and lists the model constraints and best-fit parameters.

Throughout this paper, we assume $\Lambda$CDM cosmology with $\Omega_M=0.3$, $\Omega_\Lambda=0.7$, and $H_0=70\ \mathrm{km\ s^{-1}\ Mpc^{-1}}$. All magnitudes are reported in the AB system.

\section{Archival \emph{Hubble Space Telescope} Imaging}
\label{sec:HSTimaging}

The HFF clusters are Abell 2744 ($z=0.308$), \MACSzerofour\ ($z=0.396$), \MACSzeroseven\ ($z=0.545$), \MACSeleven\ ($z=0.543$), Abell S1063 ($z=0.348$), and Abell 370 ($z=0.375$) \citep{Babyk:2012fj,Mann:2012fr,Ebeling:2007ys,Ebeling:2001rt,Bohringer:2004wd,White:2000nx,Struble:1987cr,Abell:1989ly,Abell:1958mz}. These clusters were chosen as HFF targets from a list of clusters with large lensing cross-sections and existing archival \hst\ imaging data, and satisfy several criteria (low zodiacal light, observability from the Atacama Large Millimeter Array, ability to fit within WFC3 FOV, etc.\footnote{The complete list of criteria for the selection of the HFF clusters, along with their explanations, can be found on the Frontier Fields website: \url{http://www.stsci.edu/hst/campaigns/frontier-fields/frontier-fields-high-magnification-cluster-candidate-list/}.}), which will ensure the highest quality and deepest \hst\ data and enable further studies at other wavelengths to obtain a very useful panchromatic dataset for studying these clusters and their background populations.

Multi-band archival images were used to identify galaxy cluster members and multiply imaged galaxies as inputs and constraints in the lens modeling process.  Four of the clusters, \MACSzerofour, \MACSzeroseven, \MACSeleven, and Abell S1063, were observed as part of the multi-cycle program Cluster Lensing and Supernova Survey with Hubble in 16 bands and 20 orbits each \citep[CLASH; PI: M. Postman; see][]{Postman:2012lr}. Additionally, \MACSzeroseven\ and \MACSeleven\ were observed in F555W and F814W (GO-9722, PI Ebeling), and \MACSzeroseven\ was covered in a $3\times6$ ACS mosaic in F606W and F814W (GO 10420, PI Ebeling). Abell 2744 was observed with \hst\ ACS in F435W (6 orbits) and F606W and F814W (5 orbits depth in each) as part of GO program 11689 \citep[PI: R. Dupke; ][]{Merten:2011fk}. The mosaics of Abell 370 consist of both ACS (F435W, F606W, and F814W) and WFC3 (F110W, F140W, and F160W) archival observations obtained in  as part of \hst\ SM4 ERO program 11597 (PI: K. Noll), program 11591 \citep[PI: J.-P. Kneib;][]{Richard:2010wd}, program 11582 (PI: A. Blain), and program 11108 \citep[PI. E. Hu; ][]{Cowie:2011lr}. 

The archival \hst\ imaging of all six clusters are uniformly reduced onto a common frame with $0\farcs03\ \mathrm{pixel^{-1}}$ using the ``MosaicDrizzle" pipeline \citep[Koekemoer et al 2002, ][for further details]{Koekemoer:2011fk}. The astrometry of each field has been corrected and linked to a common reference frame, to match guide star catalogs that are used for supporting ground observations. The high-level science products of these data are publicly available through the Mikulski Archive for Space Telescopes (MAST).\footnote{\url{http://archive.stsci.edu/prepds/frontier/lensmodels/}}

\section{New redshift measurements of lensed galaxies}
\label{sec:magellan}

The precision of strong lensing models depends on our ability to constrain the deflection field tensor in the image plane with the locations of multiply imaged galaxies. The deflection scales linearly with the distance to the source. Therefore, it is critical to obtain precise redshift measurements for the image systems used as strong lensing constraints. Without knowing the source redshift, lens models are vulnerable to several degeneracies between the mass distribution and the redshifts of the multiply imaged galaxies. Photometric redshifts can be used to narrow these uncertainties, but often have large uncertainties themselves. Spectroscopic redshifts fix the three-dimensional location of the source galaxy under the assumption that the cosmological parameters are measured elsewhere, constraining the mass distribution enclosed by the source galaxy's multiple images in the image plane.

Obtaining spectroscopic redshifts for these faint sources can be a challenge due to cosmic dimming of their surface brightness, but this effect can be countered with high lensing magnification. The background galaxies typically tend to be actively star forming, as  typical giant arc redshifts are $z\sim1-3$ \citep{Bayliss:2012rt,Bayliss:2011gf,Livermore:2012mz}, near the epoch of peak universal star formation, and are likely to have strong emission line spectra generated by the winds of high mass stars. Redshift measurements of lensed galaxies can be obtained within a few hours of integration at a large telescope for emission line objects, while measurements obtained from absorption features require longer integrations to attain high signal-to-noise of the galaxy continuum.

At the time the HFF clusters were selected, only a few spectroscopic redshifts were known per cluster \citep{Zitrin:2013lr,Limousin:2012fj,Smith:2009lr,Richard:2010wd}. Extensive efforts are underway to measure redshifts spectroscopically from the ground, including this work (see also \citealt{Richard:2014gf,Monna:2014lr}). Here we summarize our spectroscopic observations and redshift measurements for two of the HFF clusters: Abell S1063 and Abell 2744.

\subsection{Observations and data reduction}

\subsubsection{Magellan/LDSS3 spectroscopy}
Abell S1063 was observed on UT 11 August 2013 with the Low Dispersion Survey Spectrograph 3 (LDSS3)\footnote{\url{http://www.lco.cl/telescopes-information/magellan/instruments/ldss-3}} at the Magellan-II telescope under University of Michigan time (PI: K. Sharon). We designed a custom mask for LDSS3, placing 1\arcsec\ slits on lensed galaxy candidates surrounding the cluster selected from the archival \hst\ imaging. The remaining slit positions for the mask covering sky outside the \hst\ FOV were selected from an object catalog created from wide-field pre-imaging for mask design with the \emph{Very Large Telescope} VIsible MultiObject Spectrograph (VIMOS) (Piero Rosati, priv. comm.) We selected slit positions for the Abell S1063 mask from a photometric catalog that was made available to lens modeling teams through the HFF initiative based on $i'$ band Suburu SupremeCam imaging of the cluster \citep{Merten:2011fk}. We ensured that selected slits exclude objects observed with VIMOS, except for lensed low-surface-brightness galaxies for which redshift measurement could be expected to be extremely challenging with either telescope. The VPH-ALL grism (400 lines $\mathrm{mm^{-1}}$) was used with no order-blocking filter to allow for the largest wavelength coverage and maximum throughput -- $\Delta\lambda=4000-9800$\AA\ with $R\sim900$. The field of Abell S1063 was exposed for $6\times2400$s and $1\times1570$s for a total of 266 min, beginning at UT 02:14 as it was rising to peak altitude from airmass 1.5 to 1.04 with seeing varying between 0\farcs6-0\farcs9.

\subsubsection{Magellan/IMACS spectroscopy}
We observed Abell 2744 with the Inamori Magellan Areal Camera and Spectrograph \citep[IMACS; ][]{Dressler:2011dq} at the Magellan-I telescope under University of Michigan time (PI: K. Sharon) during the nights of UT 2013 August 2-3 and UT 2013 November 8. We designed a custom mask with similar techniques as the Abell S1063 mask. The remaining slit positions for the mask covering sky outside the \hst\ FOV were selected from an astrometric catalog that was made available to lens modeling teams through the HFF initiative based on $i'$ band Subaru SupremeCam imaging of the cluster \citep{Merten:2011fk}. We used the f/2 camera, 200 lines $\mathrm{mm^{-1}}$ grism (15 deg blaze), and no order-blocking filter to ensure the broadest wavelength coverage and maximum throughput. With our observing setup and an unbinned detector, the data have a spectral resolution $R\sim550-1200$ over the wavelength sensitivity range $\Delta\lambda=4400-9800$ \AA.

We observed Abell 2744 on UT 2013 August 3, 4 for a total exposure time of 308 minutes under clear conditions with seeing around 0\farcs6-1\farcs5 and at airmass 1.004-1.147. We resumed observations of this cluster on UT 2013 November 8 for a total of 160 minutes, again under clear conditions, seeing 0\farcs7-0\farcs8, and at airmass 1.005-1.230. The total combined exposure time between the two observing runs is 468 minutes (7.5 hours).

\subsubsection{Data reduction}
We used the COSMOS\footnote{\url{http://code.obs.carnegiescience.edu/cosmos}} reduction package to calibrate wavelengths, bias-subtract, flat-field, sum exposures, and reject cosmic rays from the two-dimensional spectra of each slit for all IMACS and LDSS3 masks. As the purpose of our observations was redshift measurements, the data have not been flux calibrated. The one-dimensional spectra were extracted from the 2D data using custom IDL scripts. We rely on both the 1D and 2D spectra as well as color information and photometric redshifts (from CLASH \hst\ imaging, see below) to determine the redshifts from the galaxy spectra.

\begin{deluxetable}{lcccc}
\tablecaption{Abell S1063 new redshift measurements}
\tabletypesize{\scriptsize}
\tablehead{
    \colhead{image \#} &
    \colhead{$\alpha$ (J2000.0)} &
    \colhead{$\delta$ (J2000.0)} &
    \colhead{$z$} &
    \colhead{$\sigma_z$}
}
\startdata
1.3 & 22:48:44.724 & -44:31:16.09 & 1.229 & 0.004 \\
2.1/2 & 22:48:46.205 & -44:31:49.93 & 1.260 & 0.004 \\
2.3 & 22:48:43.138 & -44:31:17.75 & 1.260 & 0.004 \\
5.2 & 22:48:45.074 & -44:31:38.40 & 1.398 & 0.004 \\
5.3 & 22:48:46.354 & -44:32:11.61 & 1.398 & 0.004 \\
6.1/2 & 22:48:41.808 & -44:31:41.88 & 1.429 & 0.005 \\
6.3 & 22:48:45.235 & -44:32:23.93 & 1.429 & 0.005 \\
11.1 & 22:48:42.012 & -44:32:27.76 & 3.117 & 0.004 \\
A & 22:48:42.161 & -44:32:44.30 & 1.269 & 0.004 \\
B & 22:48:38.179 & -44:32:20.25 & $>1.6$ & \nodata \\
C & 22:48:48.480 & -44:31:22.01 & $>1.6$ & \nodata \\
D & 22:48:47.292 & -44:30:47.85 & $>1.6$ & \nodata
\enddata
\label{tab:as1063_z}
\tablecomments{Lower limits on redshifts are set by the absence of [OII] 3726,3729 \AA\ emission line at $<9800$ \AA. The reported spectroscopic redshifts are consistent with the galaxy colors and photometric redshift measurements.}
\end{deluxetable}

\begin{deluxetable}{lccccc}
\tablecaption{Abell 2744 new redshift measurements}
\tabletypesize{\scriptsize}
\tablehead{
    \colhead{image \#} &
    \colhead{$\alpha$ (J2000.0)} &
    \colhead{$\delta$ (J2000.0)} &
    \colhead{$z$} &
    \colhead{$\sigma_z$} &
    \colhead{confidence}
}
\startdata
1.1		& 00:14:23.414 & -30:24:14.22 & $>1.6$	& \nodata & \nodata \\
1.3		& 00:14:20.576 & -30:24:36.44 & $>1.6$ 	& \nodata & \nodata \\
2.1		& 00:14:19.890 & -30:24:11.71 & 2.2 	& \nodata	& possible \\
2.4		& 00:14:20.635 & -30:24:08.09 & $>1.6$ 	& \nodata & \nodata \\
3.1/2		& 00:14:21.459 & -30:23:37.71 & 3.98 	& 0.02 	& secure \\
4.5		& 00:14:22.559 & -30:24:17.10 & 3.579 	& 0.005 & secure \\
6.3		& 00:14:20.701 & -30:24:34.32 & $>1.6$ 	& \nodata & \nodata \\
10.3		& 00:14:24.171 & -30:23:49.59 & 2.735 	& \nodata	& likely 
\enddata
\tablecomments{Lower limits on redshifts are set by the absence of [OII] 3726,3729 \AA\ emission line at $<9800$ \AA. Possible and likely redshifts are based on low signal-to-noise features. The reported spectroscopic redshifts are consistent with the galaxy colors and photometric redshift measurements.}
\label{tab:a2744_z}
\end{deluxetable}

\subsection{Results}

We report spectroscopic measurements of 16 lensed galaxies in the fields of Abell S1063 and Abell 2744 as detailed below.  The redshift measurements of other galaxies (cluster members and field galaxies that are not strongly lensed) will be reported on in a future publication.

\citet{Richard:2014gf} observed Abell S1063 and Abell 2744 and measured spectroscopic redshifts of several multiply imaged galaxies. These redshifts were generously shared prior to publication with the lens modeling teams as part of the preliminary lens modeling effort, and were used in producing version 1 of the lensing models that were released to the public in September 2013. Where available, we explicitly compare our results to those of \citet{Richard:2014gf}.

We also compare our measured spectroscopic redshifts to the photometric redshifts from the CLASH Bayesian Photometric Redshift \citep[BPZ;][]{Benitez:2000uq,Benitez:2004fj,Coe:2006qy} measurements for Abell S1063 and the new estimates of photometric redshifts for Abell 2744 that were produced from BPZ for the purpose of the HFF preliminary lens models. 

\subsubsection{Abell S1063 lensed galaxy redshifts}

We report redshift measurements for nine of 12 targeted lensed galaxies that were targeted. Table \ref{tab:as1063_z} lists the redshifts and errors for each of these targets.

\textit{Image 1.3}: We measure $z=1.23$ for image 1.3, which we confirm as a faint counter image of the main arc, images 1.1 and 1.2 \citep[$z=1.229$,][]{Richard:2014gf}. We identify the single, bright emission feature at $\sim8300$~\AA\ to be [O II] 3726,3729~\AA\ emission. The one-dimensional and two-dimensional spectra for image 1.3 are shown in Figure \ref{fig:as1063_spec1_3} in the Appendix.

\textit{Images 2.1/2, 2.3}: The [O II] emission line appears in both the spectra (see Figure \ref{fig:as1063_spec2} in the Appendix) of the merging pair of images 2.1 and 2.2 and the counter image 2.3 at $\sim8420$~\AA, thus providing spectroscopic confirmation that they are all images of the same background source at $z=1.26$. Our measurements are consistent with those of  \citet{Richard:2014gf}, who measured $z=1.260$ for one of the three arcs, image 2.3.

\textit{Images 5.2 and 5.3}: We measure [O II] emission at $\sim8940$~\AA\ for both images, confirming these galaxies as images of the same source at $z=1.39$, as shown in Figure \ref{fig:as1063_spec5} in the Appendix. Our redshift measurements are consistent with the spectroscopic redshift for 5.2 by  \citet{Richard:2014gf}.

\textit{Images 6.1 and 6.3}: We identify the emission line at $\sim9050$~\AA\ in the spectra (see Figure \ref{fig:as1063_spec6} in the Appendix) of both images as [O II] at $z=1.43$. This result is consistent with the redshift measurement of image 6.1 by  \citet{Richard:2014gf}. We confirm the identification of image 6.1 and 6.3 as images of the same source.

\textit{Image 11.1}: We identify image 11.1 as an image of a Lyman $\alpha$-emitting galaxy at $z=3.117$. This spectroscopic redshift is the first measurement for this galaxy and for this multiply imaged system. The Ly$\alpha$ 1215~\AA\ line redshifted to 5006 \AA\ in the spectrum of image 11.1, as shown in Figure \ref{fig:as1063_spec11_1} in the Appendix, is the only emission line we can confirm in the wavelength range $4000-9800$~\AA. We note that this result is consistent with $\zphot=3.08\pm0.1$ (95\% confidence level) measured from CLASH imaging \citep{Jouvel:2014qy}. There is weak detection for C IV 1549 \AA\ emission, however, at much lower signal-to-noise ($\sim2$). We base our redshift measurements on the Ly$\alpha$ line only.

\textit{Other lensed galaxies, in the HST FOV outside the cluster core}: We place slits on four candidate lensed galaxies, which have not been identified as part of multiply imaged systems. The object we label in Table \ref{tab:as1063_z} as arc A has an emission feature at $\sim8450$~\AA\ which we identify as [O II] at $z=1.27$ (see Figure \ref{fig:as1063_specA} in the Appendix). Our spectroscopic redshift measurement is consistent with the measured CLASH photometric redshift, $\zphot=1.12^{+0.16}_{-0.03}$. No emission features seen in the spectra of arcs B, C, and D in the wavelength range $4000-9800$~\AA. These galaxies may lie in the redshift desert where few emission lines are shifted to wavelengths accessible from ground based observatories. Based on the absence of [O II] emission, we place a lower limit on the redshift of these galaxies $z>1.6$. The BPZ measurements for arcs B, C, and D ($\zphot\sim2.7,2.4,\mathrm{and}\ 1.6$, respectively) are consistent with this limit.

\subsubsection{Abell 2744 lensed galaxy redshifts}

We report two secure redshifts and two possible redshifts of eight strongly lensed galaxies in Abell 2744. One of the newly-measured galaxy redshifts in this cluster is a new redshift measurement of the entire image system of the background source galaxy. We place lower limits on the redshifts of lensed galaxies for which we cannot determine redshifts due to the lack of spectral features. These redshift measurements and errors are listed in Table \ref{tab:a2744_z}.

\textit{Image 2.1}: We report a possible low-confidence solution of $z=2.2$ based on the strong absorption features at 4890 and 4950 \AA\ corresponding to Si II 1527~\AA\ and C IV 1549~\AA, and a possible low signal-to-noise emission line at 6110~\AA\ matching C III] 1909~\AA. This solution is consistent with the BPZ range ($1.1<\zphot<2.4$); however, the Bayesian probability distribution for the photometric redshift is bimodal, with peaks around $\zphot\sim1.2,2.2$. The one-dimensional and two-dimensional spectra of this galaxy are shown in Figure \ref{fig:a2744_spec2} in the Appendix. Spectral templates of Lyman break galaxies at $z=3$ from \citet{Shapley:2003fk} are plotted with the spectra to guide the eye to common spectral features in these galaxies. As the data are not flux calibrated, it is expected that the shape of the continua of the spectra and templates may not match precisely.

\textit{Image 3.1/2}: We measure $z=3.98$ for the merging pair of images 3.1/2 based on the detection of several absorption line features (Si II 1260~\AA, O I + Si II 1302,1304~\AA, C II* 1334~\AA, and the Si IV doublet 1393,1402~\AA) and the Ly $\alpha$ break around 6100~\AA\ (see Figure \ref{fig:a2744_spec3} in the Appendix). The spectroscopic redshift is consistent with the photometric redshift obtained from BPZ, $\zphot=4.02^{+0.4}_{-0.1}$ (95\% confidence range) for image 3.1 and $\zphot=3.96^{+0.4}_{-0.2}$ for image 3.2. This is the first spectroscopic redshift measurement for this system of lensed images mapping to the same source galaxy.

\textit{Image 4.5}: We detect an emission line around 5575~\AA, which may correspond to Ly $\alpha$ and a redshift of $z=3.58$; however, this emission line is close to the sky line at 5577~\AA, and with the resolution of our spectrum, it is difficult to distinguish the line from the residuals of the sky subtraction (see Figure \ref{fig:a2744_spec4} in the Appendix). We also find a feature around 7100~\AA\ possibly corresponding to C IV 1550~\AA.  \citet{Richard:2014gf} measure a spectroscopic redshift of $z=3.58$ for the counter image of this galaxy, image 4.3. The agreement between the spectroscopic redshifts of these two objects confirms their identification as multiple images of the same background source.

\textit{Image 10.3}: We find a likely redshift solution of $z=2.735$ for image 10.3 based on the two absorption features around 5600 and 5700~\AA, which we identify as Si II 1527~\AA\ and C IV 1550~\AA\ (see Figure \ref{fig:a2744_spec10} in the Appendix). This solution matches with the possibility of Ly$\alpha$ emission at 4542~\AA\  and several likely absorption lines (O I + Si II 1302,1304~\AA, C II* 1334~\AA, the Si IV doublet 1393,1402~\AA). The photometric redshift distribution for this galaxy covers this redshift value, $\zphot=3.08^{+0.10}_{-0.46}$.

\textit{Other lensed galaxies, in the HST FOV outside the cluster core}: The spectra of images 1.1, 1.3, 2.4, and 6.3 did not yield redshift measurements, as no emission or absorption features are identified over the wavelength range $\Delta\lambda=4400-9800$ \AA. Based on the absence of emission lines and assuming they are likely star-forming, we can place a lower limit $z>1.6$ for these galaxies.


\section{Lens models}
\label{sec:models}
Since we applied a uniform modeling procedure to all the fields in this work, we outline the methods and assumptions in this section and follow with the results, constraints, and assumptions that are more specific to each cluster in \S \ref{sec:results}.

Our lens modeling method is parametric in nature. We use the publicly available software \texttt{LENSTOOL} \citep{Jullo:2007lr}, which utilizes a Markov Chain Monte Carlo (MCMC) to find the best-fit model parameters weighted by Bayesian evidence. The modeling is done in an iterative process. We start by placing masses typical of most clusters near the center of the light distribution, and then build up the model in complexity with each iteration by adding more image constraints and more mass components. The process continues until all the image constraints  have been included, and the model rms can no longer improve significantly by adding another halo. The early iterations are done under source plane optimization, where the rms scatter used to rank models is computed when the images are traced back to the source plane. The source plane optimizations serve as an adequate starting point. The final model was computed under image plane optimization, where the rms scatter is computed for each image by tracing through the lens back to the source plane and back out to the image plane. The latter method is more computationally intensive, and thus is not carried out until the final iteration.

\subsection{Modeling constraints: multiply imaged galaxies}

The HFF clusters were selected based on their known large lensing cross sections. As such, these clusters are rich in lensing constraints in the form of multiply lensed galaxies, and have been previously studied \citep{Zitrin:2013lr,Zheng:2012fk,Limousin:2012fj,Merten:2011fk,Zitrin:2011qy,Richard:2010wd,Smith:2009lr,Zitrin:2009qy,Zitrin:2009kx}. As part of the lens modeling initiative, teams agreed upon a list of identified multiply imaged galaxies that were either published or identified by one or more of the teams and were shared with all the teams. In what follows, we refer to unpublished information that was contributed to this collaborative effort as ``private communication." We generally rely on these image identifications, but modify slightly the exact image position for better centering on identical morphological features in multiple images, as explained below. 

For each lens model, we use the families of multiply imaged background galaxies to constrain the distribution of the lensing mass in the clusters. We fix the redshifts of the source galaxies to the spectroscopic redshifts of the images whenever available. Image families which have been identified by color and morphology but may not have been spectroscopically confirmed are included in the models with redshifts left as free parameters. We use BPZ photometric redshifts for these galaxies as Bayesian priors. For the clusters with CLASH data (\MACSzerofour, \MACSzeroseven, \MACSeleven, and Abell S1063), we use the photometric redshifts computed by \citet{Jouvel:2014qy}. We compute BPZ redshift measurements for the images in Abell 2744 and Abell S1063 from preliminary \hst\ imaging. We set the redshift prior in the lens models to the 95\% confidence interval of the BPZ probability distribution. The majority of free-redshift image systems in our lens models converge to redshift values within this range. A handful of image systems do not converge to redshift solutions within the photometric redshift range, which could be the result of either incorrect photometric redshifts, from bimodal or irregular BPZ probability distributions, or lensing by substructure along the line of sight to these images unaccounted for in our models. In these circumstances, we relax the range of redshifts on the last iteration of the model to allow those image systems to converge on best-fit redshift values. 

The positional constraints from the multiply imaged galaxies are set by eye, by identifying and matching the same distinguishing features in each image that map to the same galaxy. The error in position with this method ($\lesssim1$ pixel) is much smaller than the error induced by small unseen line of sight substructure that is not accounted for in the model, for which we assign a positional error of $0\farcs3$ \citep[following][]{Jullo:2007lr,Limousin:2007fk}.\footnote{We also ran an identical set of models using a positional error of $0\farcs6$ and found very little change in the final image plane rms and best fit model parameters between the two iterations (i.e., the parameter values are within the statistical errors due to random sampling).} Moreover, identifying features by eye works better than automated measurement in scenarios where images are close to the critical curves and the magnification gradient across the image is large, in which case the light baricenters from each image may not map to the same physical location in the source galaxy.

A list of the arcs and their redshifts (spectroscopic, photometric, and model-derived with priors) is given in Tables \ref{tab:a2744_arcs}-\ref{tab:a370_arcs} in the Appendix.

\subsection{Lens model components}

Our lens models include both cluster-scale halos and halos assigned to red-sequence cluster member galaxies, all represented by pseudo-isothermal elliptical mass distributions \citep[PIEMD;][]{Jullo:2007lr,Limousin:2005cr}. The PIEMD is parameterized by a two-dimensional location in the lens plane, a lens plane redshift, ellipticity and position angle, a fiducial velocity dispersion, core radius, and cut radius. The cut radius for these cluster-scale halos is much larger than the strong-lensing regime ($<100\arcsec$ of fiducial center of cluster) and cannot be constrained in the model, so we fix the cut radius arbitrarily at 1500 kpc. Unless otherwise noted below, all of the other parameters of these cluster-scale halos are left as free parameters.

The cluster member galaxies used in each model are selected by color; those falling on the red-sequence at the cluster redshift are considered members. The PIEMD halo for each galaxy is scaled by flux relative to the light of an $\Lstar$ galaxy at the cluster redshift according to the relationships in \citet{Limousin:2005cr}. To limit the number of galaxy-scale halos in the model, we exclude faint cluster galaxies at the outskirts of the cluster; we impose a selection criterion based on a combination of brightness and projected distance from the cluster core, so that the deflection caused by an omitted galaxy is much smaller than the typical uncertainty due to unseen structure along the line of sight. We set the parameters of an $\Lstar$ galaxy to $\sigma_0^\star=120\ \mathrm{km\ s^{-1}}$, $\rcore^\star=0.15\ \mathrm{kpc}$, $\rcut^\star=30\ \mathrm{kpc}$, following \citet{Limousin:2008lr}. In most cases, these parameters cannot be constrained by the lens model, since these halos have small, localized effects on the lensing potential. When the lensed galaxies in the image plane are in close proximity to a single cluster galaxy, we allow the lensing evidence to constrain the parameters of that galaxy, using the scaling relationship as a first guess.

In a few cases, we include foreground galaxies in the model when the lensing evidence indicates that they significantly perturb the lensing potential. Because of the degeneracy between lens mass and lensing geometry, we can solve for these galaxies' contributions to the deflection at a fiducial lens redshift (however, their masses may not be reliably derived by the model), and for simplicity, place these galaxies at the cluster redshift to restrict the lensing mass to a single redshift plane. We stress here that this is an approximation that ignores minor effects of multiple lensing planes \citep[see, e.g.,][]{McCully:2014lr,DAloisio:2013ul}.

We discuss the details of each cluster lens model in \S \ref{sec:results} and list the model parameters and priors for each cluster in Tables \ref{tab:a2744_params}-\ref{tab:a370_params} in the Appendix. A summary of the model results is given in Table \ref{tab:model_summary}.

\begin{deluxetable*}{lcccc}
\tablecaption{Frontier Fields lens model summaries}
\tablehead{
& \colhead{\#} \vspace{-0.15cm} & \colhead{\# free} & \colhead{image plane} & \colhead{total \# systems} \\
\colhead{cluster} \vspace{-0.15cm} & & & & \\
& \colhead{constraints} & \colhead{parameters} & \colhead{rms (\arcsec)} & \colhead{(\# spec $z$ systems)}}

\startdata
Abell 2744		& 64		& 38		& 0.40	& 15 (3) 	\\
\MACSzerofour		& 50		& 21		& 0.51	& 15 (10)	\\
\MACSzeroseven	& 56		& 38 		& 0.38	& 14 (5)	\\
\MACSeleven		& 46		& 25		& 0.52	& 12 (3)	\\
Abell S1063		& 58		& 26		& 0.64	& 16 (6)	\\
Abell 370			& 44		& 19		& 0.82	& 9 (5)
\enddata
\label{tab:model_summary}
\end{deluxetable*}

\subsection{Magnification maps and model outputs}

To fully utilize the HFF clusters as cosmic telescopes, one needs to account for the lensing magnification. In order to provide the community with lensing magnification in the most useful form, we make all the outputs of our lens models available through MAST. We provide the magnification maps for the best-fit lens models of each HFF cluster for sources at $z=1,2,4,9$, and the convergence, shear, and deflection maps at $z=9$. The $z=9$ magnification maps are displayed in Figures \ref{fig:crit_a2744}-\ref{fig:crit_a370}. These maps give the value of the magnification $\mu$ at any location in the image plane and cover the full \hst\ ACS field of view ($300\times300\arcsec$).

Since the magnification depends non-linearly on source redshift, one can compute these maps for other redshifts using the convergence and shear. In this section, we outline the formalism of gravitational lensing needed to compute the magnifications from our lens models for any source redshift.

The magnification is computed from the second-order derivatives of the lensing potential,

\begin{equation}
\psi(\vec\theta) = \frac{2}{c^2} \frac{\dls}{\dl \ds} \int \Phi(\overrightarrow{\dl \theta},l) \mathrm{d}l,
\end{equation}

\noindent which is the projection of the Newtonian potential $\Phi$ along the line of sight, scaled by the angular diameter distances from lens to source $\dls$, observer to lens $\dl$, and observer to source $\ds$, which depend on the redshifts of the source and lens, $z_s$ and $z_l$, respectively. The translation between observed image plane position $\theta_{ij}$ and unlensed source plane position $\beta_{ij}$ is determined by the lensing Jacobian,

\begin{equation}
A_{ij} \equiv \frac{\partial \beta_{ij}}{\partial \theta_{ij}} = \delta_{ij} - \frac{\partial^2 \psi}{\partial \theta_i \partial \theta_j}.
\end{equation}

\noindent The magnification is the inverse of the determinant of this matrix,

\begin{equation}
\mu = \frac{1}{|\mathrm{det}\ A_{ij}|} = \frac{1}{|(1-\kappa)^2-\gamma^2|}, 
\end{equation}

\noindent where we have introduced the convergence $\kappa$ and shear $\gamma$, defined as

\begin{equation}
\kappa \equiv \frac{1}{2}\nabla^2 \psi, \hspace{5pt} \gamma \equiv \sqrt{\left(\frac{1}{2}\frac{\partial^2 \psi}{\partial{\theta_i}^2}-\frac{1}{2}\frac{\partial^2 \psi}{\partial {\theta_j}^2}\right)^2+\left(\frac{\partial^2 \psi}{\partial \theta_i \partial \theta_j}\right)^2}.
\end{equation}

\noindent These two tensors, $\kappa$ and $\gamma$, scale linearly with the distance fraction $f(\zl,\zs)\equiv \dls(\zl,\zs)/\ds(\zs)$. The maps of $\kappa$ and $\gamma$ we include for each cluster on MAST (at $z=9$) can be scaled to any source redshift,

\begin{equation}
\kappa(\zl,\zs) = \frac{f(\zl,\zs)}{f(\zl,9)}\kappa(\zl,9), \hspace{5pt} \gamma(\zl,\zs) = \frac{f(\zl,\zs)}{f(\zl,9)}\gamma(\zl,9),
\end{equation}

\noindent and used to compute the magnification at any desired source redshift. The errors on the magnification may also be derived from these tensors. We provide $\kappa$ and $\gamma$ maps for 100 models selected periodically from the MCMC chain, covering the same FOV as the best-fit lens model, from which to determine the distributions and errors of magnifications. The errors presented in this paper are at the 95\% confidence level from these same 100 models. The MAST website includes an online tool for computing the magnification and uncertainties from our lens models, as well as those computed by other teams.

We also provide the deflection field tensor, $\alpha$, the gradient of the lensing potential in the image plane for a given $z_s$,

\begin{equation}
\alpha_{ij} = \nabla_{ij}{\psi} = \theta_{ij}-\beta_{ij}
\end{equation}

\noindent which maps the image plane to source plane. This lens equation can have multiple solutions of $\theta$ for a single $\beta$, which corresponds to multiple images of the same source. The deflection tensor (provided as two maps, the deflection angle broken into $i$ and $j$ components), can be used to determine the multiplicity of any lensed object in the image plane and to find its counter images. The deflection angle maps are for $z=9$ and scale linearly with $\dls/\ds$ in the same manner as $\kappa$ and $\gamma$.

\subsection{Cluster masses}

\begin{deluxetable*}{lcccccc}{h}
\tablecaption{Frontier Fields cluster aperture masses}
\tabletypesize{\scriptsize}
\tablehead{
\colhead{cluster} & \colhead{cluster} & \colhead{$M(r<250\ \mathrm{kpc})$} & \colhead{$M(r<500\ \mathrm{kpc})$} & \colhead{$M(a<250\ \mathrm{kpc})$} & \colhead{$M(a<500\ \mathrm{kpc})$} & \colhead{$z$} \\
\colhead{} & \colhead{$z$} & \colhead{$(10^{14} \ \mathrm{M_\odot})$} & \colhead{$(10^{14} \ \mathrm{M_\odot})$} & \colhead{$(10^{14} \ \mathrm{M_\odot})$} & \colhead{$(10^{14} \ \mathrm{M_\odot})$} & \colhead{reference}
}
\startdata
Abell 2744 & 0.308 & $ 2.43^{+0.04}_{-0.07}$ & $6.45^{+0.22}_{-0.23}$ & $1.77^{+0.02}_{-0.04}$ & $4.90^{+0.14}_{-0.17}$ & 1 \\[4pt]
$\alpha$=00:14:21.20,$\delta$=-30:24:6.27 & & & & & & \\
$e=0.42,\theta=49.9^\circ$ & & & & & & \\ \hline
\MACSzerofour & 0.396 & $ 1.77^{+0.31}_{-0.13}$ & $4.05^{+0.90}_{-0.32}$ & $1.32^{+0.21}_{-0.09}$ & $3.29^{+0.66}_{-0.24}$ & 2 \\[4pt]
$\alpha$=04:16:8.29,$\delta$=-24:04:25.70 & & & & & & \\
$e=0.46,\theta=138.1^\circ$ & & & & & & \\ \hline
\MACSzeroseven & 0.545 & $ 2.77^{+0.07}_{-0.05}$ & $8.68^{+0.27}_{-0.13}$ & $1.77^{+0.05}_{-0.03}$ & $6.47^{+0.16}_{-0.10}$ & 3 \\[4pt]
$\alpha$=07:17:31.23,$\delta$=037:45:10.61 & & & & & & \\
$e=0.51,\theta=24.2^\circ$ & & & & & & \\ \hline
\MACSeleven & 0.543 & $ 2.57^{+0.13}_{-0.07}$ & $5.98^{+0.59}_{-0.25}$ & $2.04^{+0.07}_{-0.04}$ & $5.07^{+0.43}_{-0.18}$ & 3 \\[4pt]
$\alpha$=11:49:35.16,$\delta$=022:24:6.48 & & & & & & \\
$e=0.33,\theta=28.6^\circ$ & & & & & & \\ \hline
Abell S1063 & 0.348 & $ 2.68^{+0.03}_{-0.05}$ & $6.39^{+0.14}_{-0.32}$ & $1.88^{+0.02}_{-0.03}$ & $4.61^{+0.08}_{-0.15}$ & 4 \\[4pt]
$\alpha$=22:48:43.94,$\delta$=-44:31:52.85 & & & & & & \\
$e=0.53,\theta=143.5^\circ$ & & & & & & \\ \hline
Abell 370 & 0.375 & $ 3.48^{+0.02}_{-0.04}$ & $7.84^{+0.05}_{-0.15}$ & $2.83^{+0.01}_{-0.02}$ & $6.69^{+0.04}_{-0.11}$ & 5\\[4pt]
$\alpha$=02:39:53.10,$\delta$=-01:34:36.34 & & & & & \\
$e=0.32,\theta=82.7^\circ$ & & & & &
\enddata
\label{tab:masses_aperture}
\tablecomments{Masses are computed within sky apertures centered on $\alpha,\delta$. Aperture shape include a circle of radius $r=250,500$ kpc and an ellipse of semi-major axis $a=250,500$ kpc, ellipticity $e=(a^2-b^2)/(a^2+b^2)$, and position angle $\theta$ measured north of west. Foreground galaxies are not included in the cluster mass computation for \MACSzerofour\ and \MACSzeroseven.}
\tablerefs{(1) \citet{White:2000nx}, (2) \citet{Mann:2012fr}, (3) \citet{Ebeling:2007ys} (4) \citet{Bohringer:2004wd}, (5) \citet{Struble:1987cr}}
\end{deluxetable*}

\begin{deluxetable*}{lccc}
\tablecaption{Frontier Fields cluster critical curves}
\tabletypesize{\scriptsize}
\tablehead{
\colhead{cluster} & \colhead{mass within $z=2$ critical curve} & \colhead{area within $z=2$ critical curve} & \colhead{$\theta_{E,\mathrm{eff}}(z=2)$} \\
\colhead{} & \colhead{$(10^{14} \ \mathrm{M_\odot})$} & \colhead{(\sq\arcmin)} & \colhead{(\arcsec)}
}
\startdata
Abell 2744 & $0.62^{+0.01}_{-0.01}$ & $0.38^{+0.01}_{-0.01}$ & $20.9^{+0.2}_{-0.2}$ \\[4pt]
\MACSzerofour & $0.80^{+0.12}_{-0.06}$ & $0.58^{+0.01}_{-0.01}$ & $25.7^{+0.2}_{-0.3}$ \\[4pt]
\MACSzeroseven & $5.27^{+0.20}_{-0.11}$ & $2.19^{+0.07}_{-0.03}$ & $50.1^{+0.8}_{-0.3}$ \\[4pt]
\MACSeleven & $1.12^{+0.01}_{-0.04}$ & $0.40^{+0.01}_{-0.02}$ & $21.5^{+0.2}_{-0.4}$ \\[4pt]
Abell S1063 & $1.35^{+0.02}_{-0.03}$ & $0.77^{+0.01}_{-0.01}$ & $29.8^{+0.1}_{-0.3}$ \\[4pt]
Abell 370 & $2.36^{+0.02}_{-0.05}$ & $1.13^{+0.01}_{-0.02}$ & $36.0^{+0.1}_{-0.4}$
\enddata
\label{tab:masses_criticcurve}
\tablecomments{Masses and areas enclosed by the $z=2$ critical curves of the best-fit lens models and the effective Einstein radii for sources at $z=2$, $\theta_{E,\mathrm{eff}}=\sqrt{\mathrm{area}/\pi}$. Foreground galaxies are not included in the cluster mass computation for \MACSzerofour\ and \MACSzeroseven.}
\end{deluxetable*}

The HFF, in addition to being used as cosmic telescopes, will be helpful for understanding the construction of such galaxy clusters and exactly what about their nature makes them efficient lenses. Strong lens modeling provides us with a well constrained profile of the mass at the core of the cluster, typically on the same scale as the image plane separation between the multiply imaged galaxies. In pre-HFF imaging, the HFF clusters have on the order of a dozen image systems per cluster and are more or less evenly distributed in the image plane, which can constrain well the inner slope of the mass profile out to about 1 Mpc. The surface mass density $\Sigma$ of the lens can be computed from the convergence maps, 

\begin{equation}
\kappa = \frac{1}{2} \nabla^2 \psi = \frac{\Sigma}{\Sigma_\mathrm{crit}} = \frac{4\pi G}{c^2} \frac{D_\mathrm{L}D_\mathrm{S}}{D_\mathrm{LS}} \Sigma.
\end{equation}

\noindent The convergence $\kappa$ is defined as the surface mass density in units of the critical surface mass density $\Sigma_\mathrm{crit}\equiv c^2 D_\mathrm{LS}/4\pi G D_\mathrm{L} D_\mathrm{S}$. The $\kappa$ maps we provide on MAST are computed for $z_l$ of the cluster and $z_s=9$.

In Figures \ref{fig:crit_a2744}-\ref{fig:crit_a370}, we show the contours of the surface mass density of each cluster. We overlay these contours on the light distribution of the cluster member galaxies. We find that the projected mass distributions of the HFF clusters are highly-elongated on the sky. This elongation is common in merging clusters and makes them powerful lenses for magnifying background galaxies (e.g., \textit{the Bullet Cluster}, \citet{Clowe:2006fj,Clowe:2004uq}; MACSJ0025.5-1222, \citet{Bradac:2008kx}; Abell 520, \citet{Clowe:2012vn}; Abell 3667, \citet{Joffre:2000yq}). It is also likely, based on the number of subhalos required to construct the mass distributions, that these clusters are nodes in the cosmic web \citep{Limousin:2012fj,Jauzac:2012ab,Ma:2009gf,Ebeling:2004ai} composed of merging sub clusters, allowing for these clusters to have large cross-sections for lensing based on total mass alone.

We exclude the contributions from the foreground galaxies in the mass models when computing the masses of \MACSzerofour\ and \MACSzeroseven. The masses of the foreground galaxies are unrealistically high when placed at the same redshift plane as the galaxy cluster. The lensing deflection due to a foreground perturber depends on several factors (e.g., mass, redshift, impact parameter at the plane of the interloper), and thus its true mass cannot be easily estimated or scaled from the best-fit model parameters. More realistic masses could be estimated for these galaxies when multiple lensing planes are considered. This type of analysis is beyond the scope of this paper, and so we exclude the masses of these galaxies in Figures \ref{fig:crit_m0416} and \ref{fig:crit_m0717}.

We list the aperture masses of each cluster in Table \ref{tab:masses_aperture}. We select elliptical apertures roughly matching the shape and orientation of the contours of the mass distribution as well as circular apertures, and compute the enclosed mass as a function of aperture semi-major axis (or radius) of 250 kpc and 500 kpc. In Table \ref{tab:masses_criticcurve}, we list the masses enclosed by the tangential critical curves (analogous to the Einstein ring of circular potentials) for sources at $z=2$. We selected these methods in an attempt to make nearly direct comparisons with values derived from other published strong lensing models and to facilitate comparison to simulations. We remind the reader that direct comparisons can be done using the output files that are publicly available through MAST. Cluster mass estimates which use other mass proxies (e.g. Sunyaev-Zel'dovich effect, weak lensing, x-ray,  dynamics) often quote a virial mass or $M_{200}$ of clusters. While we can compute these values from strong lensing models, doing so requires extrapolating the strong lensing models far beyond the region of strong lensing constraints, resulting in a crude estimate of the total mass of the cluster. Our models are likely missing some of the large scale structure outside the \hst\ FOV which may not contribute significantly to the strong lensing. We explore lens model uncertainties in \S \ref{sec:precision}.

\section{Results by cluster}
\label{sec:results}

In this section we detail the lensing constraints, lens model component, and derived mass for each of the HFF clusters. Lists of constraints, priors and best-fit parameters are tabulated in the Appendix. 

\subsection{Abell 2744}
\label{sec:results_a2744}

We use the images identified by \citet{Merten:2011fk} as constraints for the lens model, supplemented by identifications made by Johan Richard (priv. comm.),\footnote{\label{note}Unpublished multiple image systems, photometric and spectroscopic redshifts were shared as part of the HFF preliminary lens modeling initiative.} as shown in Table \ref{tab:a2744_arcs} in the Appendix. We fix the redshifts of image systems \# 4 and 6 to the spectroscopic redshifts obtained by  \citet{Richard:2014gf}. We include our new spectroscopic redshift measurement of image system \#3.

\begin{figure*}[h]
\centering
\includegraphics[width=0.8\textwidth]{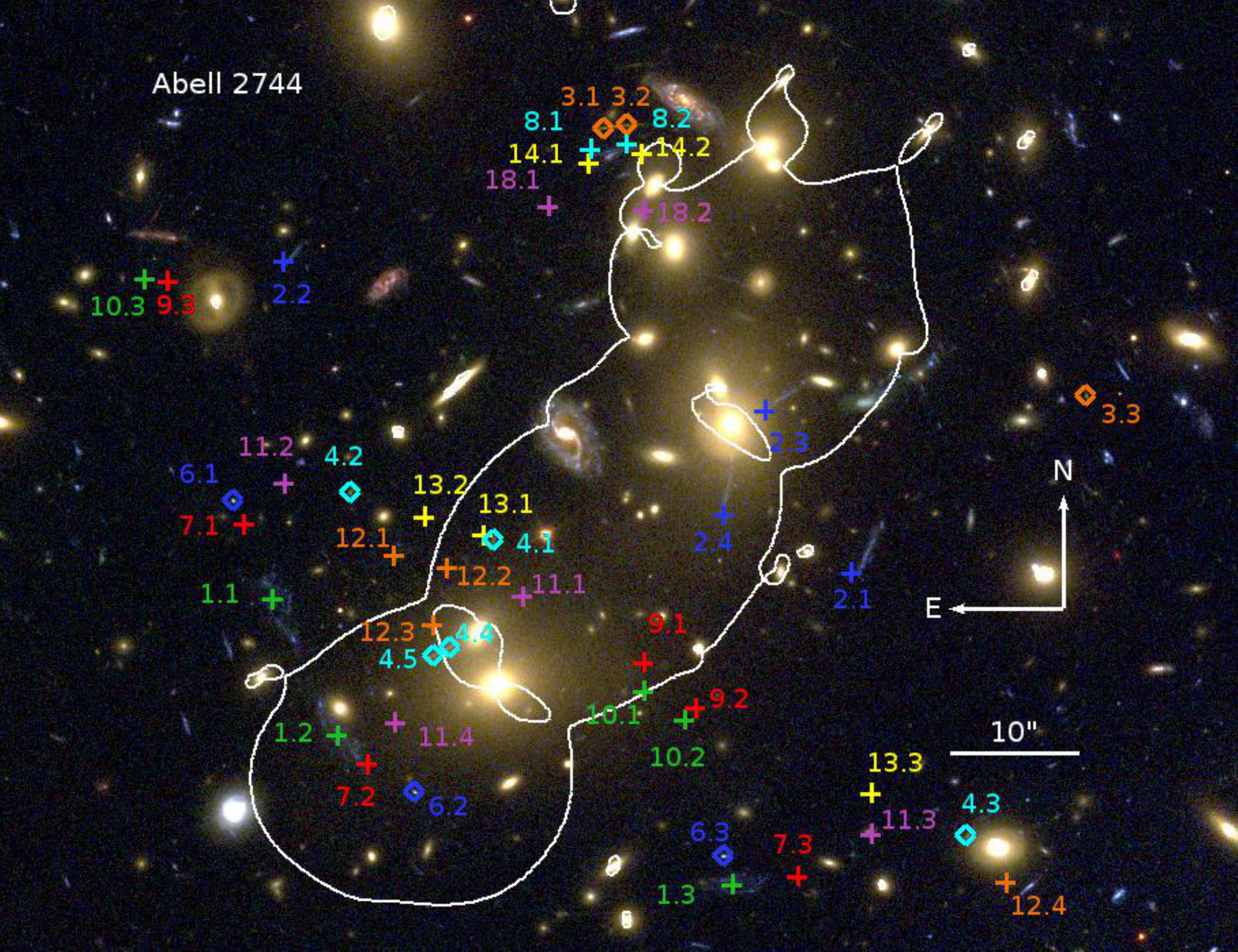} \\
\includegraphics[height=0.36\textheight]{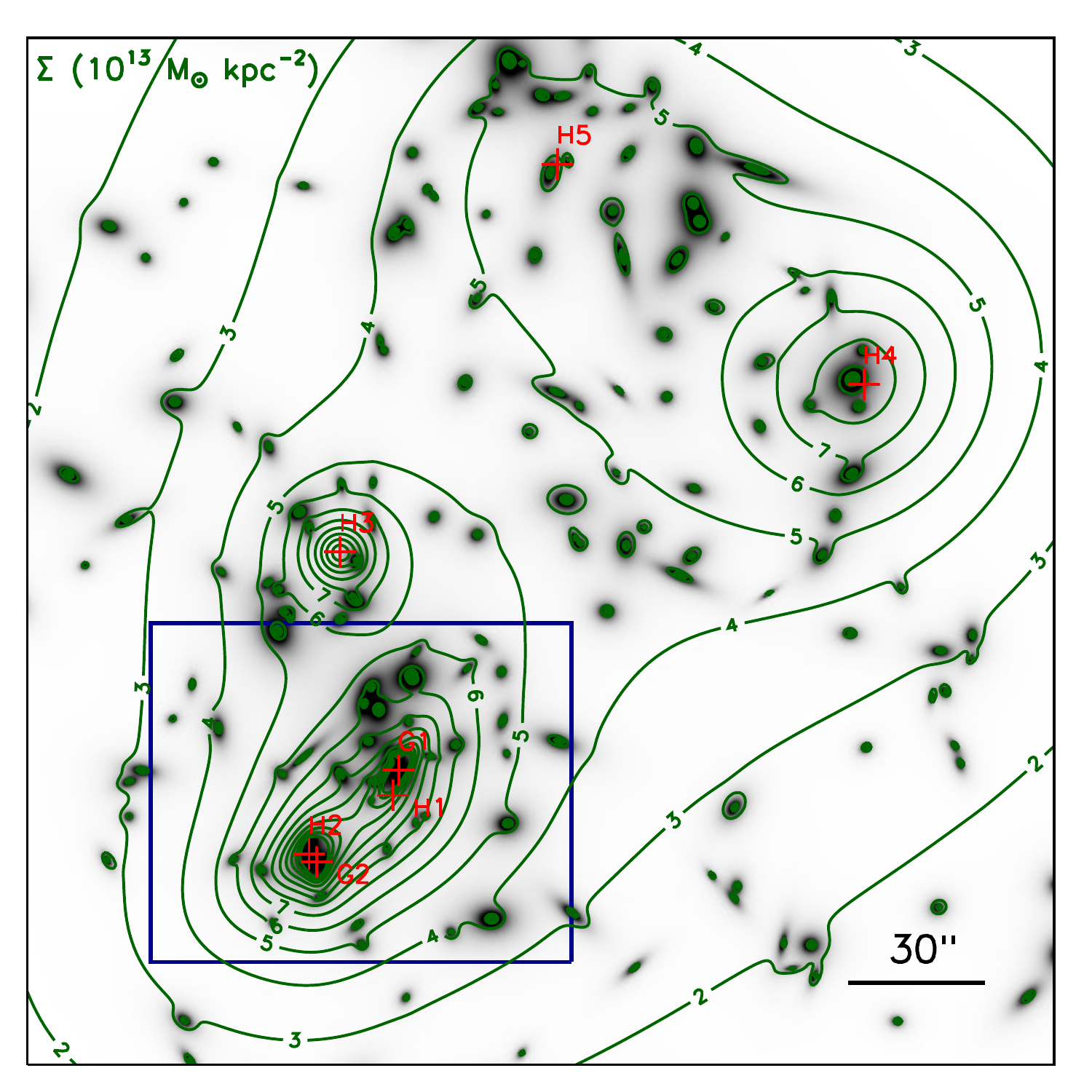}
\includegraphics[height=0.36\textheight]{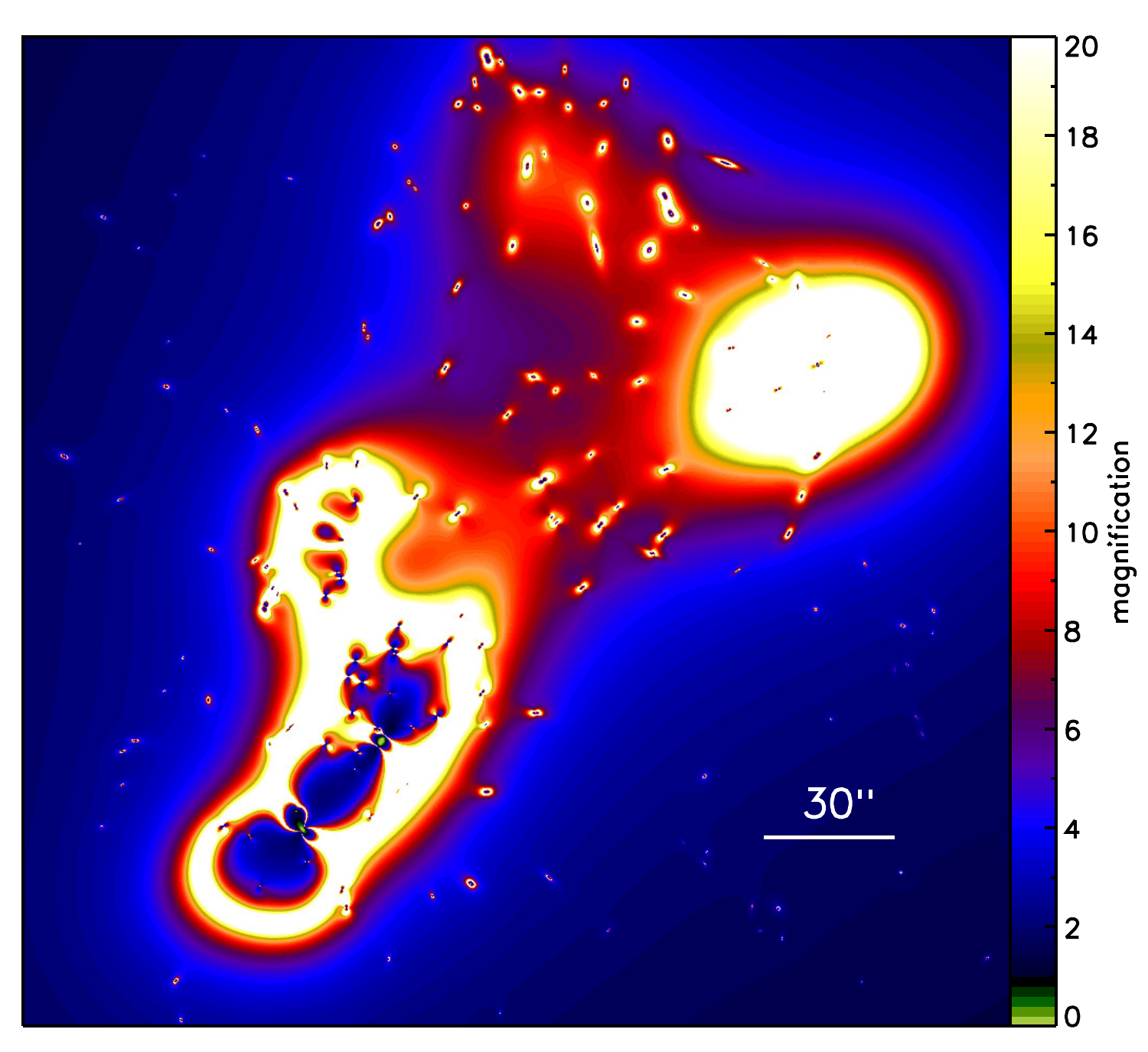}
\caption[Abell 2744 image constraints and critical curves]{Top: False color image of Abell 2744 from archival ACS imaging (red, F814W; green, F606W; blue, F435W) obtained as part of GO program 11689 (PI: R. Dupke). The locations of multiply imaged galaxies used as constraints in the model of Abell 2744 are overlaid. Diamonds designate image systems with spectroscopic redshifts, where crosses indicate that the redshift of the image system is left as a free parameter in the model. The critical curves (white) trace the region of high magnification for a source at $z=2$. Left: Contour map of the total surface mass distribution (in units of $10^{13}\ \mathrm{M_\odot \ kpc^{-2}}$) overlaid on the mass contained in cluster member galaxies. The spacing of the contours is linear. The locations and labels of each optimized halo (see Table \ref{tab:a2744_params}) are shown in red. The blue box indicates the field of view of the top image. Right: Absolute value of magnification in the image plane for a source at $z=9$.}
\label{fig:crit_a2744}
\end{figure*}

The majority of the image constraints are positioned within $<40$\arcsec of the cluster core (the region enclosed by HFF \hst\ WFC3/IR imaging), however, we include constraints for image \#16, which is about 2\arcmin\ northwest from the core to help constrain mass in that region. Since there are no known spectroscopic redshifts in this region of the lensing map, the redshift of this image system is not well constrained and is degenerate with other model parameters. Therefore, we fix the redshift to $z=3$, which lies in the middle of the range of its photometric redshifts ($2.6<\zphot<3.6$).

Our preliminary model (version 1, released in September 2013) included an image system, \#5, a low surface brightness arc stretching nearly 20\arcsec\ in the northern part of the cluster core. The image identifications were placed on brighter parts of the arc, but based on the monotonic color of the arc and overall low surface brightness, it is uncertain if these locations map to the same part of the source. There are also several galaxies along the line of sight to this arc that could potentially influence its lensing deflection. We exclude this image system from the model presented here based on its ambiguous identification. We find that the overall shape of this arc can be approximately replicated with the new best-fit model -- it is highly elongated and may be affected by lensing of cluster member galaxies. With deeper HFF imaging of this cluster, we may better understand this image system and include it in future models.

The complex mass distribution of Abell 2744 at $z=0.308$ is composed of five cluster or group-scale halos. We use two cluster-scale halos (H1, H2) to shape the mass distribution in the cluster core. In early model iterations, we find that the ellipticity of the halo (H1) lying closest to the brightest cluster galaxy (BCG) was near zero and not well constrained by the lensing evidence, so we assign it a circular halo. A third halo (H3) is located at $\sim50$\arcsec\ north-northeast of the cluster core close to an over density of cluster member galaxies, and is assigned a circular potential. We place a halo near image system \#16 (H4), which lies $\sim130$\arcsec\ northwest of the cluster core. A fifth halo (H5) is placed 140\arcsec\ north-northwest from the cluster core, which corresponds to an overdensity of cluster member galaxies in this region and a possible sub group of this cluster. The purpose of this halo is to add external shear to the lensing in the cluster core, and so its position and velocity dispersion are the only parameters that we can constrain, thus, we assign a circular potential to this halo and fix the $\rcore=150$ kpc. Weak lensing analyses by \citet{Merten:2011fk} and the Merten et al. preliminary HFF lens model reveal high surface mass densities in the regions outside of the HFF FOV, roughly in the areas where we place secondary halos far from the cluster core.

For cluster members, we assign halos at the cluster redshift and with parameters scaled by their magnitude in ACS F814W, such that an $\Lstar$ galaxy at the cluster redshift ($z=0.308$) has a magnitude $m_\star=18.50$. We freed the velocity dispersion, ellipticity, position angle, and $\rcut$ of the two brightest galaxies in the core for optimization, the central galaxy at $\alpha$=0:14:20.702, $\delta$=-30:24:00.62 and another galaxy at $\alpha$=0:14:22.091, $\delta$=-30:24:20.71. The BCG ellipticity tends toward unrealistically high values; in the final iteration we fix $e=0.8$.

We detect several galaxies with similar [F606W-F814W] colors, $\sim0.1$ mag blueward of the cluster red sequence. Our spectroscopic observation did not target any of these galaxies, since lensed galaxies in the core of the cluster received the highest priority. However, in examining the catalog of spectroscopic redshifts presented by \citet{Owers:2011rr}, we find one galaxy ($\alpha$=0:14:17.63, $\delta$=-30:22:40.58) with $z=0.239$. The color of the interloping galaxies is consistent with that of an elliptical galaxy at this redshift. We extend our red-sequence selection cut of cluster member galaxies to include these galaxies, as it is likely that these interloping galaxies contribute to the column mass of the cluster. Although not attempted here, this cluster is a clear case where a multi-plane lensing analysis may be necessary \citep[e.g., ][]{McCully:2014lr,Bayliss:2014fk,DAloisio:2013ul} in order to model this system.

Our lens model predicts a massive halo north-northeast of the cluster (H3), needed in order to reproduce the lensing of image systems \#3, 8, 14, and 18. Due to its mass and proximity to the main cluster halo, it produces a significant secondary critical curve component.  A visual inspection of the archival data show several low surface brightness arcs around the sub halo H3 which, at the depth of these data, cannot be confirmed strong lensing features. Nevertheless, new arcs may be confirmed in this region in the near future, with the new HFF data (e.g., the deep HFF observation of \MACSzerofour\ resulted in the identification of  $\sim200$ images in the ACS field by \citealt{Jauzac:2014fj}). We note that this region of the lens plane is not as well constrained as the rest of the cluster core, and may be prone to systematic errors in mass and magnification. Future work on this cluster lens model with more images should help to map the mass of this sub halo.

Figure \ref{fig:crit_a2744} shows the best-fit critical curves for a source at $z=2$ and the multiply imaged galaxies that were used as constraints in the lens model of Abell 2744, overplotted on a color composite image of the cluster. The critical curves map regions of high magnification in the image plane. The magnification map at $z=9$ and the mass distribution of the cluster are displayed in the bottom panels.

We derive a cylindrical mass enclosing a projected radius of 250 kpc surrounding the core of Abell 2744 to be $M(r<250\ \mathrm{kpc})=2.43^{+0.04}_{-0.07}\times10^{14}\ \mathrm{M_\odot}$ (more mass measurements are given in Table \ref{tab:masses_aperture}), consistent with \citet{Merten:2011fk}, who derive $M(r<250\ \mathrm{kpc})=2.2\pm0.5\times10^{14}\ \mathrm{M_\odot}$ from weak and strong lensing evidence, which encompasses our value.

We note that since 14 of 15 image systems surround the region we refer to as the core, our strong lensing model best constrains the mass in this region of the cluster. Nonetheless, our model requires additional mass outside of the core to explain the lensing of these images. Abell 2744 is an actively merging cluster \citep{Merten:2011fk}, and the complexity of its multiple halo components and unrelaxed state make this cluster a challenge to model in the entirety of the HFF FOV. The accuracy of the model will improve after the HFF observations are complete, and additional multiple-image systems are identified in other regions of the image plane.

\subsection{\MACSzerofour}

The lensed galaxies that were used as constraints in the model were originally identified by \citet[see Table \ref{tab:m0416_arcs} in the Appendix]{Zitrin:2013lr}. We fix the redshift of image system \#1 to the spectroscopic redshift from \citeauthor{Zitrin:2013lr} and systems \#2, 3, 4, 7, 10, 13, 14, 16, and 17 to the spectroscopic redshifts obtained under VLT program 186.A-0798 \citep{Balestra:2013uq,Grillo:2014uq}.

\begin{figure*}[h]
\centering
\includegraphics[width=0.8\textwidth]{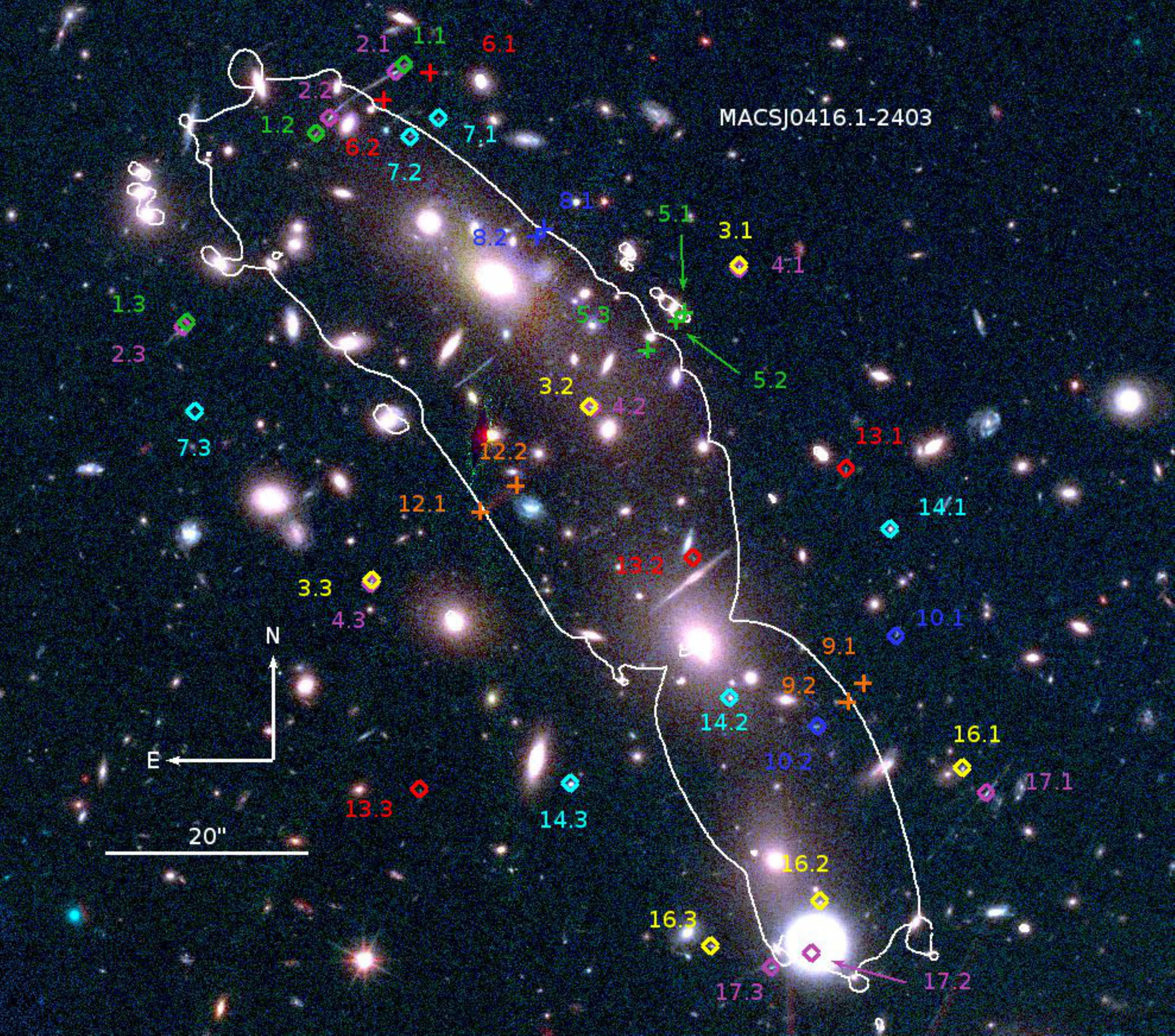}\\
\includegraphics[height=0.36\textheight]{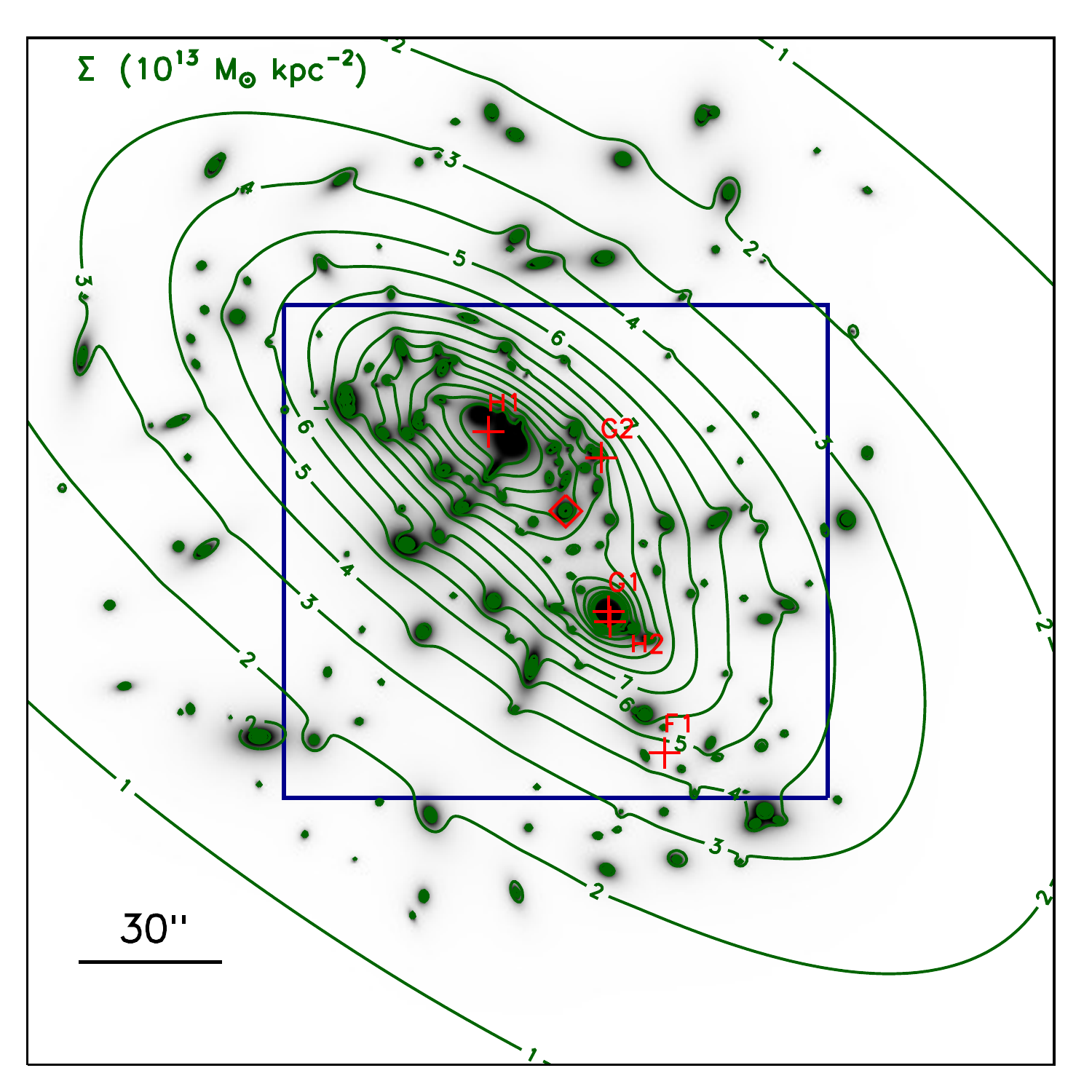}
\includegraphics[height=0.36\textheight]{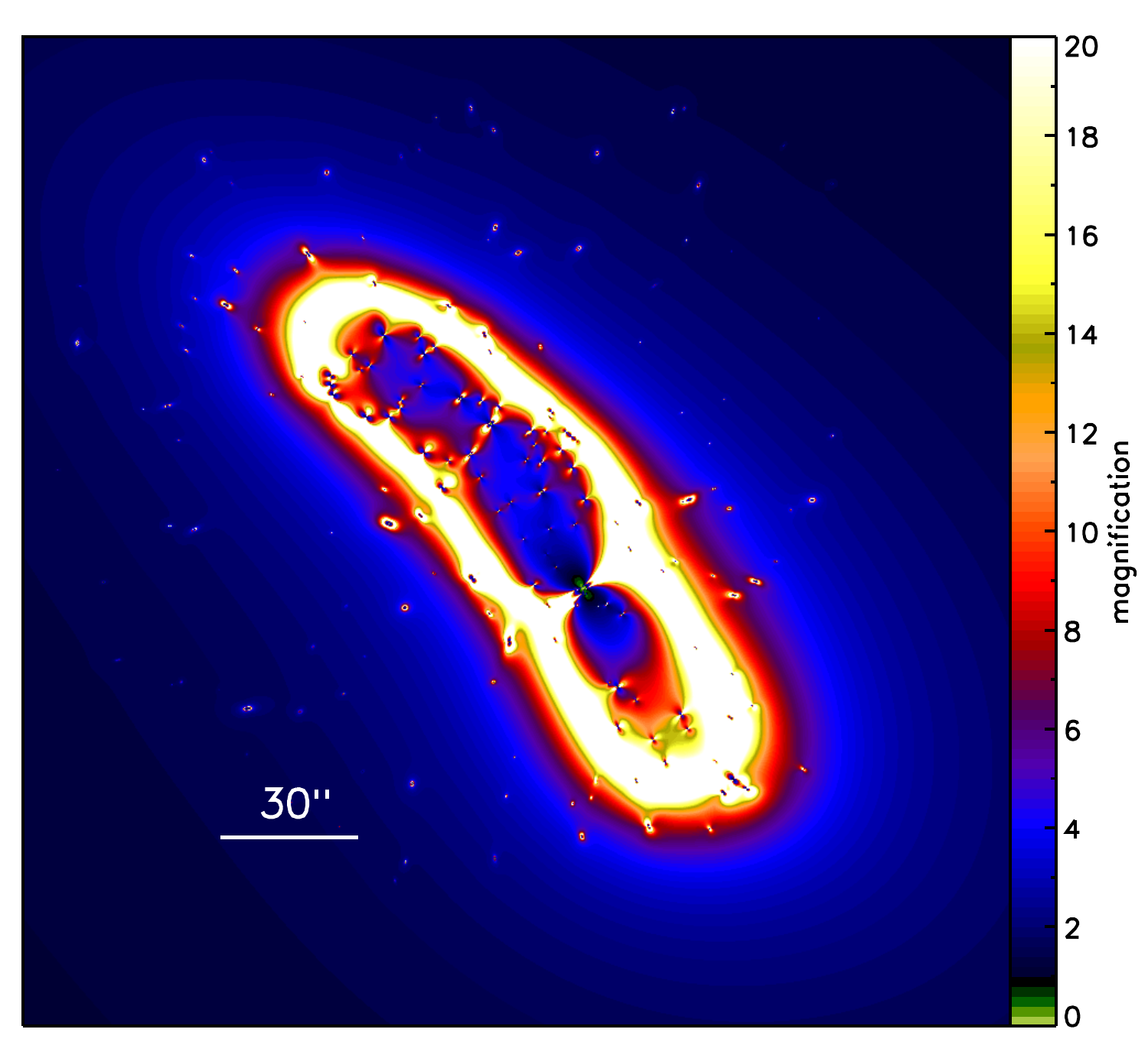}
\caption[\MACSzerofour\ image constraints and critical curves]{Top: False color image of \MACSzerofour\ from WFC3/IR (red; F105W, F110W, F125W, F140W, and F160W), ACS (green; F435W, F475W, F606W, F625W, F775W, F814W, and F850LP), and WFC3/UVIS (blue; F225W, F275W, F336W, and F390W) imaging. Labels are the same as in Figure \ref{fig:crit_a2744}. Left: Contour map of the total surface mass distribution (in units of $10^{13}\ \mathrm{M_\odot \ kpc^{-2}}$) overlaid on the mass contained in cluster member galaxies. The spacing of the contours is linear. The locations and labels of each optimized halo (see Table \ref{tab:m0416_params}) are shown in red crosses. The red diamond indicates the location of the reference point. The blue box indicates the field of view of the top image. Right: Absolute value of magnification in the image plane for a source at $z=9$.}
\label{fig:crit_m0416}
\end{figure*}

The model of \MACSzerofour\ at $z=0.396$ consists of two cluster-scale components, halos assigned to each cluster member galaxy within the ACS FOV, and a foreground galaxy. We allow all parameters of the cluster-scale halos to be optimized with the exception of cut radius. The scaling relation for the cluster member galaxies is based on their magnitude in ACS F775W with $m_\star=19.33$. The parameters of the BCG cannot be constrained during optimization, so they are fixed to the values derived from the scaling relations. We also allow the velocity dispersion of the galaxy near images 5.2 and 5.3 ($\alpha$=4:16:07.786, $\delta$=-24:04:06.51; G2) to be a free parameter in the lens model, as it contributes to the cluster-boosted galaxy-galaxy lensing in image system \#5. While this galaxy has similar color to the cluster member galaxies, it did not make the red-sequence cut on first pass. The extracted shape of the galaxy did not match well with observations due to contamination by other galaxies, so we assign a circular PIEMD halo instead. We assign a halo to the bright foreground galaxy at ($\alpha$=4:16:06.817, $\delta$=-24:05:08.44; F1) and allowed it to vary in core and cut radii and velocity dispersion, as this galaxy may deviate from the scaling relations typical of elliptical galaxies at the cluster redshift. Early iterations converged to a very large cut radius of this galaxy which is not well constrained, so we arbitrarily fixed it to 1500 kpc. Figure \ref{fig:crit_m0416} shows the best-fit critical curves, modeling constraints, mass distribution, and magnification map for the \MACSzerofour\ lens model.

We compute $M(r<250\ \mathrm{kpc})=1.77^{+0.31}_{-0.13}\times10^{14}\ \mathrm{M_\odot}$ and $M(r<500\ \mathrm{kpc})=4.05^{+0.90}_{-0.32}\times10^{14}\ \mathrm{M_\odot}$ for the mass of \MACSzerofour. We note that a weak lensing model by \citet{Gruen:2014lr} yields $M(r<250\ \mathrm{kpc})=1.8\pm0.3\times10^{14}\ \mathrm{M_\odot}$ and $M(r<500\ \mathrm{kpc})=3.8\pm0.7\times10^{14}\ \mathrm{M_\odot}$, which are in agreement with our model. We measure $M(<\mathrm{crit})=0.80^{+0.12}_{-0.06}\times10^{14}\ \mathrm{M_\odot}$ for the mass within the $z=2$ critical curve. \citet{Zitrin:2013lr} report the mass within the critical curve of the main arc (at $z=1.896$) to be $M(<\mathrm{crit})=1.25\pm0.09\times10^{14}\ \mathrm{M_\odot}$. The critical curves for both models enclose nearly the same amount of area (0.58 \sq\arcmin). The discrepancy between the two models can possibly be explained by the fact that the \citet{Zitrin:2013lr} model was computed prior to the spectroscopic confirmations of several lensed galaxies in this cluster; therefore, leaving the critical curve less constrained in some regions of the map. The redshift predictions of the image systems in the \citet{Zitrin:2013lr} model are systematically higher than the spectroscopic redshifts we included in the model we present here. The source redshifts of the images place a strong constraint on the slope of the mass distribution, and higher redshift predictions require larger mass within the same critical curve area. We discuss the implications of including spectroscopic redshifts on the derived lens model of clusters later in \S \ref{sec:specz}.

\subsection{\MACSzeroseven}

We use the multiple images initially identified by \citet{Zitrin:2009qy} and later revised by \citet{Limousin:2012fj} (Table \ref{tab:m0717_arcs} in the Appendix). The coordinates of the lensed galaxies have been matched to the CLASH imaging data. We fix the redshifts of image systems \#1, 3, 13, 14, and 15 to the spectroscopic redshifts reported by \citet{Limousin:2012fj}.

\begin{figure*}[h]
\centering
\includegraphics[width=0.8\textwidth]{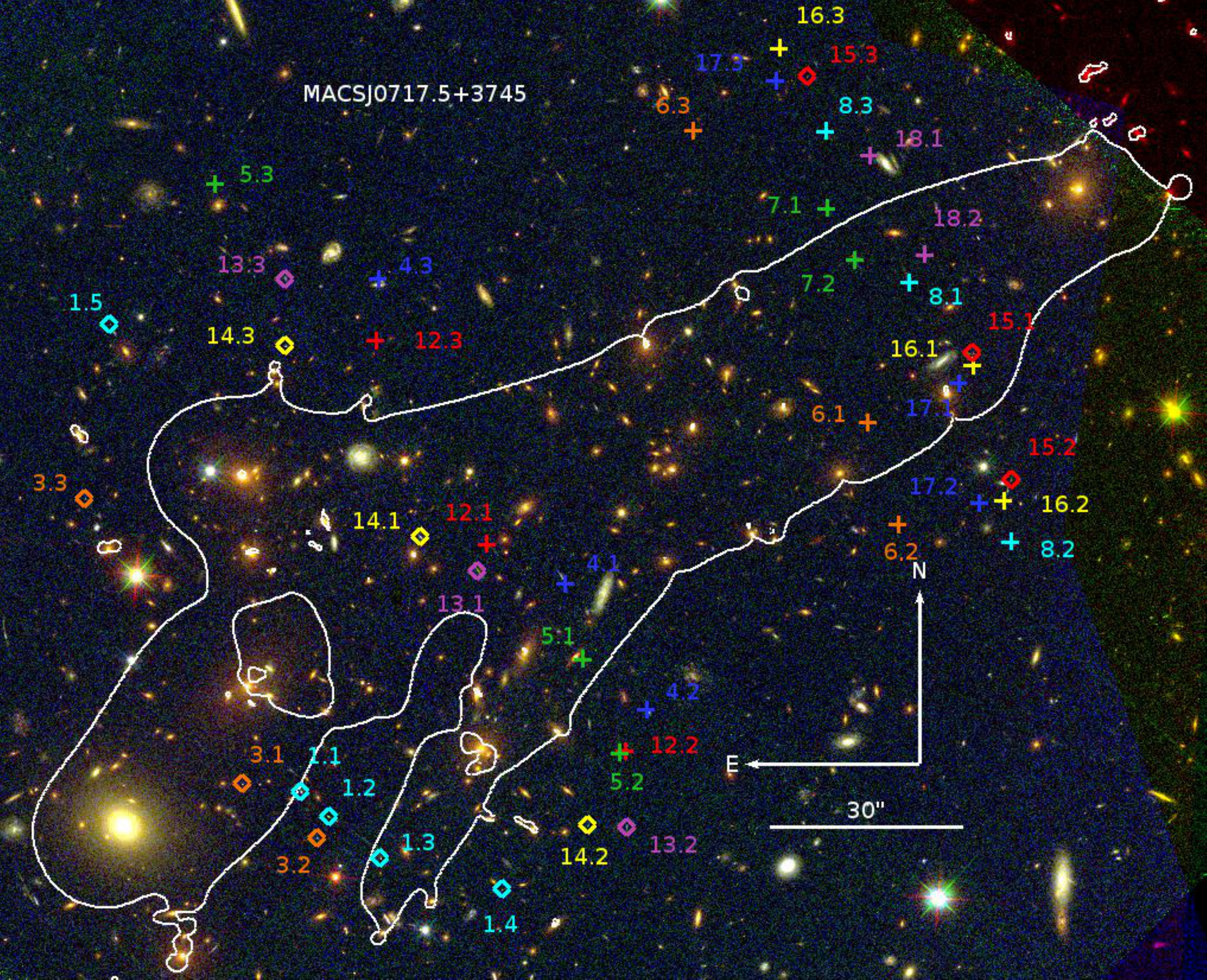} \\
\includegraphics[height=0.36\textheight]{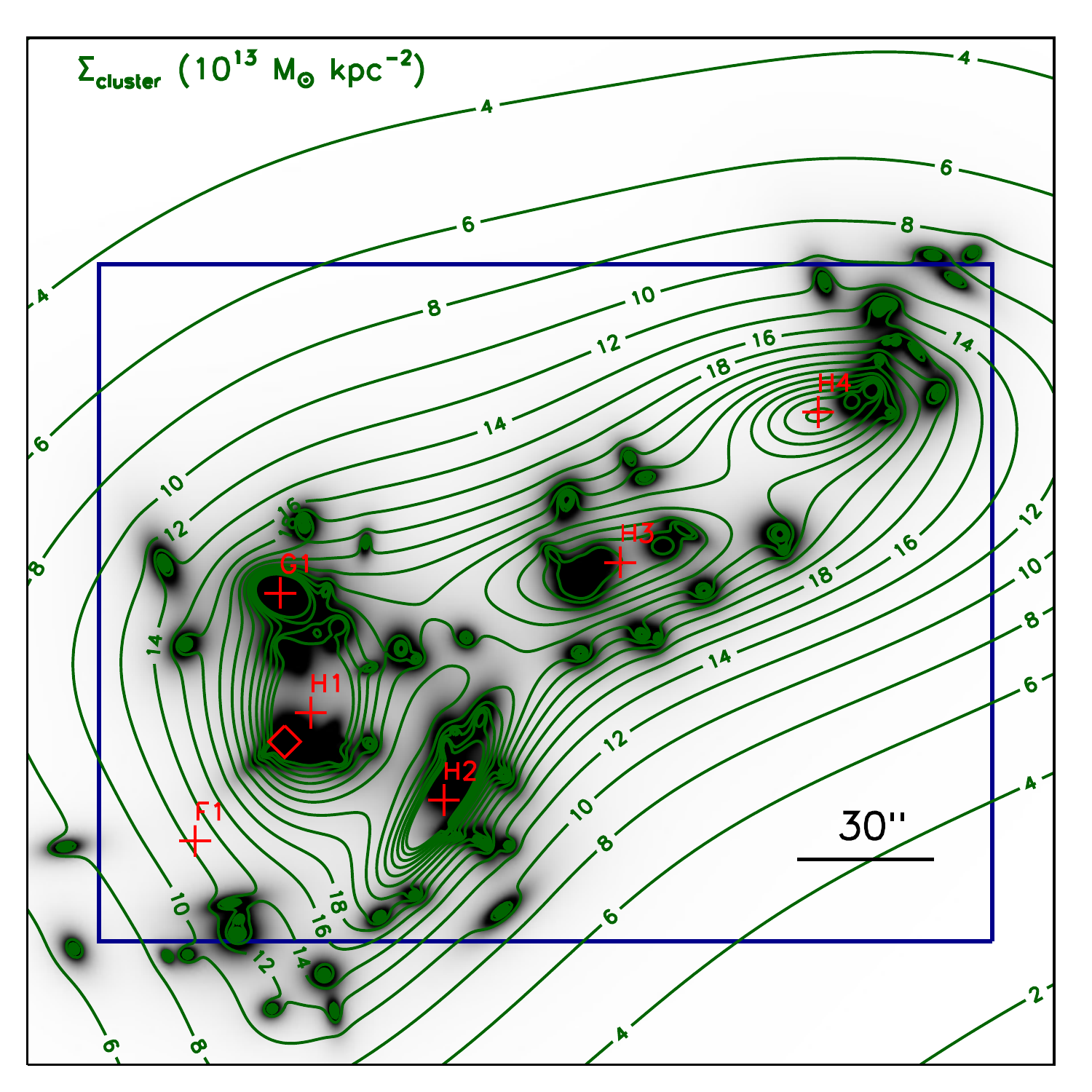}
\includegraphics[height=0.36\textheight]{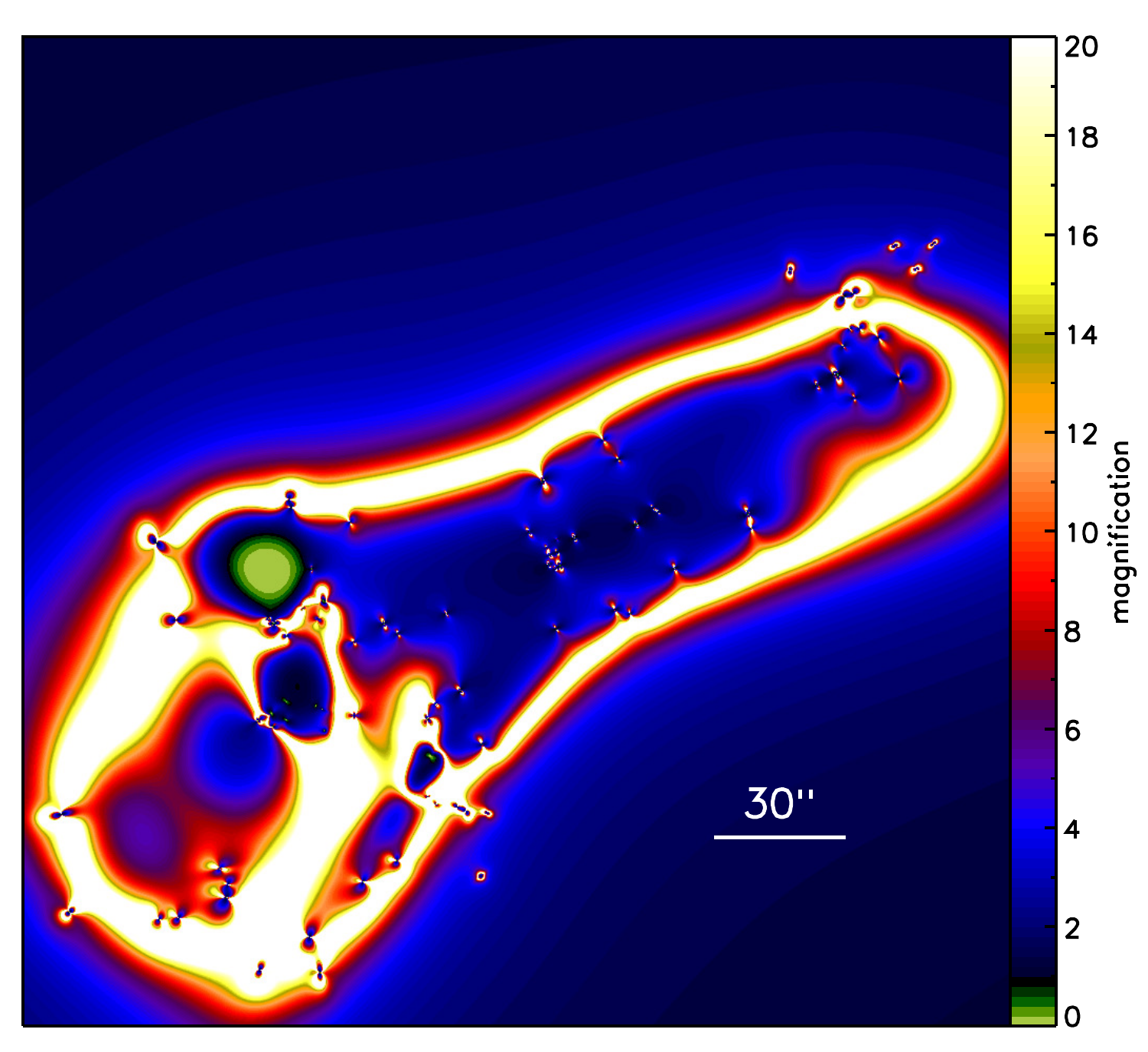}
\caption[\MACSzeroseven\ image constraints and critical curves]{Top: False color image of \MACSzeroseven\ from ACS imaging (red, F814W; green, F606W; blue, F435W). Labels are the same as in Figure \ref{fig:crit_a2744}. Left: Contour map of the total surface mass distribution (in units of $10^{13}\ \mathrm{M_\odot \ kpc^{-2}}$) of the cluster overlaid on the mass contained in cluster member galaxies. The spacing of the contours is linear. The locations and labels of each optimized halo (see Table \ref{tab:m0717_params}) are shown in red crosses. The red diamond indicates the location of the reference point.} The blue box indicates the field of view of the top image. Right: Absolute value of magnification in the image plane for a source at $z=9$.
\label{fig:crit_m0717}
\end{figure*}

The mass distribution of \MACSzeroseven\ at $z=0.545$ is best represented by several separate halo components, consistent with the findings of an X-ray / optical study by \citet{Ma:2009gf}. Besides the lensing evidence, the locations of the halos are observationally motivated as they lie close to overdensities of cluster member galaxies. We include cluster member galaxies in the model, as selected by \citet{Limousin:2012fj}, with $m_\star=20.66$ in ACS F814W. We allow the velocity dispersion and cut radius of the cluster galaxy at ($\alpha$=7:17:35.646, $\delta$=+37:45:17.40; G1) to be optimized by the model. A foreground galaxy at ($\alpha$=7:17:37.224,$\delta$=+37:44:22.99; F1) is also included in the model, with a circular PIEMD halo, and free cut radius and velocity dispersion parameters. The core radius of this galaxy could not be constrained, so it is fixed arbitrarily to the value at the cluster redshift based on the scaling relation. Figure \ref{fig:crit_m0717} shows the inputs to and details of the lens model for this cluster.

It is likely that this merging system has a more complex mass distribution that cannot be accurately represented by parameterized halos; nevertheless, the resulting image plane rms for the constraints used in this model is good ($0\farcs38$). We note that a direct, halo-to-halo comparison with the \citet{Ma:2009gf} findings is not meaningful, for two reasons: the PIEMD velocity dispersion is not defined as the measured velocity dispersion \citep[see][]{Eliasdottir:2007ve}, and the halos have different geometry, positions, and fragmentation. Nevertheless, we can directly measure the mass from our model within each area associated with the subclusters defined by \citet{Ma:2009gf} We find the mass within the regions labeled A, B, C, and D in \citet{Ma:2009gf} are $M = 5.5^{+0.3}_{-0.4}, 6.9\pm0.2, 16.6^{+0.4}_{-0.6}, 6.5^{+0.1}_{-0.2}\times10^{13}\ \mathrm{M_\odot}$, respectively. The values for cores A, C, and D correspond well with the observed velocity dispersions within errors, taking for simplicity their viral masses. The derived mass for core B is slightly higher than what is inferred from the observations. \citet{Ma:2009gf} find that this core is moving at a high radial velocity ($\Delta v>3000 \ \mathrm{km\ s^{-1}}$) relative to the cluster core, near the infall velocity. Estimating the mass from virial assumptions may not be best in this scenario, which could account for the discrepancy in the mass estimates. The agreement between the completely independent lensing evidence and dynamics suggests that future modeling efforts can benefit from including the measured velocity dispersion as a Bayesian constraint. 

\MACSzeroseven\ is by far the most massive cluster in the HFF, with a mass $M(r<500\ \mathrm{kpc})=8.68^{+0.27}_{-0.13}\times10^{14}\ \mathrm{M_\odot}$. We find $M(<\mathrm{crit})=5.27^{+0.20}_{-0.11}\times10^{14}\ \mathrm{M_\odot}$ for the mass enclosed by the $z=2$ critical curve, yielding an impressive effective Einstein radius of $\theta_E=50\farcs1^{+0.8}_{-0.3}$. This critical curve encloses a factor of more than two greater area than any other cluster in the HFF, and should prove to be an efficient lens of background galaxies. Additionally, we find the mass within the $z=2.5$ critical curve to be $M(<\mathrm{crit})=5.91^{+0.20}_{-0.08}\times10^{14}\ \mathrm{M_\odot}$. Our measurement differs significantly from \citet{Zitrin:2009qy}, who find the mass enclosed by the critical curve at $z=2.5$ to be $M(<\mathrm{crit})=7.4\pm0.5\times10^{14}\ \mathrm{M_\odot}$. Their lens model uses fewer image constraints and is not supported by spectroscopic or photometric redshifts. As a result, the critical curve for $z=2.5$ may change positions in the image plane based on the input redshifts of the image constraints, causing the discrepancy between the two models. We discuss the effects of model redshifts as inputs in \S \ref{sec:specz}.

\subsection{\MACSeleven}

We use as constraints the strongly lensed image lists from \citet{Smith:2009lr}, \citet{Zitrin:2009kx}, and \citet{Zheng:2012fk} supplemented by unpublished identifications made by Adi Zitrin (priv. comm.).$^{\ref{note}}$ We consolidate all lists of images; the complete list is shown in Table \ref{tab:m1149_arcs} in the Appendix.  We fix the redshifts of systems \#1, 2, and 3 to the spectroscopic redshifts reported by \citet{Smith:2009lr}. Our preliminary model (version 1, released September 2013) included an image system labelled \#12 whose high image-plane rms of 4\farcs6 indicates a potential misidentification. Excluding this system as a constraint, the model presented here predicts a redshift of $z>3$ for the two outer-most images of this image set, in stark contrast to the photometric estimate of $z\sim1$. The nature of this system may be better understood with the full HFF depth and spectroscopic confirmation.

\begin{figure*}[h]
\centering
\includegraphics[width=0.8\textwidth]{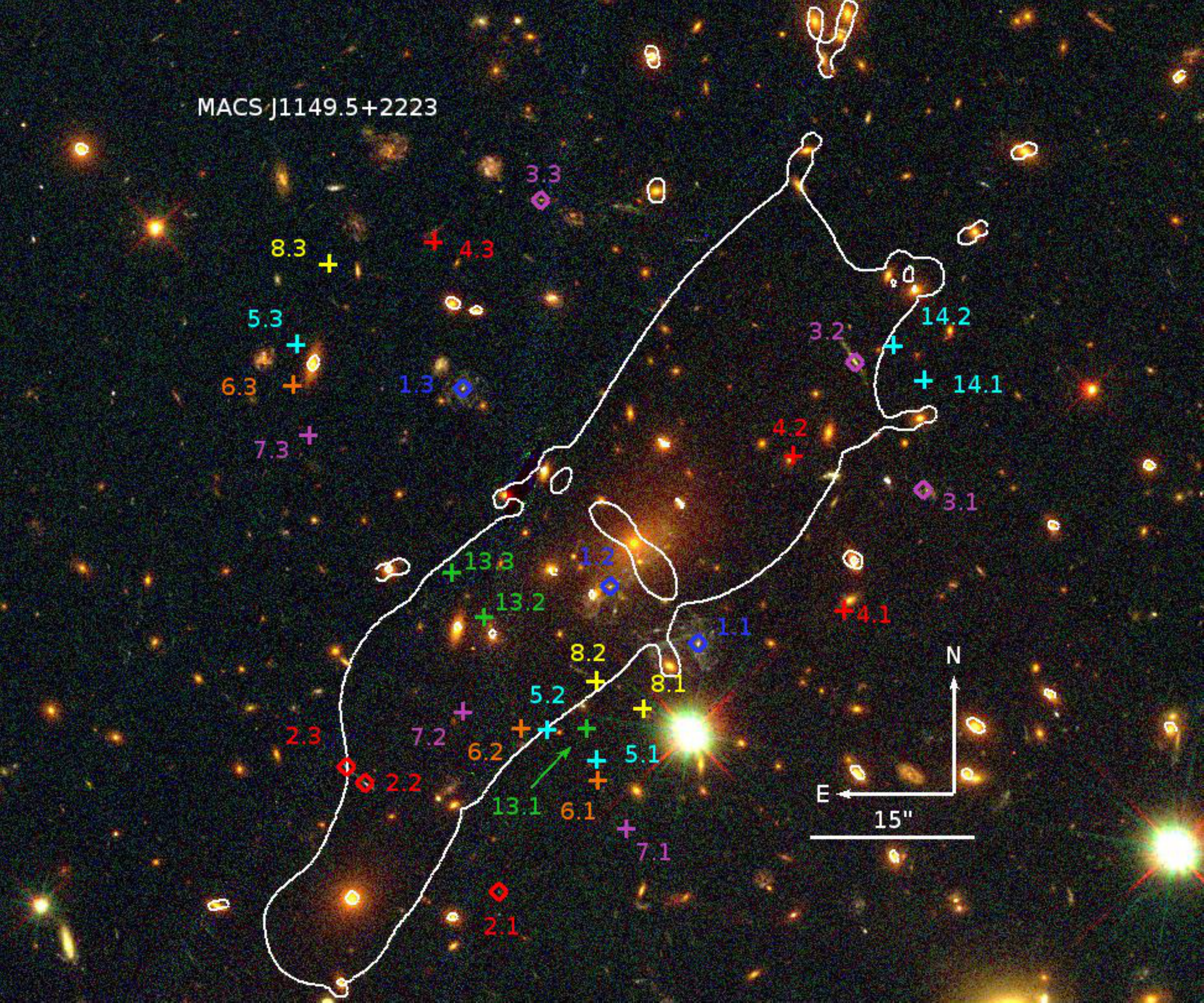} \\
\includegraphics[height=0.36\textheight]{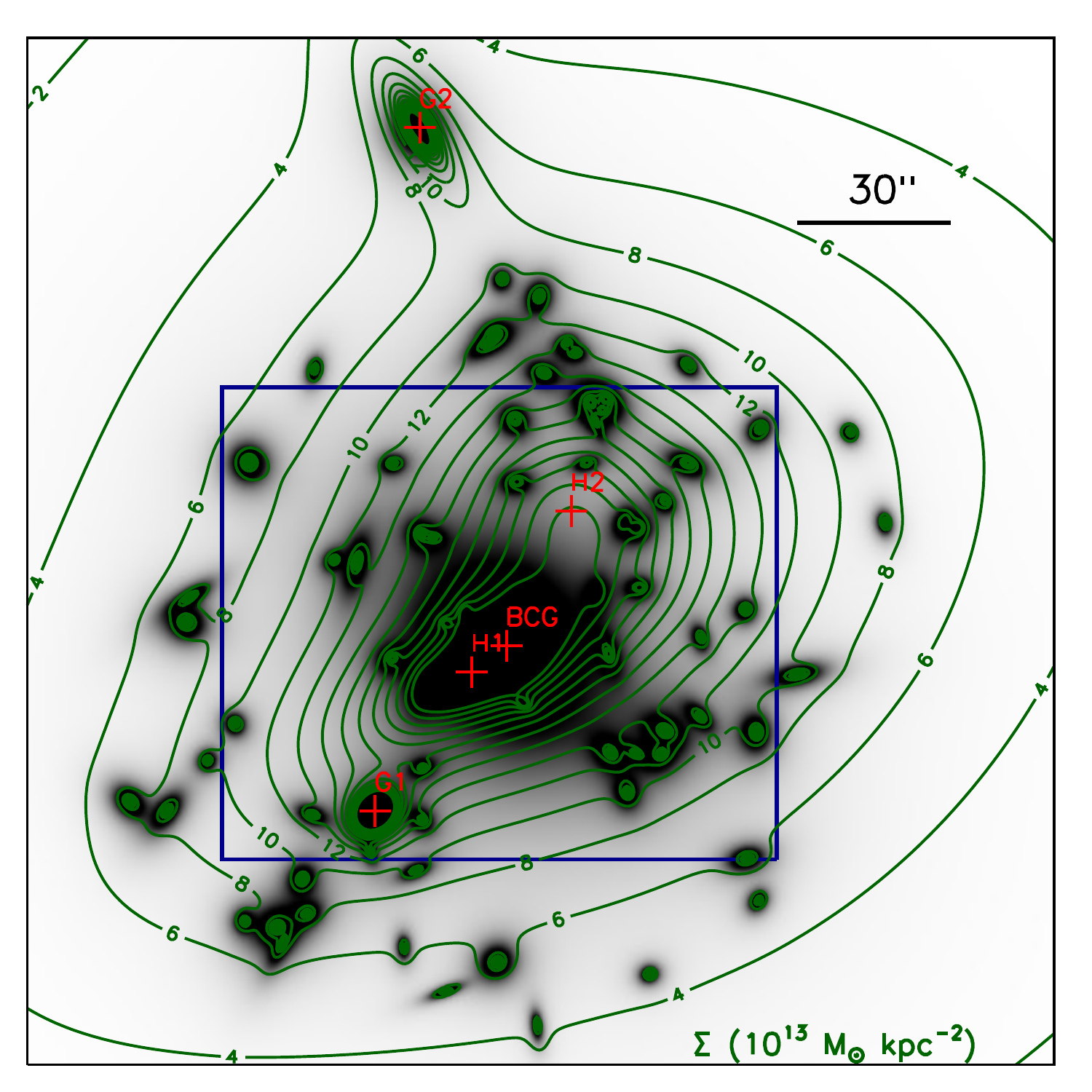}
\includegraphics[height=0.36\textheight]{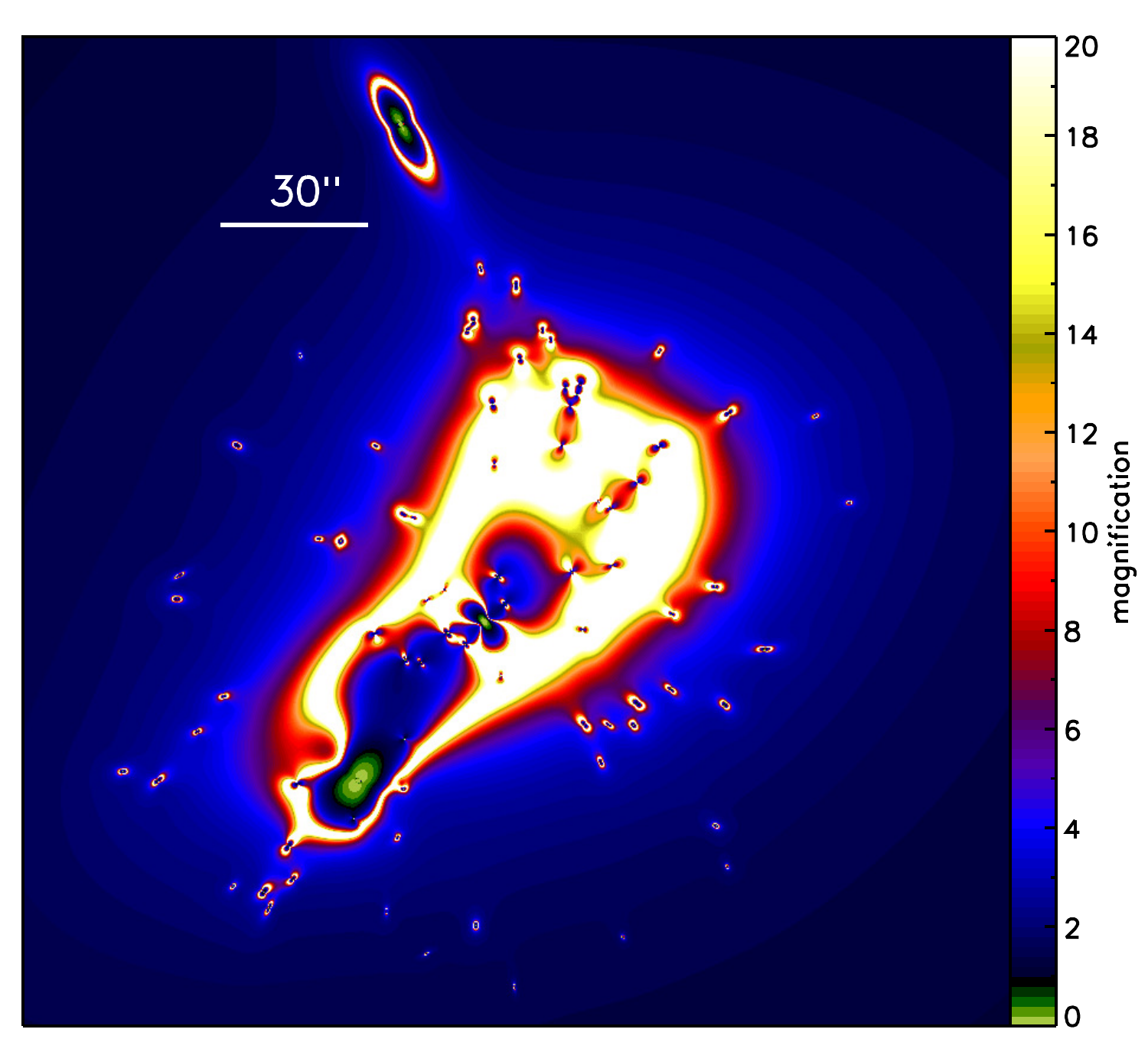}
\caption[\MACSeleven\ image constraints and critical curves]{Top: False color image of \MACSeleven\ from ACS imaging (red, F814W; green, F606W; blue, F435W). Labels are the same as in Figure \ref{fig:crit_a2744}. Left: Contour map of the total surface mass distribution (in units of $10^{13}\ \mathrm{M_\odot \ kpc^{-2}}$) overlaid on the mass contained in cluster member galaxies. The spacing of the contours is linear. The locations and labels of each optimized halo (see Table \ref{tab:m1149_params}) are shown in red. The blue box indicates the field of view of the top image. Right: Absolute value of magnification in the image plane for a source at $z=9$.}
\label{fig:crit_m1149}
\end{figure*}

The lens model of \MACSeleven\ at $z=0.543$ consists of two dark matter halos, one lying close to the BCG (H1) and the other located near an overdensity of cluster galaxies 100\arcsec\ north of the cluster center (H2). We include the cluster member galaxies selected by \cite{Smith:2009lr}, with $m_\star=20.3$ from K band imaging. We allow only the position, velocity dispersion, and cut radius of the second halo to vary in the model. We include the velocity dispersion and cut radii of the BCG ($\alpha$=11:49:35.695,$\delta$=+22:23:54.70) and cluster member galaxy at ($\alpha$=11:49:37.541,$\delta$=+22:23:22.51; G1) as free parameters. We also include a galaxy-scale halo north of the cluster (G2) accounting for the lensing of image systems \#9 and \#10, due to the galaxy-galaxy lensing boosted by the mass from the dark matter halo of the cluster. Since neither of these two systems have spectroscopic redshifts, we do not have enough constraints to attempt to model both the individual galaxy plus other substructure in that vicinity, which is essentially isolated from the rest of the cluster. Instead, we use a single halo with position priors matching the galaxy at ($\alpha$=11:49:36.926,$\delta$=+22:25:35.82). The model requires an unrealistically high ellipticity, indicating that more substructure may be needed; we thus fix it to $e=0.8$ and leave the position angle, velocity dispersion, and cut radius as free parameters. The necessity of optimizing this halo in the model far away from the majority of modeling constraints suggests the presence of significant substructure in part of the lens plane. The critical curves, image constraints, mass distribution, and magnification map for this cluster are shown in Figure \ref{fig:crit_m1149}.

We compute a cylindrical mass at the core of \MACSeleven\ of $M(r<500\mathrm{kpc})=5.98^{+0.59}_{-0.25}\times10^{14}\ \mathrm{M_\odot}$. We can directly compare this value with the model by \citet{Smith:2009lr}, who find $M(r<500\mathrm{kpc})=6.7\pm0.4\times10^{14}\ \mathrm{M_\odot}$. In fact, the \citet{Smith:2009lr} and our model were constructed independently with \texttt{LENSTOOL} and resulted in similar locations of cluster halo components in the lens plane. However, the previous model was built with fewer identified image systems. Our model includes 35 images from 12 uniques sources, whereas \citet{Smith:2009lr} identified 19 images from 6 unique, multiply imaged sources. We find $M(<\mathrm{crit})=1.12^{+0.01}_{-0.04}\times10^{14}\ \mathrm{M_\odot}$ for the mass enclosed by the $z=2$ critical curve ($0.40^{+0.01}_{-0.02}$ \sq\arcmin), which does not agree with \citet{Zitrin:2011qy}, who find $M(<\mathrm{crit})=1.71\pm0.20\times10^{14}\ \mathrm{M_\odot}$ (0.63 \sq\arcmin). We note that the \citet{Zitrin:2011qy} model does again not include any spectroscopic or photometric redshifts, have a different set of multiple image identifications, and do not treat their image redshift constraints as free parameters. We refer the reader to the discussion in \citet{Smith:2009lr}, where they rule out the inner slope of the surface mass density profile of the \citet{Zitrin:2011qy} model by $7\sigma$. This example demonstrates how different modeling inputs can result in significantly different lens models. We will discuss this further in \S \ref{sec:specz}.

\subsection{Abell S1063}
\label{sec:results_as1063}

We constrain the lens model of Abell S1063 with a combination of the images identified by \citet{Monna:2014lr} and Johan Richard (priv. comm.).$^{\ref{note}}$ For features common to both catalogs, we use the \citet{Monna:2014lr} coordinates. We fix the redshifts of systems \#1, 2, 5, 6 to the spectroscopic redshifts measured in this work and by  \citet{Richard:2014gf}, \#12 to the spectroscopic redshift measured by \citet{Balestra:2013uq,Boone:2013lr}, and the redshift of \#11 to the spectroscopic redshift we measured in this work.

\begin{figure*}[h]
\centering
\includegraphics[width=0.8\textwidth]{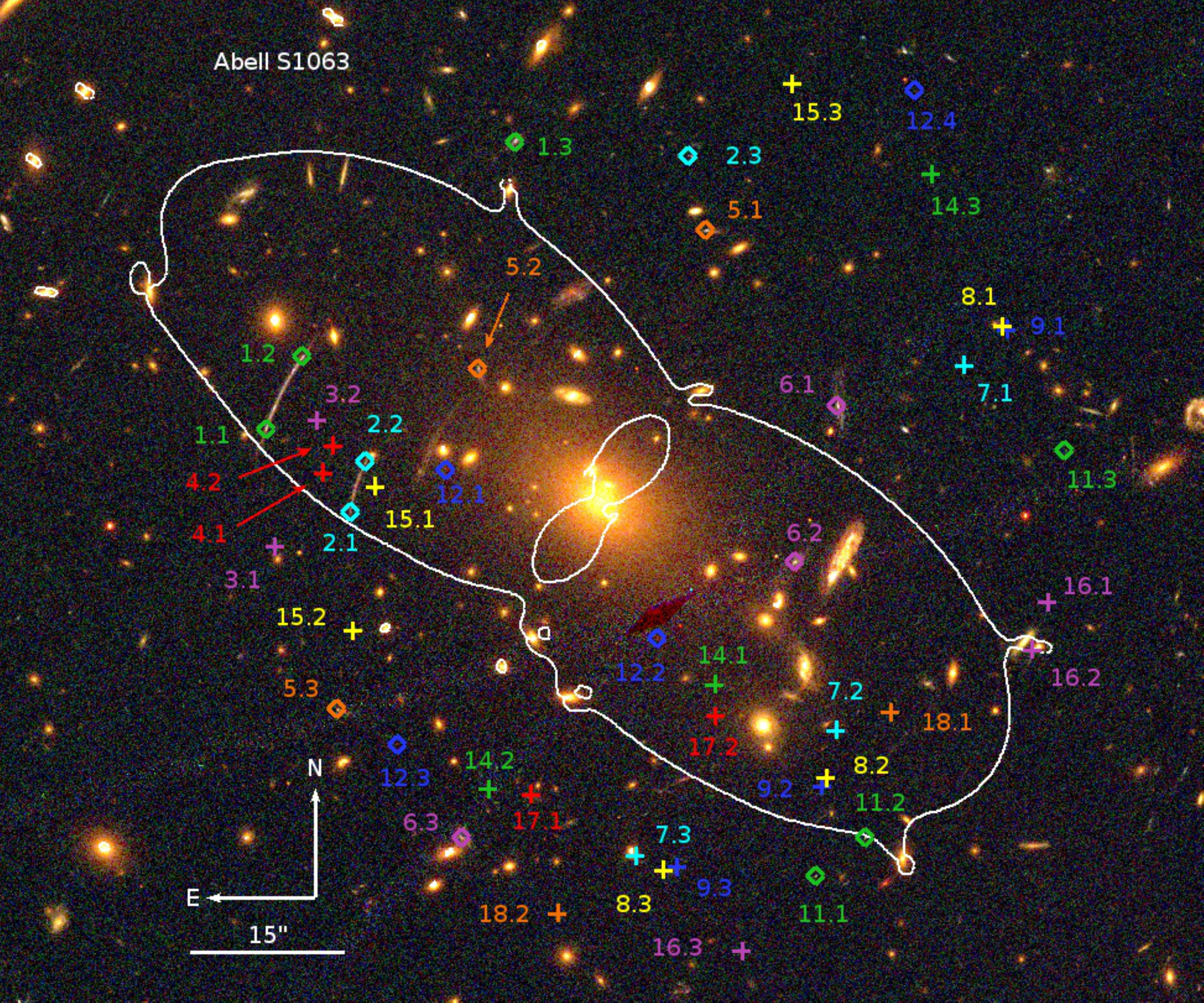} \\
\includegraphics[height=0.36\textheight]{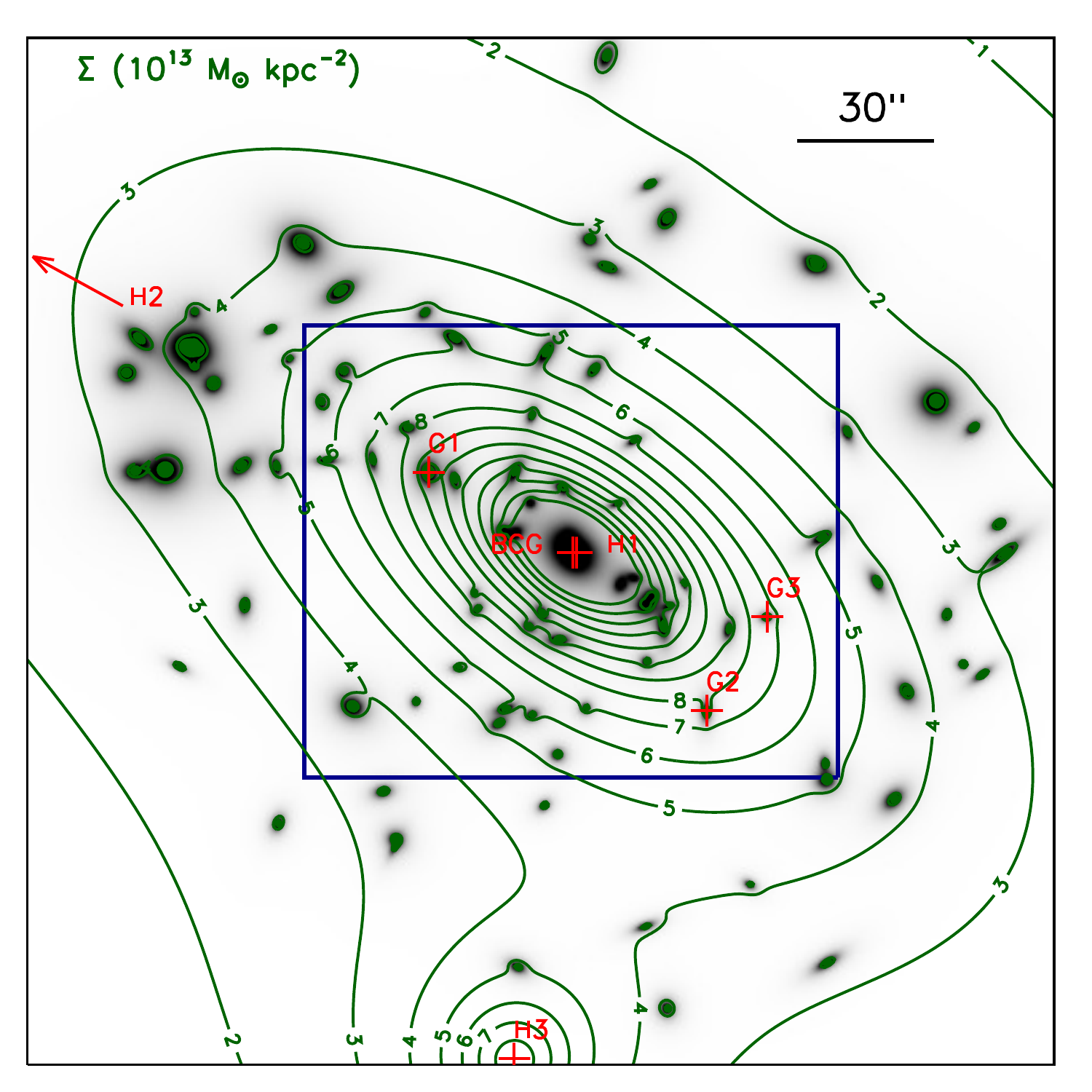}
\includegraphics[height=0.36\textheight]{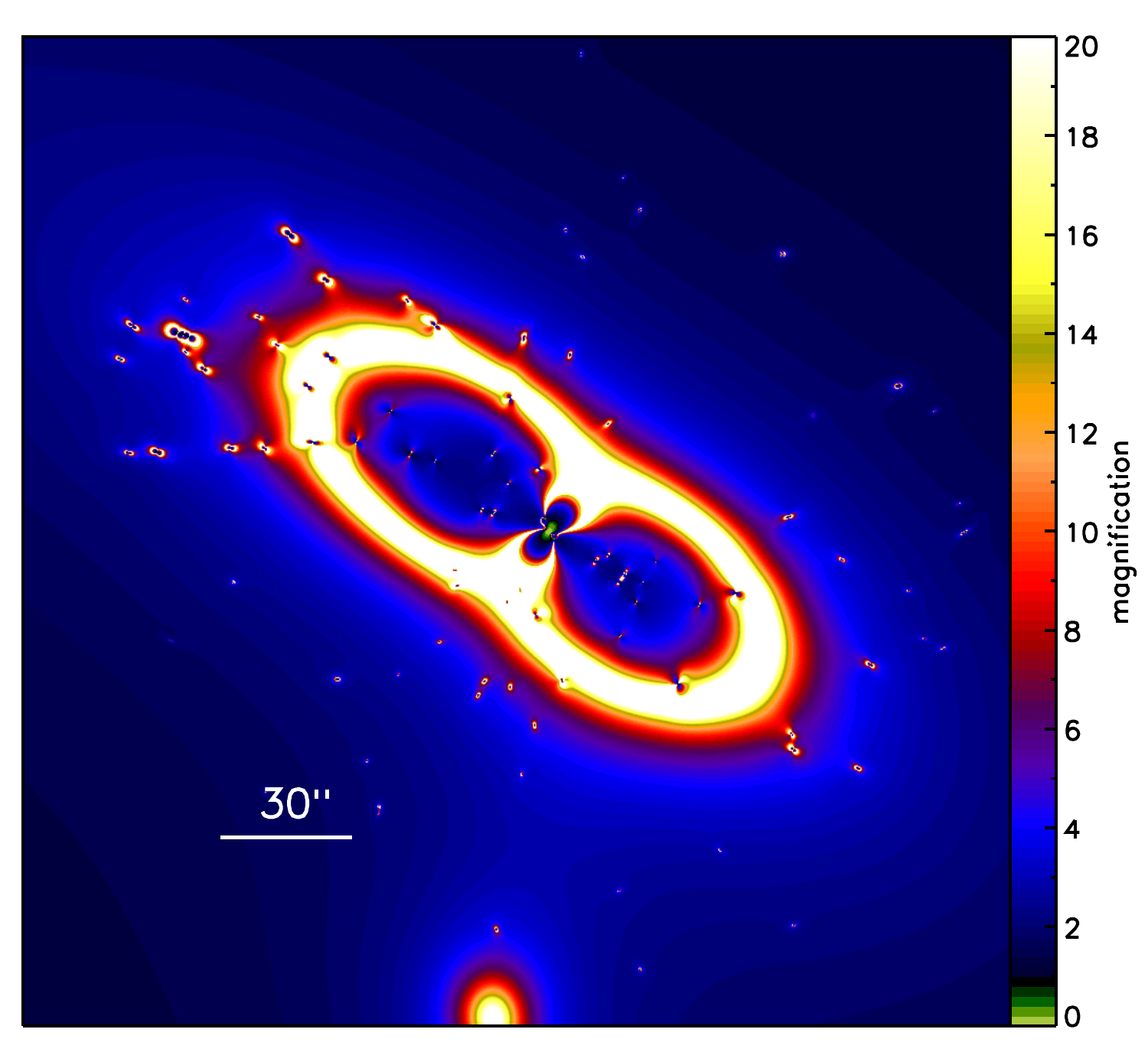}
\caption[Abell S1063 image constraints and critical curves]{Top: False color image of Abell S1063 from ACS imaging (red, F814W; green, F606W; blue, F435W).  Labels are the same as in Figure \ref{fig:crit_a2744}. Left: Contour map of the total surface mass distribution (in units of $10^{13}\ \mathrm{M_\odot \ kpc^{-2}}$) overlaid on the mass contained in cluster member galaxies. The spacing of the contours is linear. The locations and labels of each optimized halo (see Table \ref{tab:as1063_params}) are shown in red. The blue box indicates the field of view of the top image. Right: Absolute value of magnification in the image plane for a source at $z=9$.}
\label{fig:crit_as1063}
\end{figure*}

\begin{figure*}[h]
\centering
\includegraphics[width=0.55\textwidth]{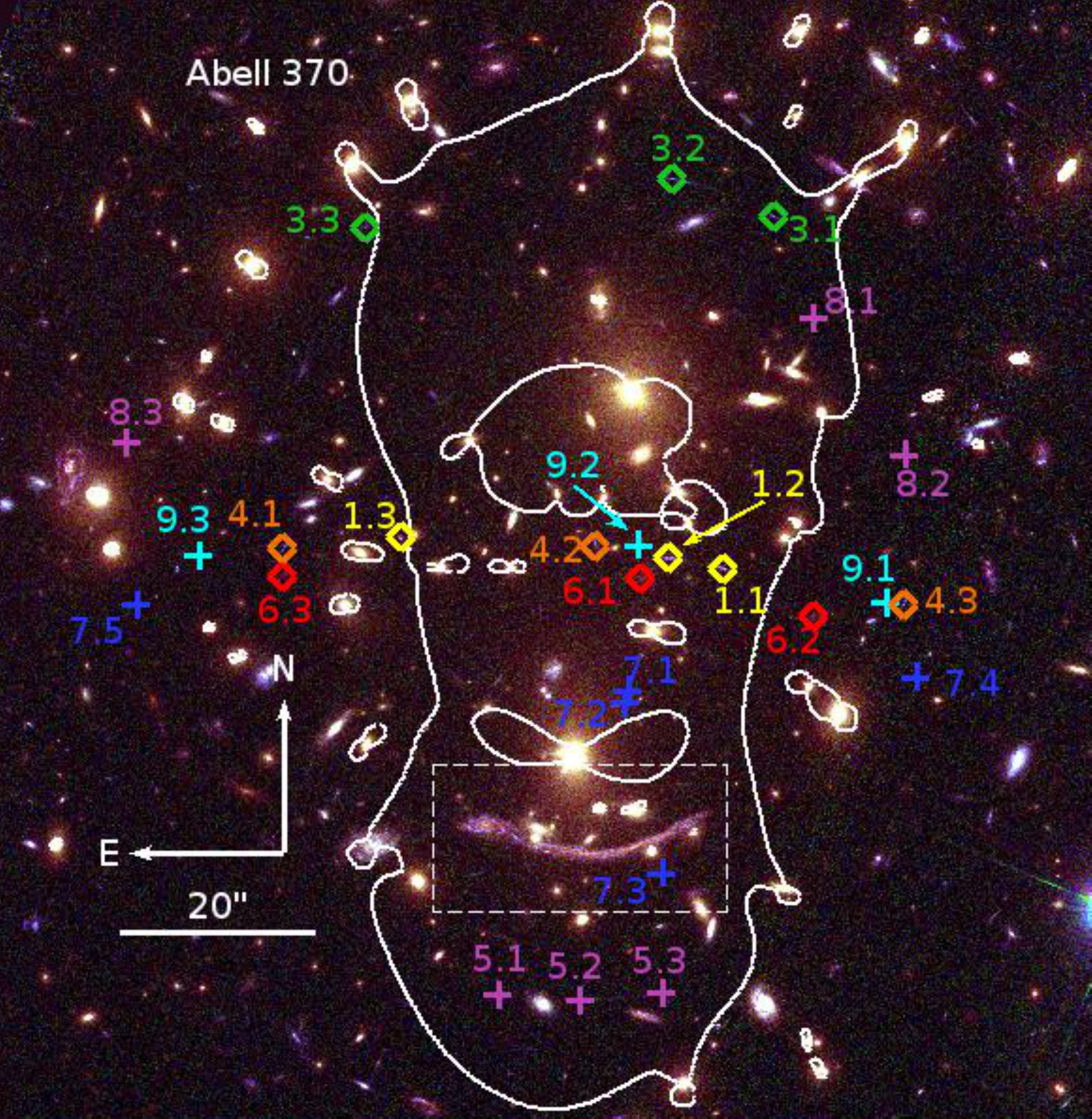}
\includegraphics[width=0.4\textwidth]{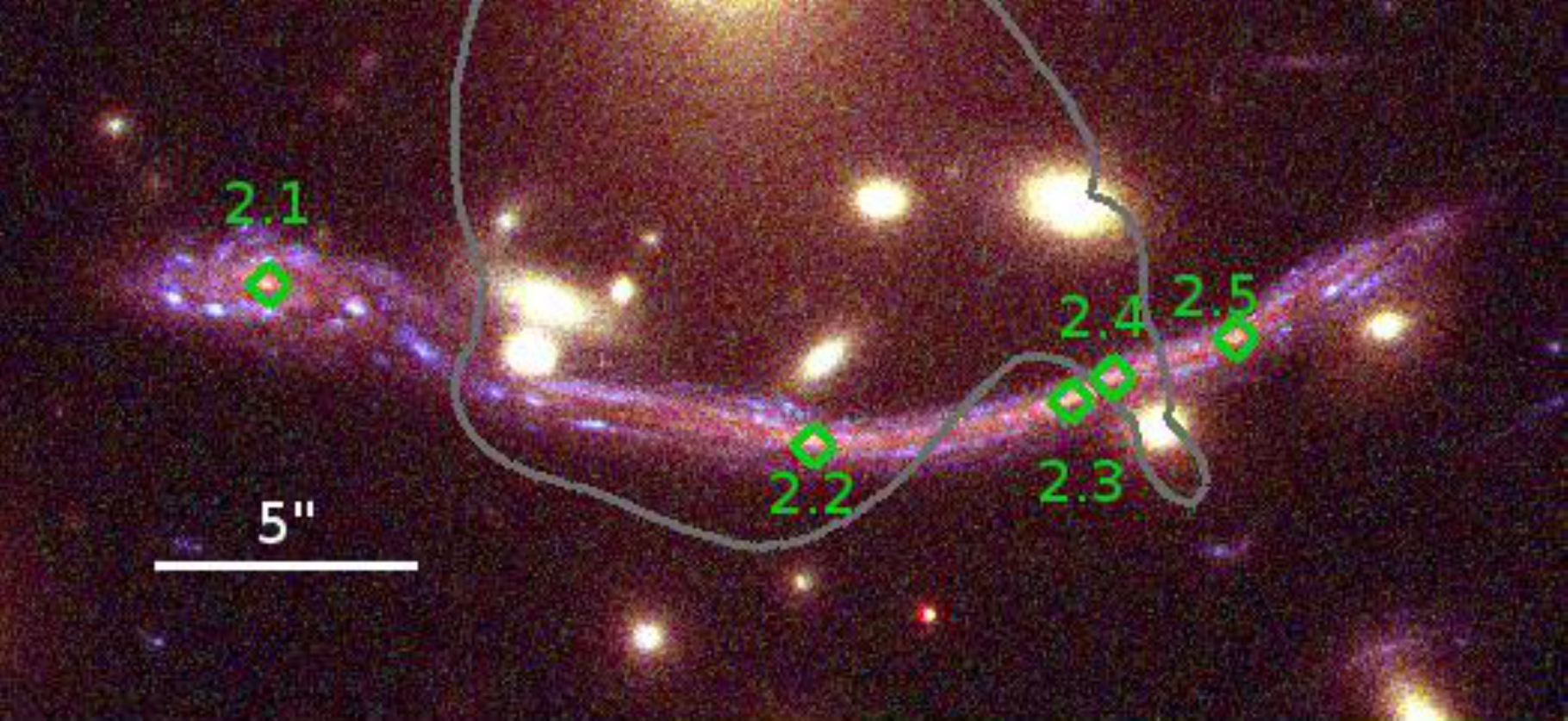}
\\
\includegraphics[height=0.36\textheight]{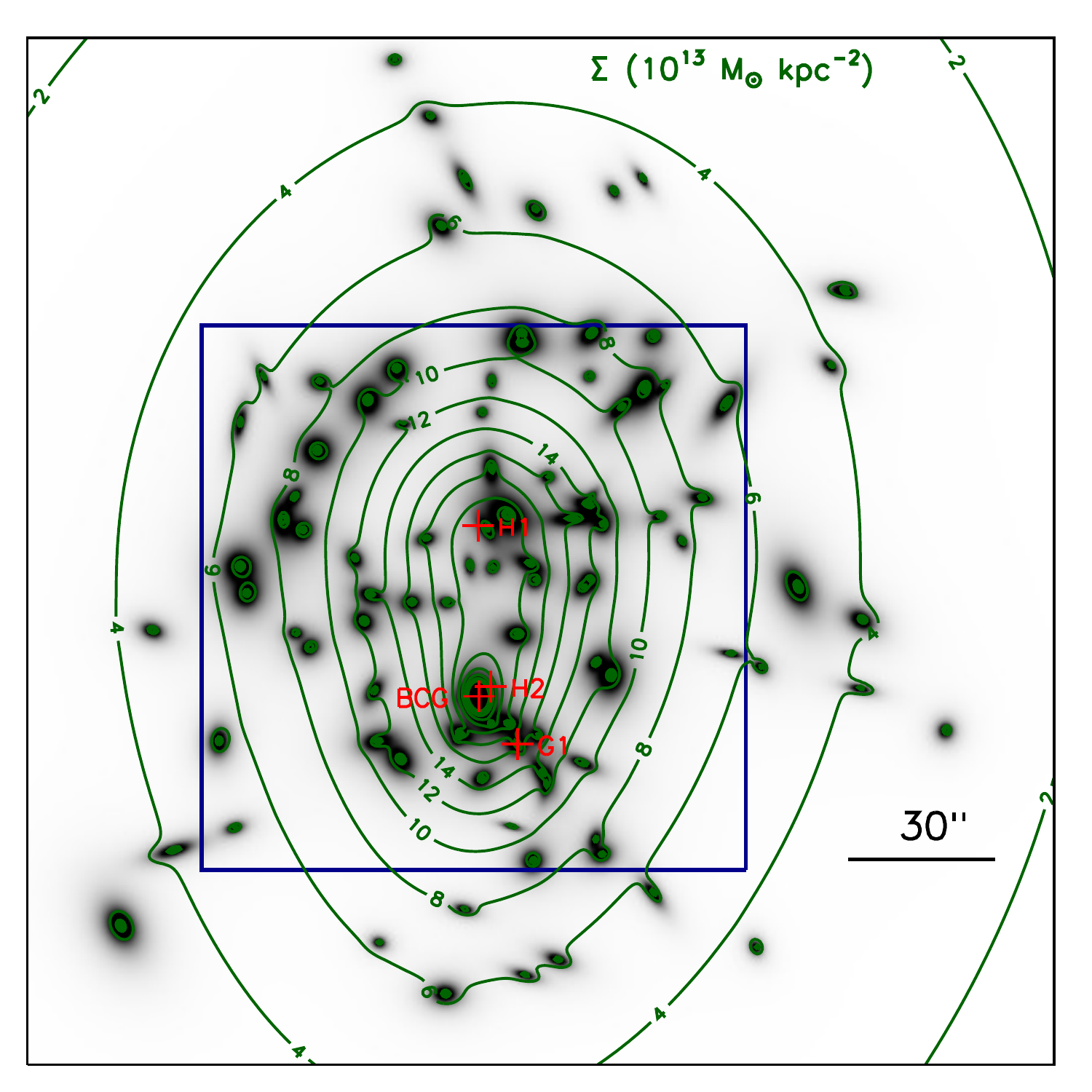}
\includegraphics[height=0.36\textheight]{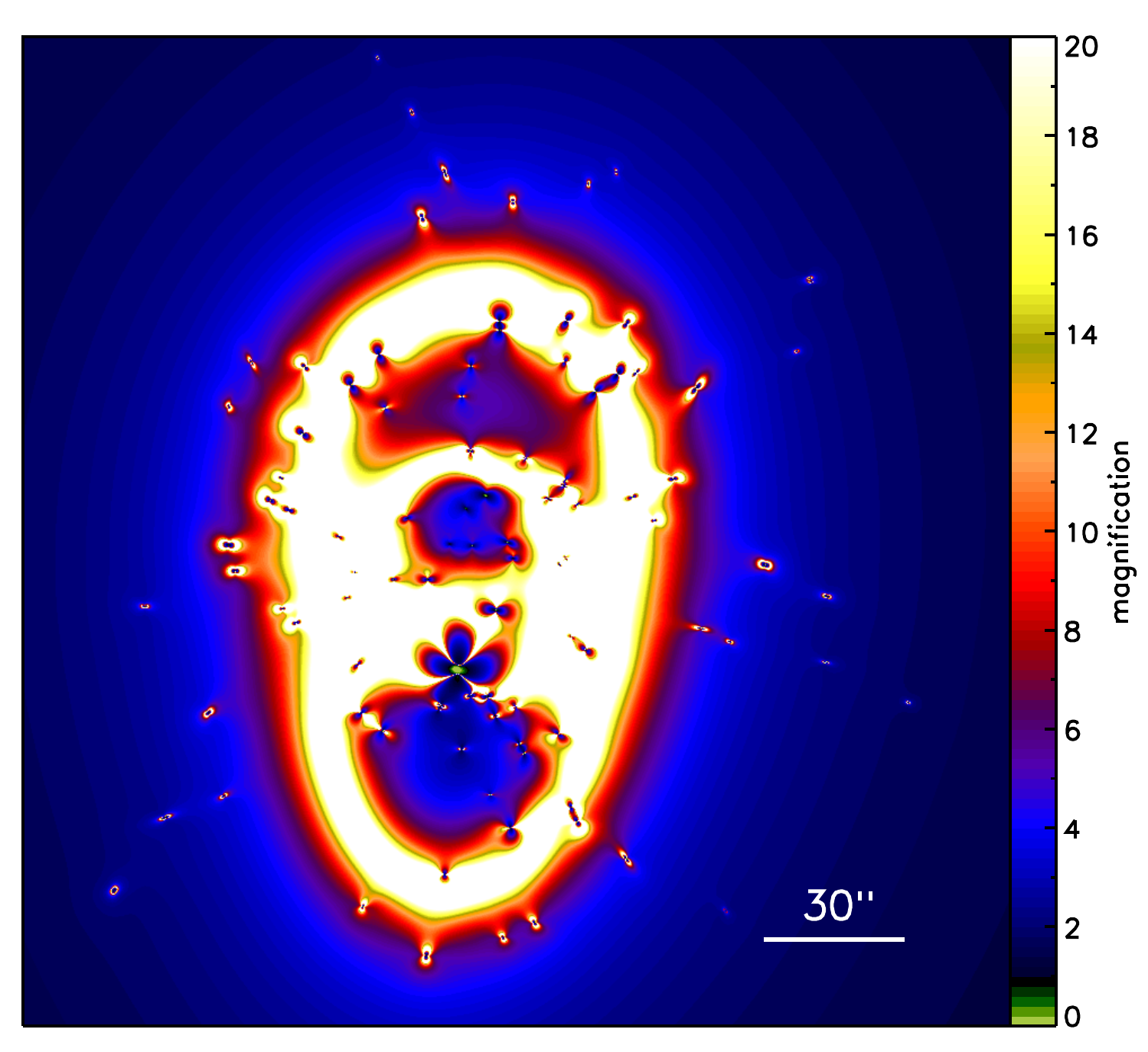}
\caption[Abell 370 image constraints and critical curves]{Top left: False color image of Abell 370 from archival ACS imaging (red, F814W; green, F606W; blue, F435W) by HST SM4 ERO program 11597 (PI: K. Noll), program 11591 (PI:. J.-P. Kneib), program 11582 (PI: A. Blain), and program 11108 (PI: E. Hu).  Labels are the same as in Figure \ref{fig:crit_a2744}. Top right: The inset shows the image constraints of the giant arc (image system \#2); the gray line is the critical curve corresponding to the spectroscopic redshift of this arc, $z=0.725$. Bottom left: Contour map of the total surface mass distribution (in units of $10^{13}\ \mathrm{M_\odot \ kpc^{-2}}$) overlaid on the mass contained in cluster member galaxies. The spacing of the contours is linear. The locations and labels of each optimized halo (see Table \ref{tab:a370_params}) are shown in red. The blue box indicates the field of view of the top image. Bottom right: Absolute value of magnification in the image plane for a source at $z=9$.}
\label{fig:crit_a370}
\end{figure*}

The lens model for Abell S1063 at $z=0.348$ consists of three cluster-scale halos: a central halo located near the BCG (H1), a halo $\sim400\arcsec$ northeast of the cluster center (H2), and another $\sim100\arcsec$ to the south (H3). We assign circular potentials with fixed $r_\mathrm{core}=50\ \mathrm{kpc}$ to the secondary halos since there are no strong lensing constraints in this vicinity to constrain the inner slope of the density profiles. We can only constrain the mass and position of these halos and the slope of the density profiles at the location of the multiple images, which allows us to place constraints on the velocity dispersion and position of the halos. We included all cluster member galaxies within the ACS FOV, with $m_\star=18.82$ in ACS F814W. We set the velocity dispersions of the BCG and three other cluster member galaxies as free parameters in the model. These three galaxies lie close to image systems \#1 ($\alpha$=22:48:46.925,$\delta$=-44:31:33.60; G1), \#11 ($\alpha$=22:48:41.223,$\delta$=-44:32:25.96; G2), and \#16 ($\alpha$=22:48:39.984,$\delta$=-44:32:05.33; G3). Image constraints, critical curve, mass distribution, and magnification map of the lens model are shown in Figure \ref{fig:crit_as1063}.

The two outer cluster-scale halos (H2 and H3) are new additions to the first version of the model released in September 2013, and are motivated by our spectroscopic redshift measurement of image system \#11. In version 1 of this model, the soft prior that was set by the photometric redshifts of system \#11 and other images in its vicinity allowed the model to converge to a solution with less complexity, by predicting lower source redshifts than the photometric redshift estimates.  We discuss this further in \S \ref{sec:specz}.  We note that the outer halos, which are strictly motivated by the lensing constraints, coincide with higher densities of galaxies in the northeastern most part of the \hst\ FOV. Independent weak lensing models (\citet{Gruen:2013lr} and the Merten et al. preliminary HFF lens model) show evidence for mass in the same regions as these new halos. These structures are outside the \hst\ FOV, but may correspond to galaxy over densities in the wide field imaging used by \citet[][Figure 15 of that publication]{Gruen:2013lr}.

We compute cylindrical masses of $M(r<250\ \mathrm{kpc})=2.68^{+0.03}_{-0.05}\times10^{14}\ \mathrm{M_\odot}$ and $M(r<500\ \mathrm{kpc})=6.39^{+0.14}_{-0.32}\times10^{14}\ \mathrm{M_\odot}$. This cluster has a large effective Einstein radius, $\theta_E=29\farcs8^{+0.1}_{-0.3}$, making it a very efficient lens of background sources. In the first strong lensing analysis of this cluster, \citet{Monna:2014lr} find $\theta_E=29\farcs9^{+1.7}_{-1.9}$ and $M(<\mathrm{crit})=1.24\pm0.01\times10^{14}\ \mathrm{M_\odot}$ for $z=2$, and also compute $M(r<250\ \mathrm{kpc})=2.8\pm0.1\times10^{14}\ \mathrm{M_\odot}$ and $M(r<500\ \mathrm{kpc})=6.3\pm0.3\times10^{14}\ \mathrm{M_\odot}$. From weak lensing analysis, \citet{Gruen:2013lr} find $M(r<250\ \mathrm{kpc})=2.3^{+0.3}_{-0.2}\times10^{14}\ \mathrm{M_\odot}$ and $M(r<500\ \mathrm{kpc})=6.1^{+0.6}_{-0.7}\times10^{14}\ \mathrm{M_\odot}$. All three of these analyses are in excellent agreement. This cluster has been studied in detail in the optical and x-ray \citep{Cruddace:2002vn,Maughan:2008rt,Comis:2011fr,Gomez:2012yq} and in exploring its SZ effect \citep{Plagge:2010ys}. Many of these studies indicate a complex mass distribution beyond the FOV of existing \hst\ data for this cluster.

\subsection{Abell 370}

We use the multiple images identified by \citet{Richard:2010wd} and  \citet{Richard:2014gf}$^{\ref{note}}$ as constraints in our lens model, and fix the redshifts of systems \#1, 2, 3, 4, and 6 to the spectroscopic redshifts measured by these groups. The list of image constraints used in the model can be found in Table \ref{tab:a370_arcs} in the Appendix.

We model Abell 370 at $z=0.375$ with two cluster-scale dark matter halos and cluster member galaxies, for which we scale the parameters using the ACS F814W magnitudes with $m_\star=19.04$. We allow the velocity dispersion of the BCG ($\alpha$=2:39:53.125, $\delta$=-1:34:56.42) to be optimized. Early model iterations indicate that the core and cut radii of this galaxy cannot be constrained by the lensing evidence, so we leave these parameters fixed to the scaled values. The perturbing galaxy at ($\alpha$=2:39:52.595, $\delta$=-1:35:06.22; G1) is responsible for creating the swallowtail caustic which produces the quintuply-imaged giant arc (image system \#2). We began by allowing all parameters of the galaxy except position to vary in the model; however, we found that only velocity dispersion, position angle, and ellipticity could be constrained. The ellipticity of the galaxy is forced by the constraints to be unrealistically high for an elliptical galaxy, so we fixed the value to $e=0.8$ as opposed to the value matching the light distribution and leave velocity dispersion and position angle as free parameters. Figure \ref{fig:crit_m0416} shows the critical curve, image constraints, mass distribution, and magnification map of this cluster.

We compute masses of $M(r<250\ \mathrm{kpc})=3.48^{+0.02}_{-0.04}\times10^{14}\ \mathrm{M_\odot}$ within 250 kpc of the cluster center and $M(<\mathrm{crit})=2.36^{+0.02}_{-0.05}\times10^{14}\ \mathrm{M_\odot}$ within the $z=2$ critical curve for Abell 370, respectively. The strong lensing \texttt{LENSTOOL} model by \citet{Richard:2010wd} yields $M(r<250\ \mathrm{kpc})=3.8\pm0.2\times10^{14}\ \mathrm{M_\odot}$ and $M(<\mathrm{crit})=2.82\pm0.15\times10^{14}\ \mathrm{M_\odot}$  ($z=2$ critical curve). Our values differ slightly from those of \citeauthor{Richard:2010wd} Both models used similar image constraints; however, our model is up to date with the latest spectroscopic redshifts by  \citet{Richard:2014gf}, which may account for the discrepancy. These measurements are responsible for producing a narrower range of uncertainties on the derived masses in our model compared to the earlier \citeauthor{Richard:2010wd} model.

\section{Discussion and future work}
\label{sec:discussion}

The detailed lens models in this work are based on archival \hst\ data that exist prior to the deep imaging of the HFF in \hst\ Cycles 21-23, and on all the known spectroscopic redshifts of lensed galaxies as of this publication. These models can be used for deriving magnification estimates for background sources as well as for studying the mass distribution approximately within the footprint of these archival data (within 3\arcmin\ of the cluster center). In this section, we highlight a few caveats and discuss the implications of some possible uncertainties and systematics on the model parameters and magnification estimates; we also outline future work that would advance our understanding of these issues. 

\subsection{Precision in lensing maps}
\label{sec:precision}

Our models of the HFF derive magnifications most precisely within the strong lensing regime, approximately the region enclosed by the strongly lensed galaxies (typically within $\sim100\arcsec$ of cluster center). Therefore, the regions within a few arcseconds of any images used as constraints in our model will have the most precise magnifications, especially if those images had spectroscopic redshifts.

The regions of the map that are most vulnerable to high statistical and modeling errors are generally near the critical curves and far from any image constraints. The critical curve in the image plane shows regions of the lensing map where the magnification diverges to extremely high values. The magnification values drop off quickly with projected distance from the critical curves and converge to magnifications of unity far from the cluster center. Slight changes in the lens parameters may cause a small shift in the location of the critical curve, and significantly change the magnification values near the critical curves. Nevertheless, since the critical curves map lines of reflective symmetry within the lensing map, the location of the critical curve is well constrained between the multiply imaged galaxies.

While our model can be extrapolated as far as the location of the parallel fields, roughly $6\arcmin$ away from the center of the cluster fields, we do not recommend the use of our models for computing the magnifications in these regions, primarily because the outer slope of the mass distribution is not observationally constrained outside the strong lensing regime. Additionally, we may not account for all the mass outside of the combined FOV of existing \hst\ data (additional cluster members, large-scale structure, etc.) which could boost the lensing magnification in these regions. Because there are no strong lensing constraints here, the extrapolation results in a crude, unconstrained estimate of the magnification. For the parallel fields, we suggest that one uses maps generated by other lensing techniques, which include weak lensing as constraints (e.g., the preliminary HFF models by Merten cover the parallel fields).

\subsection{The importance of spectroscopic redshift confirmation}
\label{sec:specz}
Our revised model for Abell S1063 demonstrates the necessity of spectroscopic follow-up of the multiply lensed galaxies for improving the accuracy of strong lens modeling. As mentioned in \S \ref{sec:results}, we found several inconsistencies in the derived mass distribution between our models, which use the full extent of spectroscopic and photometric redshifts of the lensed galaxies and models, and models which do not include observational constraints on redshift. In these few studied cases, we find that models with limited lensing constraints produce enclosed masses 10-20\% higher than models with more redshift constraints. We have yet to explore these findings, and determine whether the systematic discrepancies cannot be attributed to differences in modeling techniques or assumptions.

In this section, we investigate how adding a new spectroscopic redshift to the model of Abell S1063 affects the model-predicted redshifts of multiple image systems and the magnification of background sources by comparing two different lens models with and without the spectroscopic redshift constraint for image \#11. Our preliminary lens model (hereby referred to as Model A) of this cluster included image system \#11 as a constraint, using the BPZ range as a redshift prior. The model for Abell S1063 we present in this paper (hereby referred to as Model B) uses identical image constraints as Model A, except that we use the newly measured spectroscopic redshift of this image system. The lens model components of the two models are similar, except that cluster halos \#2 and \#3 (see Table \ref{tab:as1063_params}) were not included in Model A. The details of Model B, as well as the reasons for including the additional halos, are explained in \S \ref{sec:results_as1063}.

\begin{figure}[h]
\epsscale{1.2}
\plotone{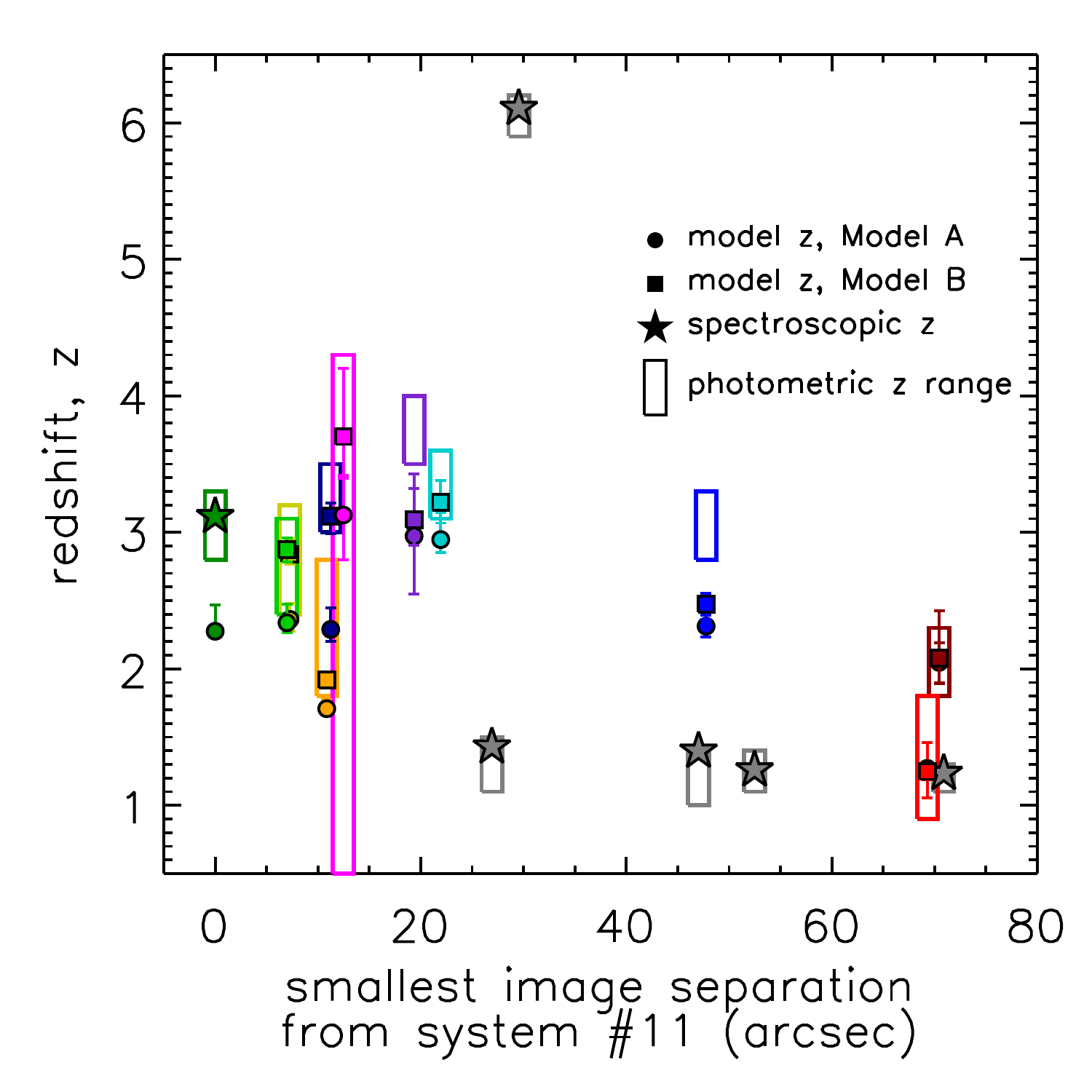}
\caption{We show the model-predicted redshifts of image systems included as constraints in our preliminary lens model (Model A in text, circles) of Abell S1063 and the model we present here (Model B in text, squares) plotted versus their shortest image plane separation from one of the images in system \#11. The gray stars indicate image systems fixed to their spectroscopic redshifts in both models. In Model A, the redshift of image system \#11 is left as a free parameter, where in model B, we fix the redshift to the spectroscopic redshift. There is a systematic increase in the model-predicted redshifts from Model A to Model B for the images closest ($<50\arcsec$) to the new spectroscopic redshift system \#11. We also plot the 95\% confidence range for photometric redshifts.}
\label{fig:as1063_free_z_dist}
\end{figure}

When the redshifts of image constraints are left as free parameters, the best-fit model predicts the most likely redshift for each source.  In Figure \ref{fig:as1063_free_z_dist}, we plot the model-predicted redshifts of all the multiple image systems used as constraints in both Models A and B against the minimum angular separation from an image in system \#11. We find that the model-derived redshifts of image systems within $\sim25$\arcsec\ of image \#11 were systematically lower in Model A. Nevertheless, this model formally converged to a ``good" solution with small image plane rms (1\farcs2). The change in predicted redshift between models becomes less significant with increasing image plane separation; however, this may be tied to a closer proximity to other spectroscopic redshift systems. Images of systems \#3 and \#4 are close to system \#1, which has a spectroscopic redshift constraint in both models; the model-derived redshifts of these systems did not change between models within the errors. As discussed in \S\ref{sec:results_as1063}, the new spectroscopic redshift constraint was inconsistent with the model-predicted redshift for Model A and forced us to include two secondary cluster-size halos in order to converge on a new solution in model parameter space. This new mass drives the free redshift parameters in that part of the image plane to higher values. The photometric redshifts of lensed galaxies behind this cluster \citep[derived from 16-band CLASH data, ][]{Jouvel:2014qy} are generally in good agreement with the spectroscopic redshifts measured so far. This may not always be the case, especially when the photometric redshifts are based on only a few bands. The output redshift probability distributions may have multiple peaks, converge on the wrong redshift, or have large uncertainties. Therefore, the photometric redshift constraints for an individual galaxy are treated with more suspicion if they are inconsistent with the lensing geometry or other evidence. Nevertheless, if there appears to be a systematic offset between the model-predicted redshifts and photometric redshifts of several lensed galaxies (i.e., they are all significantly higher or lower), one must consider the possibility that there may be an error in the lens modeling assumptions and the model needs to be revised, as was the case in the revision of Model A to Model B.

\begin{figure}[h]
\epsscale{1.2}
\plotone{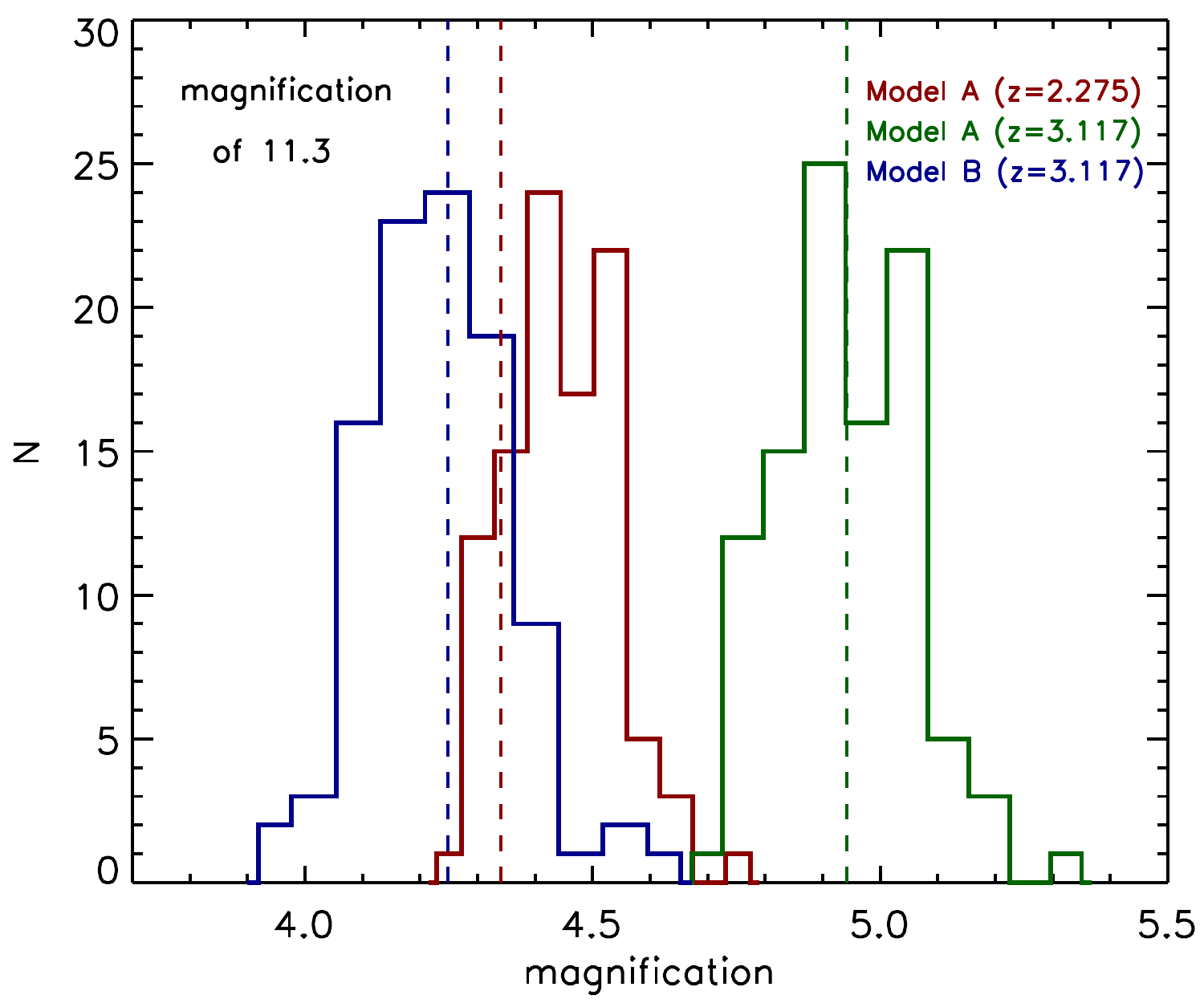}
\caption{We plot the magnifications of image \#11.3 from Model A (the redshift of image system \#11 is a free parameter in the model) and Model B (the redshift of image \#11 is fixed to the spectroscopic redshift). For Model A, we derive the magnification for two source plane redshifts: the model-predicted redshift of the image system $z=2.275$ (red) and the spectroscopic redshift $z=3.117$ (green). Model B (blue) is set to the spectroscopic redshift. The dashed lines represent the value of magnification of the best-fit model.}
\label{fig:mag11}
\end{figure}

Constraints with spectroscopic redshifts in lens models will have a profound impact on reducing the uncertainties and increasing accuracy of the magnification values computed from a lens model. The magnification is required for relating many observables to the intrinsic properties of the background galaxies, and as the primary focus of the HFF is studying the populations of the galaxies behind these clusters, achieving the most accurate magnification maps possible is imperative. We begin to investigate the effects on the magnification by adding new spectroscopic redshifts by looking at the distribution of magnifications in a single location in the image plane. In Figure \ref{fig:mag11}, we show the distribution of magnifications of image 11.3 computed from the model simulations of Model A and Model B. We selected this image because it is in a region far from the critical curve where the magnification gradient is small, but in the part of the image plane where \#11 has a strong constraint on the location of the critical curve. For Model A, we compute magnifications at two different source planes: $z=2.275$, the model-predicted redshift for image \#11 from Model A, and $z=3.117$, the spectroscopic redshift of this image we measured in this work. For Model B, we compute the magnification corresponding to the spectroscopic redshift. The magnification distributions of Model A at $z=2.275$ and Model B overlap. This is expected, because the $z=2.275$ critical curve for Model A and the $z=3.117$ critical curve for Model B overlap in the image plane for the lowest image plane rms model and redshift parameters in both scenarios. However, the magnification is 10\% higher in Model A than Model B for identical source plane redshifts and the distributions do not overlap, mimicking a scenario where one is interested in the magnification of a random galaxy at this redshift.  Typically, this statistical errors on magnification will increase in regions with higher magnification gradients (closer to critical curves). This investigation, albeit somewhat anecdotal and far from being thorough, indicates that the lensing magnification of a galaxy that is used as constraint may be a robust measurement. Further analysis is needed to determine whether this is universal to lens models.

Figure \ref{fig:mag11} addresses the concern of magnification accuracy, but does not address precision. Including spectroscopic redshifts will allow models to converge on a set of parameters which best describe the true mass distribution and lensing of the cluster; however, the precision of the model depends on how well those best-fit parameters can be constrained. Including two additional halos to the model outside the \hst\ FOV is required to explain the image configurations of the strongly lensed galaxies; however, the exact values for the parameters of these halos have wide distributions. We are finding that this can result in wider distribution of magnification values in the image plane of Model B. To investigate this further, we will need to homogenize the model inputs (e.g., redshift priors, parameter priors, number of halos, etc.) in order to isolate the effects of spectroscopic redshifts. We could also consider the outcomes of models with similar inputs, but exclude a priori knowledge on the redshifts of the multiple images. This type of analysis is beyond the scope of this paper, but is future work that we will be looking into. It would also be valuable to determine which, if any, of the properties of the lens models and their outputs are immune to lens modeling assumptions and constraints; we leave this investigation to future work as well.

\subsection{The cumulative magnification power of the HFF}
The derivation of detailed magnification maps, as presented in this work, enable the use the HFF clusters as cosmic telescopes to study the background Universe; in particular, one of the science goals of the HFF is to study the galaxy population at $z\sim9$.  The lensing magnification simultaneously acts to produce two competing effects: it increases the observed source-plane area, and it magnifies faint source above the detection limit. In a given solid angle on the sky the former effect reduces the number of bright galaxies, and the latter increases the number of faint galaxies. It is thus constructive to estimate the source-plane area (and co-moving volume) that is magnified by the cluster as a function of magnification factor. 

\begin{figure}[t]
\epsscale{1.2}
\plotone{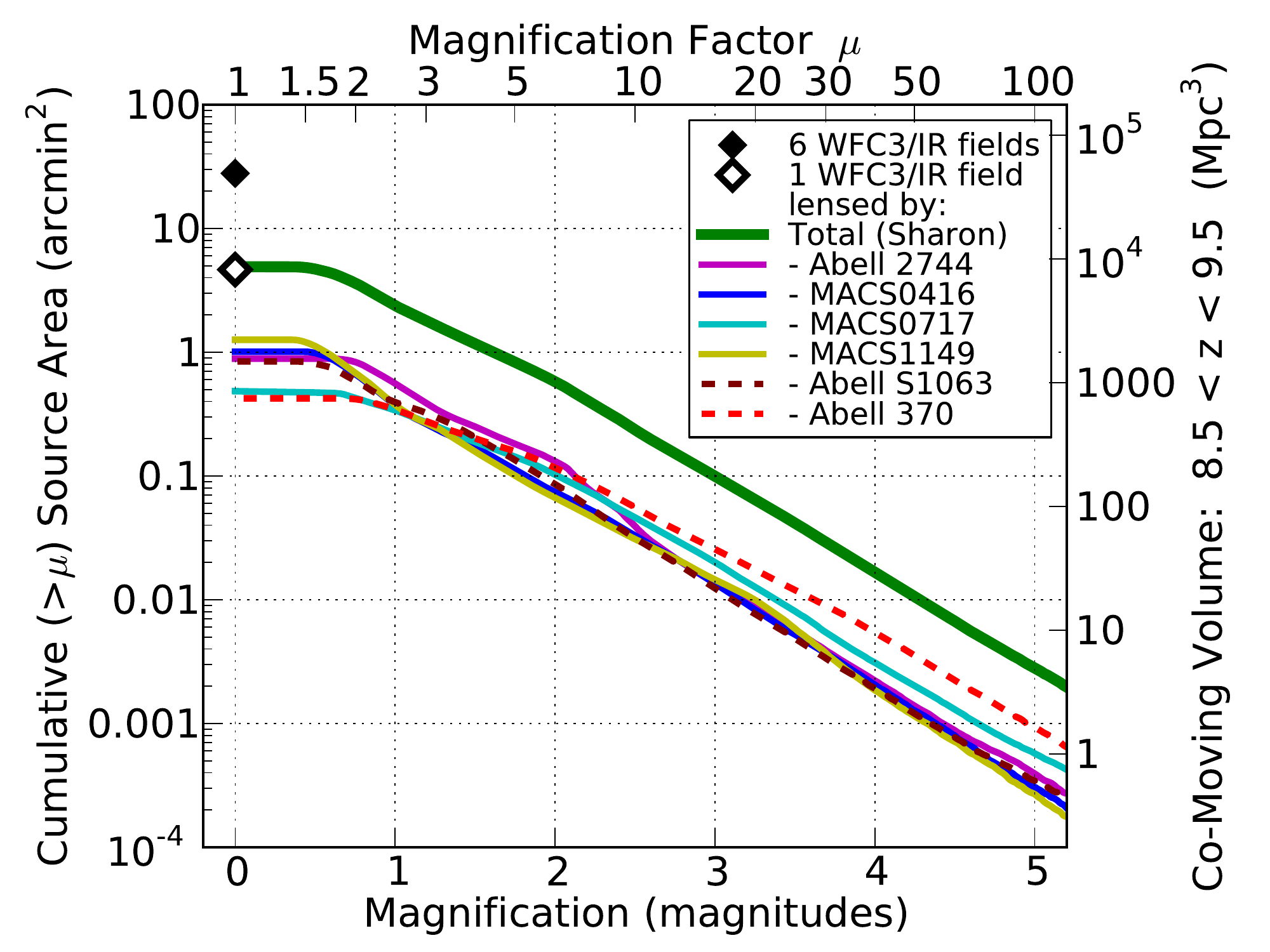}
\caption{We show the cumulative area and co-moving volume of background sources at $8.5<z<9.5$ lensed by each of the HFF clusters to magnification $\mu$ and higher. We show, for reference, the corresponding area and co-moving volume of a one and six WFC3/IR FOVs (diamonds).}
\label{fig:volumes}
\end{figure}

In Figure \ref{fig:volumes} we plot the magnification power of each cluster, as the cumulative volume and area magnified above a certain value as a function of magnification. For each cluster field, the source-plane area is computed by de-lensing the $z=9$ magnification map to the source plane, and summing the area that is magnified by more than a given magnification. For the purpose of estimating the volume we multiply this area by the co-moving distance between $z=8.5$ and $z=9.5$. We find that on average, the total $z=9$ area that is observed through each cosmic telescopes is about 20\% of the WFC3/IR field of view. However, 10\% of this area is magnified by more than a factor of 6, which is equivalent to having a limiting magnitude at least 2 magnitudes fainter in these areas.

\subsection{Model comparisons}

The coordinated parallel effort to generate multiple independent lens models for the HFF provide a unique opportunity for the lensing community to analyze in depth for the first time the systematics of various lens modeling techniques. \hst\ imaging continues to be key for developing methods for lens modeling in clusters, as the pristine image quality is crucial for the identification of multiple image systems for strong lensing and for accurately determining the shear fields for weak lensing. 

Within the past decade or so, we have seen the development of several innovative computational softwares for lens modeling (e.g., \texttt{LENSTOOL}, \citet{Jullo:2007lr}; \texttt{LensPerfect}, \citet{Coe:2008mz}; \texttt{glafic}, \citet{Oguri:2010gr}; \texttt{GRALE}, \citet{Liesenborgs:2010gf}; strong and weak lensing united, \citet{Bradac:2005ve}; SaWLens, \citet{Merten:2009ul}; light-traces-mass, \citet{Broadhurst:2005qy} and \citet{Zitrin:2009qf}; \texttt{GRAVLENS}, \citet{Keeton:2001lr}). Indirect comparisons exist in the literature when different methods were applied to the same clusters; however, direct comparisons of the models require similar modeling inputs and quantitative comparisons are few in number. In-depth analyses of the systematic errors of different lens modeling methods have yet to be done.

The five lens modeling teams for the preliminary HFF lensing maps each use unique modeling techniques, which include strong and weak lensing approaches, parametric and non-parametric constructions of the lensing potential, and differing assumptions regarding the shapes of individual mass halos to list only a few of the differences. The HFF models are perfect for model comparisons since the modeling inputs are nearly identical due of the collective agreements amongst the modeling teams to share these inputs (i.e. photometric and spectroscopic redshifts, galaxy catalogs, etc.) amongst each other. The lens modeling techniques used to map the HFF clusters will be compared directly through the lens modeling of simulated clusters prepared in a similar manner as \citet{Meneghetti:2008ve} and designed to represent the depth and image quality of a complete HFF data set. These model comparisons are ongoing and will be presented elsewhere.

\subsection{Model revisions}

As more data on the HFF clusters becomes available, the strong lensing models will be iterated on and improved. Preliminary models can be used to hunt for more multiple image systems in the deeper data sets. Ground-based campaigns are ongoing to obtain more spectroscopic redshifts to improve the precision of these models.

Future models could incorporate more free parameters to account for both visible and unseen structure along the line of sight. As is indicated in Table \ref{tab:model_summary}, all of these lens models are over-constrained, a trend which will continue as more image systems are discovered, allowing for more free parameters to be added to the models. Currently our models account for structure only at the cluster redshift (correlated) with a few exceptions in some of our models for obvious foreground interlopers, though other uncorrelated groups and cosmic variance could also contribute to the lensing. \citet{Bayliss:2014fk} show from spectroscopic observations evidence for large scale uncorrelated substructure along the line of sight to strong lensing clusters, which could become a significant systematic in lens modeling precision, and \citet{DAloisio:2013ul} estimate that such structures could account for fluctuations of $\sim30\%$ in magnification for highly-magnified sources ($\times10$). As we noted in \S \ref{sec:results_a2744}, there is clear evidence in some of the clusters that line-of-sight structure plays significant role in the overall lensing potential. These clusters would be ideal for developing and testing multi-plane lens modeling technique on observational data. 

\section{Summary}
\label{sec:conclusion}

We present high-quality strong lens models for each of the six Frontier Fields clusters. The models are based on archival \hst\ imaging, obtained prior to the deep HFF imaging in Cycle 21, and on ground-based spectroscopy of the lensed galaxies and cluster members, including the new spectroscopic redshifts for lensed galaxies in Abell S1063 and Abell 2744 we present in this work. We compute parametric, strong-lensing models for each cluster in the HFF using the systems of multiply imaged background galaxies as constraints. The HFF clusters are powerful gravitational lenses; we quantify from our lens models their cumulative magnification power of high redshift galaxies. We compare the cluster masses computed from our models to other lens models in the literature. We generally find that the early models, that did not use redshifts as inputs, are in disagreement with our results, and typically produce 10-20\% higher mass at the core of the cluster. The models presented in this work are publicly available for use in analyzing the lensing of the high-redshift Universe behind these galaxy clusters. We outline the formalism for how the outputs of our lens models can be used to derive the magnifications of any background source in the entire FOV of the HFF observations, and discuss the caveats of using our lens models in these analyses. Specifically, we note that the regions of highest-precision magnification values are those that lie closest to the image constraints used in the model. The highest statistical uncertainties in magnification values lie close to critical curves, where the gradient in magnification is high, and more than $\gtrsim1\arcmin$ from the strong-lensing region, where there are no strong lensing constraints. We begin to address systematic uncertainties in lens modeling, noting that the derived models can change significantly when the assumptions regarding the redshifts of multiply imaged background galaxies are different and plan to investigate these systematics more thoroughly in the future. As more data become available for each cluster as the \hst\ observations take place over the next three years, we will revise these models and provide the public with the most precise models possible. These revisions will be based on new image systems of lensed background galaxies, better constraints on photometric redshift priors, new spectroscopic confirmation of the lensed arcs from ground-based observatories, incorporating line of sight structure, and new inquiry into the systematic modeling errors, which will unfold after ongoing model comparisons.

\acknowledgements
Some of the data presented in this paper were obtained from the Mikulski Archive for Space Telescopes (MAST). STScI is operated by the Association of Universities for Research in Astronomy, Inc., under NASA contract NAS5-26555. Support for MAST for non-HST data is provided by the NASA Office of Space Science via grant NNX13AC07G and by other grants and contracts. This work was supported by NASA Grant \#5-26555. KS acknowledges support from the University of Michigan's President's Postdoctoral Fellowship.

We thank the anonymous referee for contributing helpful comments and feedback. We would like to thank the Frontier Fields lens modeling teams for sharing unpublished data (image identifications, photometric and spectroscopic redshift measurements, galaxy catalogs, and ground-based imaging) which we used as inputs to our lens models and useful for discussion. We thank Marceau Limousin for generously providing the catalogs of galaxy cluster members for \MACSzeroseven\ and \MACSeleven. For their help with supplementary data that were used to guide our mask design for Abell S1063, we wish to thank Daniel Gruen, Piero Rosati, and Stella Seitz). We also thank Daniel Gruen for providing the aperture masses from weak lensing models for direct comparisons with the models presented in this paper. We thank Daniel Gifford for carrying out our November 2013 Magellan/IMACS observations of Abell 2744.


\newpage

\appendix

\begin{figure*}[h]
\epsscale{1.0}
\plotone{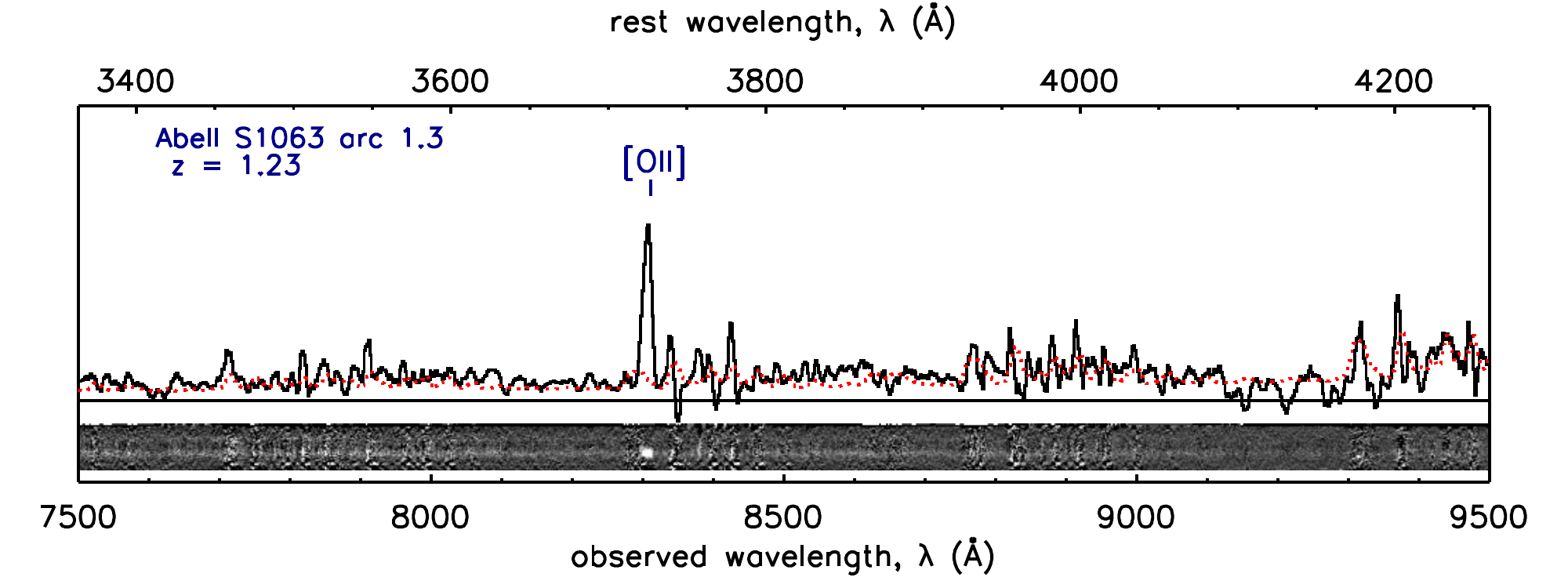}
\caption[Abell S1063, arc 1.3 spectrum]{One-dimensional and two-dimensional spectra of Abell S1063 arc 1.3. The noise level of the one-dimensional spectrum is plotted in red.}
\label{fig:as1063_spec1_3}
\end{figure*}

\begin{figure*}[h]
\epsscale{1.0}
\plotone{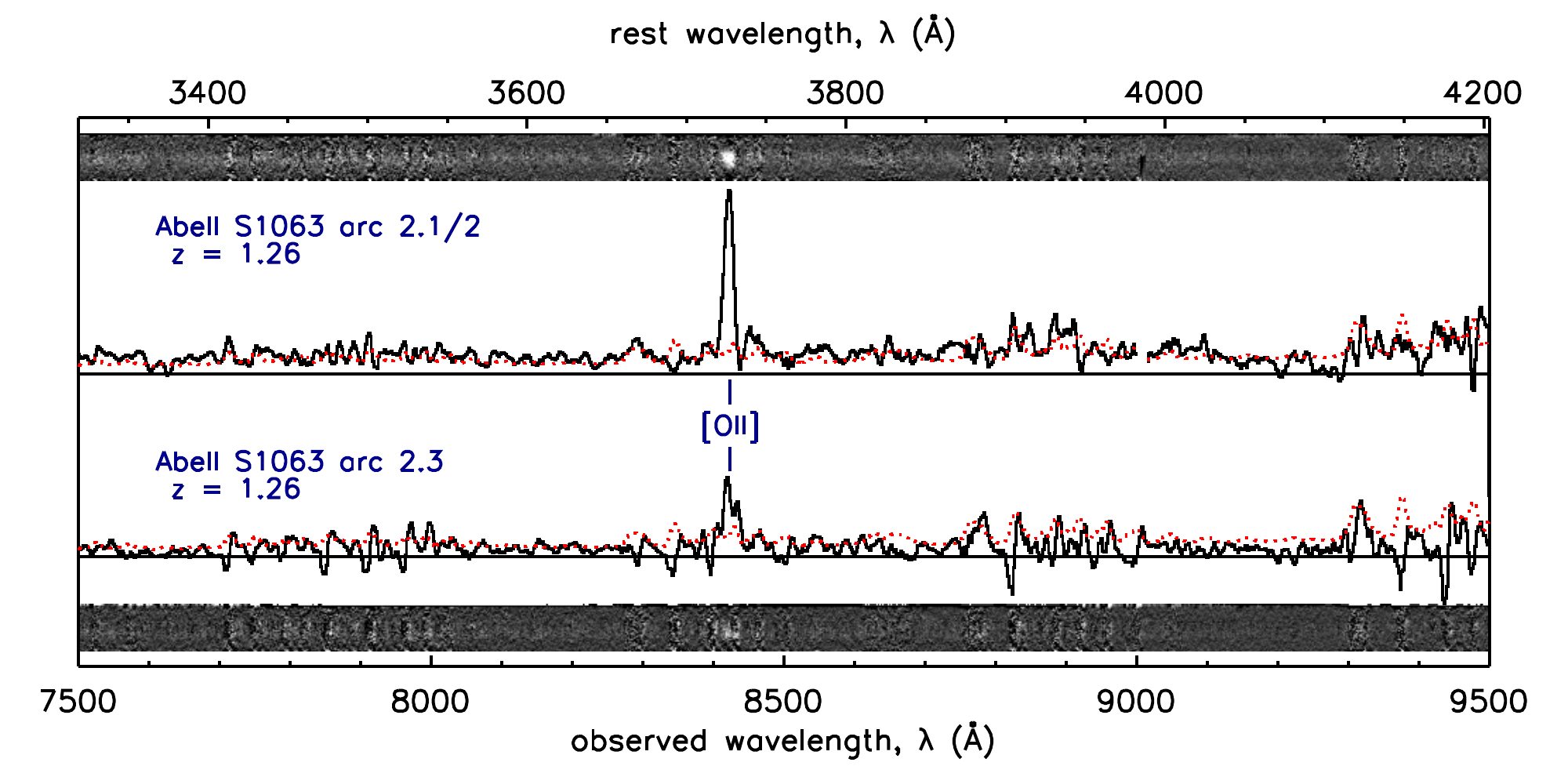}
\caption[Abell S1063, arc 2.1/2 and 2.3 spectra]{One-dimensional and two-dimensional spectra of Abell S1063 arcs 2.1/2 (merging pair) and 2.3. The noise level of the one-dimensional spectrum is plotted in red.}
\label{fig:as1063_spec2}
\end{figure*}

\begin{figure*}[h]
\epsscale{1.0}
\plotone{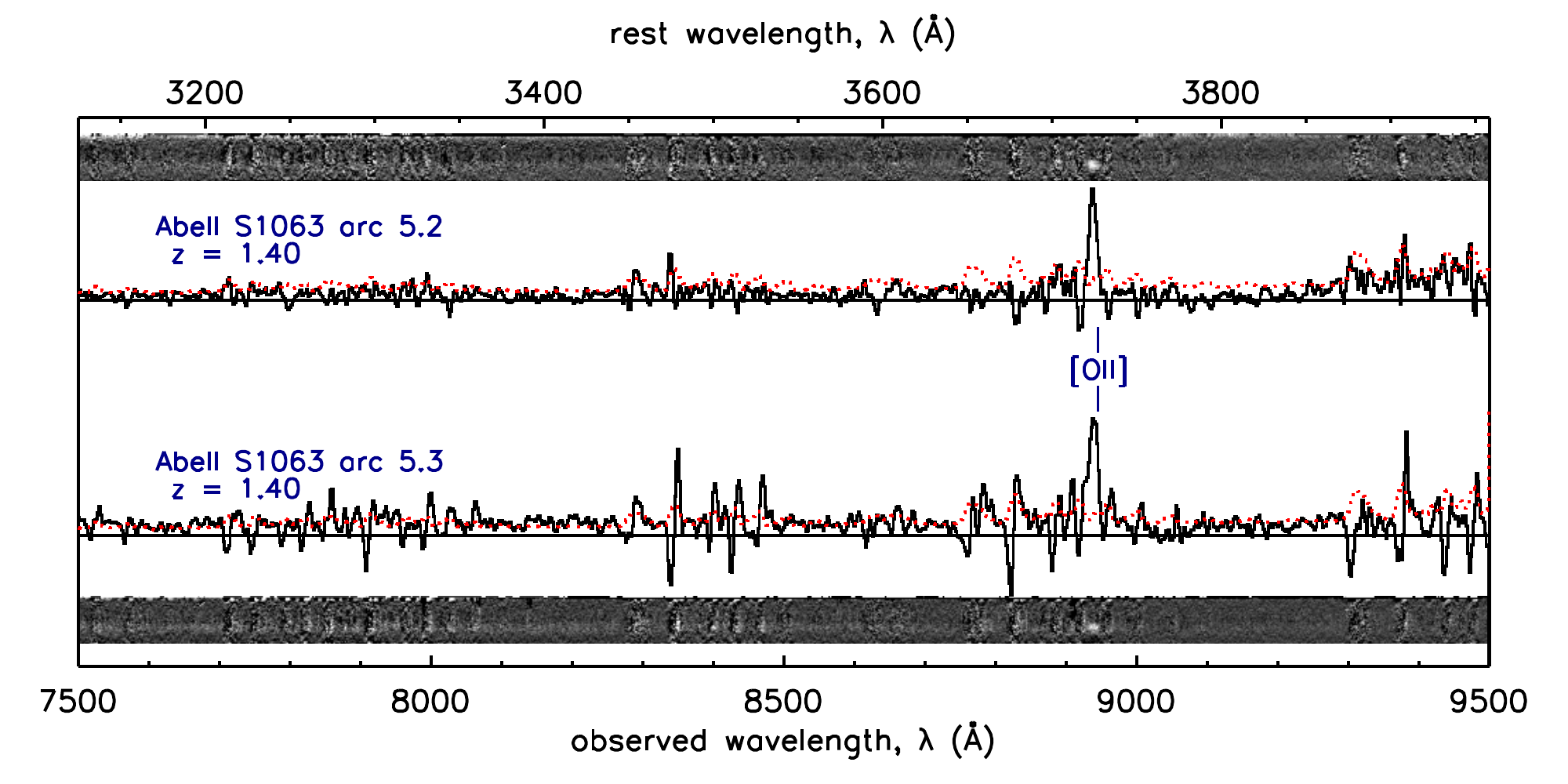}
\caption[Abell S1063, arc 5.2 and 5.3 spectra]{One-dimensional and two-dimensional spectra of Abell S1063 arcs 5.2 and 5.3. The noise level of the one-dimensional spectrum is plotted in red.}
\label{fig:as1063_spec5}
\end{figure*}

\newpage
\begin{figure*}[h]
\epsscale{1.0}
\plotone{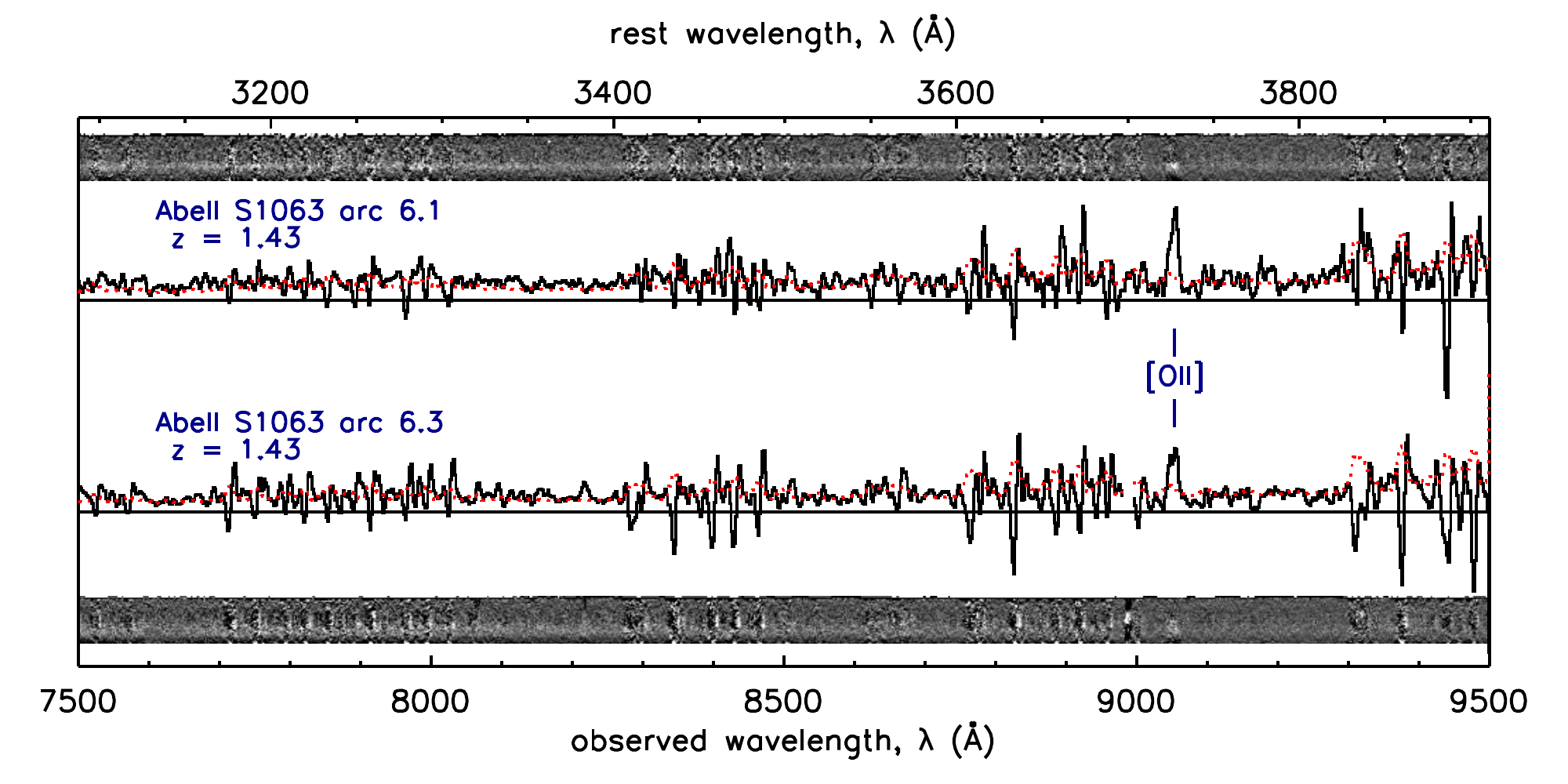}
\caption[Abell S1063 arcs 6.1 and 6.3]{One-dimensional and two-dimensional spectra of Abell S1063 arcs 6.1 and 6.3. The noise level of the one-dimensional spectrum is plotted in red.}
\label{fig:as1063_spec6}
\end{figure*}

\begin{figure*}[h]
\epsscale{1.0}
\plotone{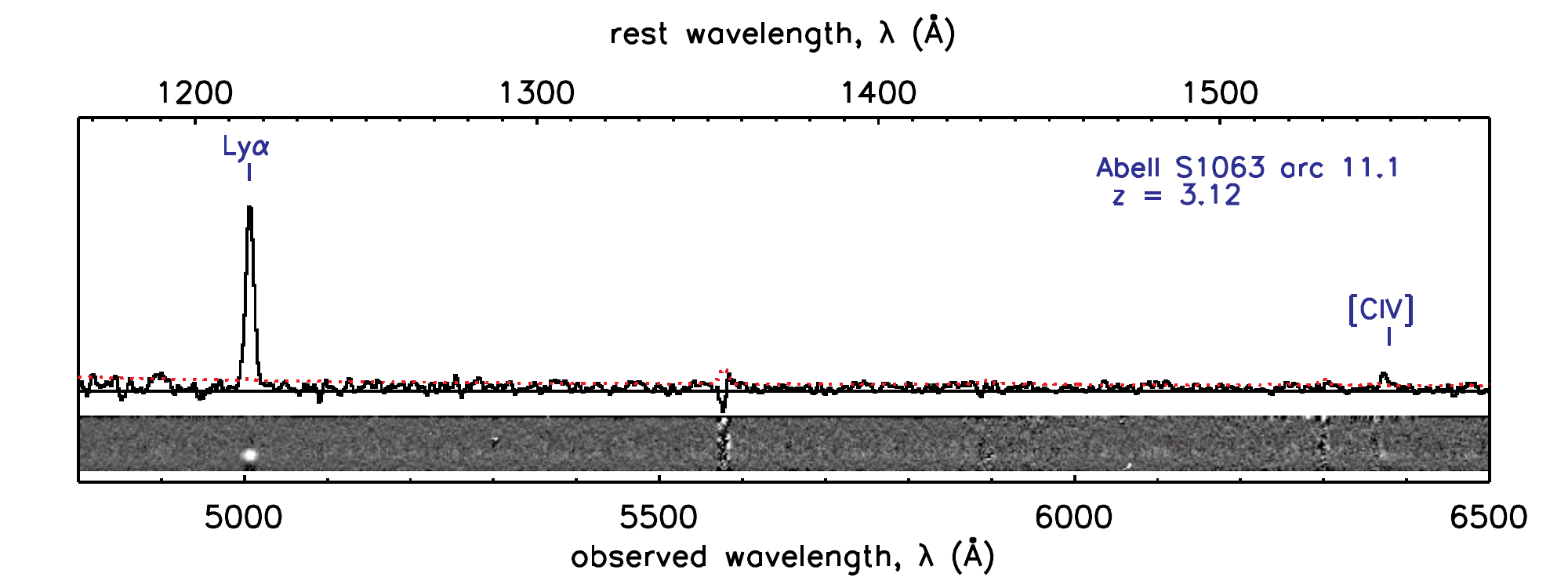}
\caption[Abell S1063 arc 11.1 spectrum]{One-dimensional and two-dimensional spectra of Abell S1063 arc 11.1. The noise level of the one-dimensional spectrum is plotted in red.}
\label{fig:as1063_spec11_1}
\end{figure*}

\begin{figure*}[h]
\epsscale{1.0}
\plotone{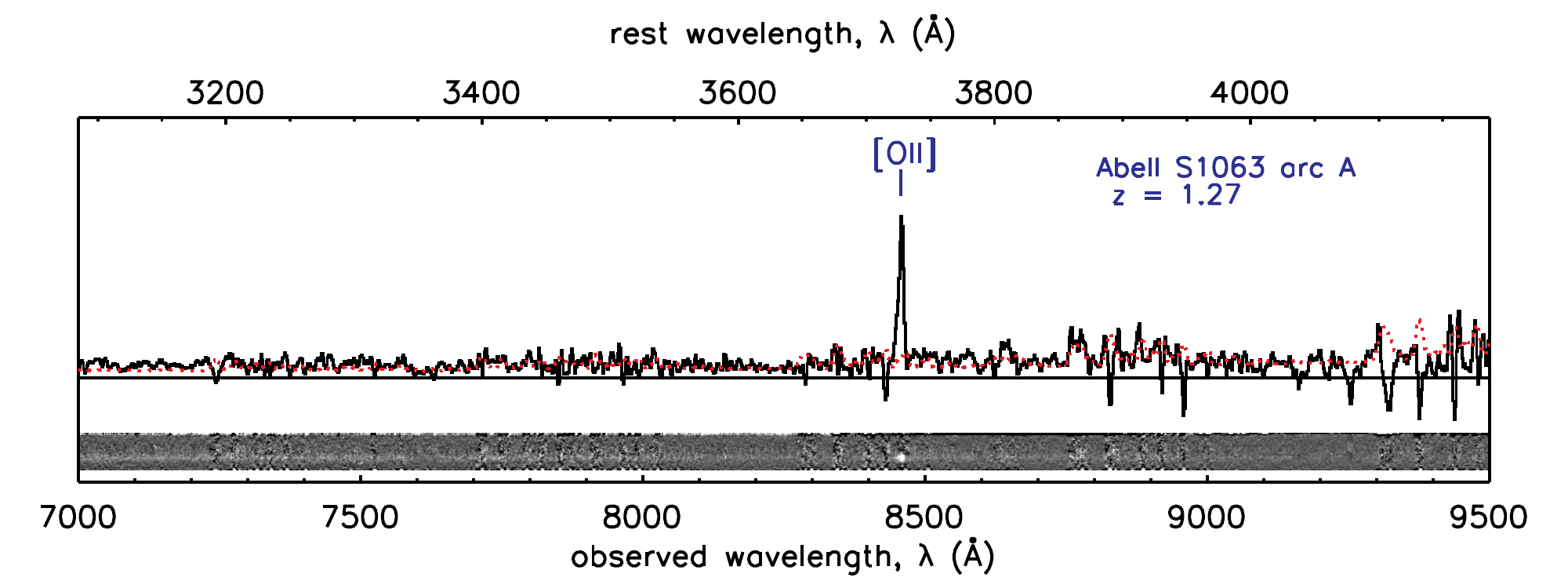}
\caption[Abell S1063 arc A spectrum]{One-dimensional and two-dimensional spectra of Abell S1063 arc A. The noise level of the one-dimensional spectrum is plotted in red.}
\label{fig:as1063_specA}
\end{figure*}

\begin{figure*}[h]
\epsscale{1.0}
\plotone{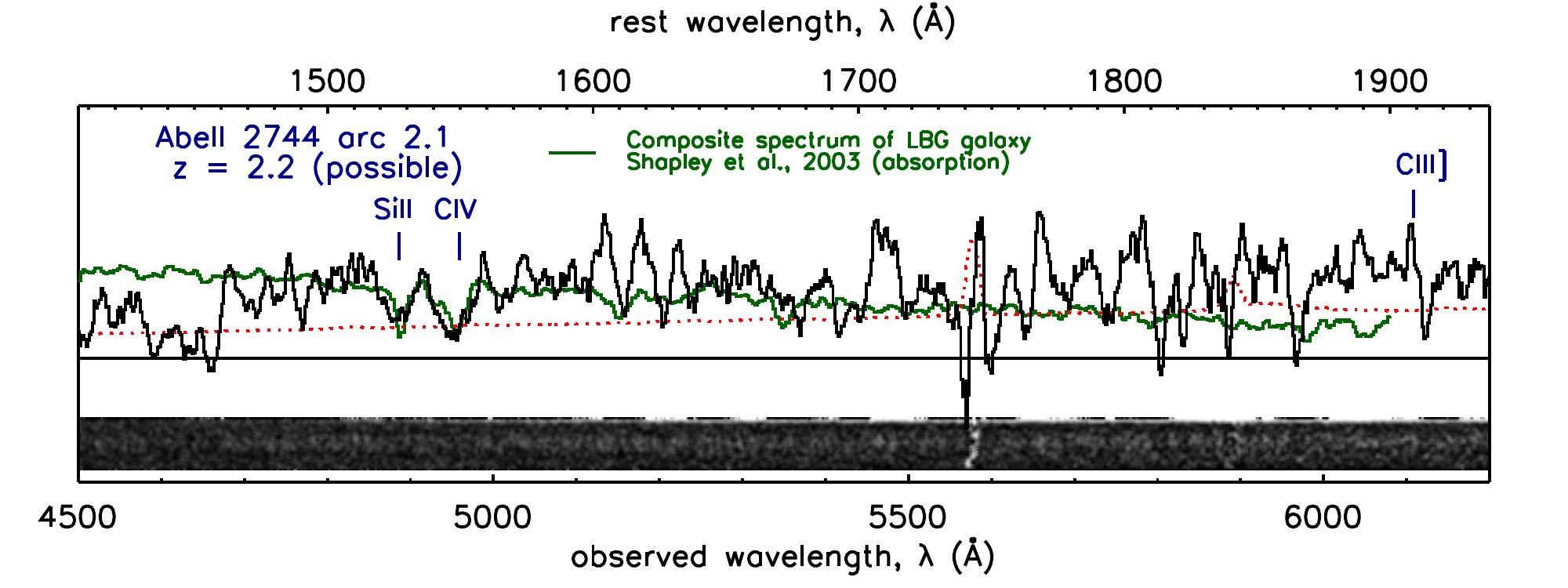}
\caption[Abell 2744, arc 2.1 spectrum]{One-dimensional and two-dimensional spectra of Abell 2744 arc 2. The noise level of the one-dimensional spectrum is plotted in red. We over plot a composite spectrum of high-redshift ($z=3$) Lyman break galaxies with strong absorption features from \citet{Shapley:2003fk} to show similarities in spectral features as well as the Lyman break in the continuum. We find a possible solution for the redshift $z\sim2.2$ for this galaxy.}
\label{fig:a2744_spec2}
\end{figure*}

\begin{figure*}[h]
\epsscale{1.0}
\plotone{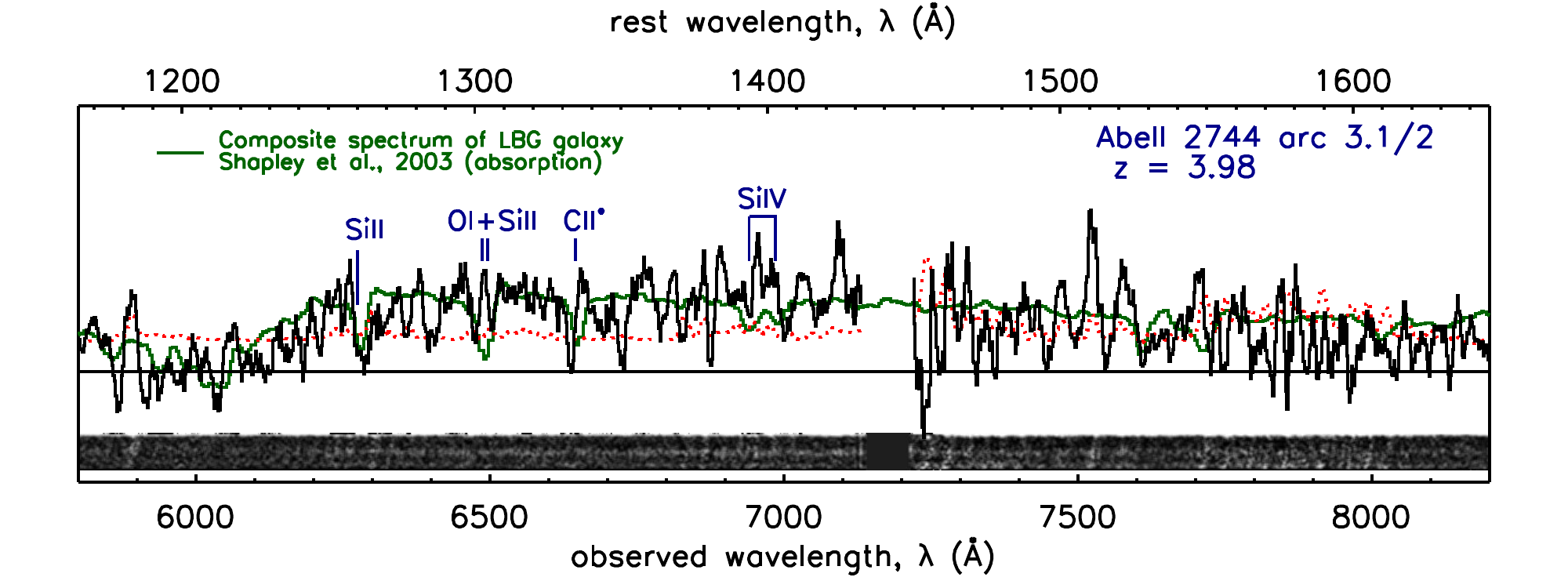}
\caption[Abell 2744, arc 3.1/2 spectrum]{One-dimensional and two-dimensional spectra of Abell 2744 arcs 3.1/2 (merging pair). The noise level of the one-dimensional spectrum is plotted in red. We over plot a composite spectrum of high-redshift ($z=3$) Lyman break galaxies with strong absorption features from \citet{Shapley:2003fk} to show similarities in spectral features as well as the Lyman break in the continuum. This lensed arc corresponds to a star-forming galaxy at $z=3.98$.}
\label{fig:a2744_spec3}
\end{figure*}

\newpage
\begin{figure*}[h]
\epsscale{1.0}
\plotone{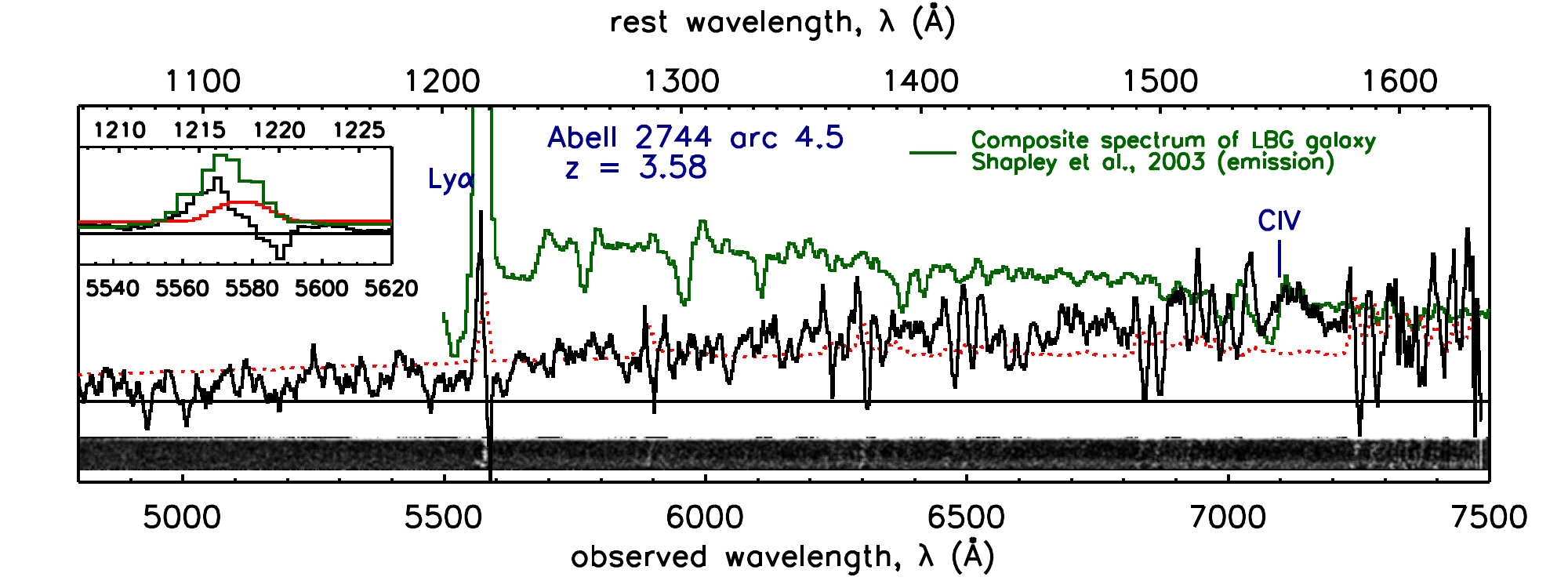}
\caption[Abell 2744, arc 4.5 spectrum]{One-dimensional and two-dimensional spectra of Abell 2744 4.5. The noise level of the one-dimensional spectrum is plotted in red. We over plot a composite spectrum of high-redshift ($z=3$) Lyman break galaxies with strong Lyman $\alpha$ emission from \citet{Shapley:2003fk} to show similarities in these spectral features as well as Lyman break in the continuum. The Ly$\alpha$ emission line lies slightly blueward of the 5577\AA\ skyline residual, as shown in the inset in the upper left. We find a likely solution of $z=3.58$ for this galaxy.}
\label{fig:a2744_spec4}
\end{figure*}

\begin{figure*}[h]
\epsscale{1.0}
\plotone{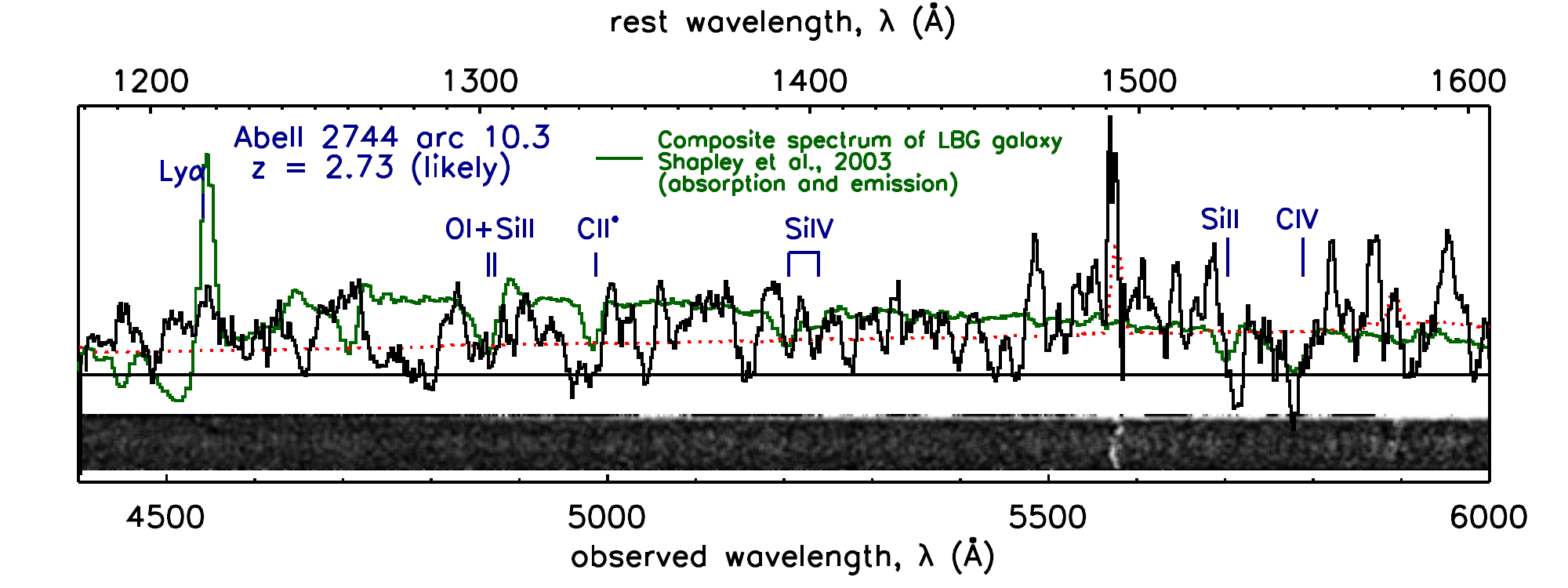}
\caption[Abell 2744, arc 10.3 spectrum]{One-dimensional and two-dimensional spectra of Abell 2744 10.3. The noise level of the one-dimensional spectrum is plotted in red. We over plot a composite spectrum of high-redshift ($z=3$) Lyman break galaxies with both strong absorption and emission features from \citet{Shapley:2003fk} to show similarities in these spectral features as well as Lyman break in the continuum. We find a possible solution for the redshift $z=2.73$ for this galaxy.}
\label{fig:a2744_spec10}
\end{figure*}

\begin{deluxetable}{ccccccccc}
\tablecaption{Abell 2744 image constraints}
\tabletypesize{\tiny}
\tablenotetext{1}{This work.}
\tablenotetext{2}{ \citet{Richard:2014gf}.}
\tablenotetext{3}{BPZ measured from \hst\ preliminary data reductions.}
\tablenotetext{4}{Image system fixed to $z=3$ (see text, \S \ref{sec:results_a2744}).}
\tablehead{
    \colhead{Image} &
    \colhead{R.A.} &
    \colhead{Dec.} &
    \colhead{Spec $z$} &
    \colhead{Photo $z$\tablenotemark{3}} &
    \colhead{Model $z$} &
    \colhead{$z$ prior} &
    \colhead{Image plane} &
    \colhead{Image plane}
    \\
    \colhead{system} &
    \colhead{} &
    \colhead{} &
    \colhead{} &
    \colhead{} &
    \colhead{} &
    \colhead{} &
    \colhead{individual rms (\arcsec)} &
    \colhead{system rms (\arcsec)}
}
\startdata
1.1 & 00:14:23.41 & -30:24:14.10 & \nodata & 0.5-2.2 & $1.74^{+0.09}_{-0.08}$ & 0.5-2.6 & 0.04 & 0.34 \\
1.2 & 00:14:23.03 & -30:24:24.56 &  &  &  &  & 0.18 &  \\
1.3 & 00:14:20.69 & -30:24:35.95 &  &  &  &  & 0.05 &  \\
\hline
2.1 & 00:14:19.98 & -30:24:12.06 & \nodata & 0.5-2.9 & $1.91^{+0.14}_{-0.05}$ & 0.4-3.0 & 0.61 & 0.61 \\
2.2 & 00:14:23.35 & -30:23:48.21 &  &  &  &  & 0.15 &  \\
2.3 & 00:14:20.50 & -30:23:59.63 &  &  &  &  & 0.19 &  \\
2.4 & 00:14:20.74 & -30:24:07.66 &  &  &  &  & 0.33 &  \\
\hline
3.1 & 00:14:21.45 & -30:23:37.95 & 3.98\tablenotemark{1} & \nodata & \nodata & \nodata & 0.27 & 0.49 \\
3.2 & 00:14:21.31 & -30:23:37.69 &  &  &  &  & 0.07 &  \\
3.3 & 00:14:18.60 & -30:23:58.44 &  &  &  &  & 0.32 &  \\
\hline
4.1 & 00:14:22.11 & -30:24:09.48 & 3.58\tablenotemark{2} & \nodata & \nodata & \nodata & 0.42 & 0.76 \\
4.2 & 00:14:22.95 & -30:24:05.84 &  &  &  &  & 0.65 &  \\
4.3 & 00:14:19.30 & -30:24:32.13 &  &  &  &  & 0.68 &  \\
4.4 & 00:14:22.37 & -30:24:17.69 &  &  &  &  & 0.57 &  \\
4.5 & 00:14:22.46 & -30:24:18.38 &  &  &  &  & 0.53 &  \\
\hline
6.1 & 00:14:23.65 & -30:24:06.48 & 2.02\tablenotemark{2} & \nodata & \nodata & \nodata & 0.28 & 0.48 \\
6.2 & 00:14:22.57 & -30:24:28.84 &  &  &  &  & 0.10 &  \\
6.3 & 00:14:20.74 & -30:24:33.74 &  &  &  &  & 0.25 &  \\
\hline
7.1 & 00:14:23.58 & -30:24:08.35 & \nodata & 0.3-3.4 & $2.69^{+0.23}_{-0.12}$ & 2.5-3.7 & 0.11 & 0.45 \\
7.2 & 00:14:22.85 & -30:24:26.73 &  &  &  &  & 0.18 &  \\
7.3 & 00:14:20.30 & -30:24:35.33 &  &  &  &  & 0.28 &  \\
\hline
8.1 & 00:14:21.53 & -30:23:39.62 & \nodata & \nodata & $4.57^{+0.87}_{-0.45}$ & 3.0-6.5 & 0.29 & 0.51 \\
8.2 & 00:14:21.32 & -30:23:39.20 &  &  &  &  & 0.24 &  \\
\hline
9.1 & 00:14:21.21 & -30:24:18.98 & \nodata & 0.6-2.8 & $2.99^{+0.39}_{-0.13}$ & 1.0-4.0 & 0.77 & 0.87 \\
9.2 & 00:14:20.91 & -30:24:22.47 &  &  &  &  & 0.81 &  \\
9.3 & 00:14:24.04 & -30:23:49.75 &  &  &  &  & 0.67 &  \\
\hline
10.1 & 00:14:21.22 & -30:24:21.16 & \nodata & 1.8-3.2 & $3.83^{+0.51}_{-0.27}$ & 1.8-5.0 & 0.64 & 0.66 \\
10.2 & 00:14:20.97 & -30:24:23.33 &  &  &  &  & 0.36 &  \\
10.3 & 00:14:24.17 & -30:23:49.56 &  &  &  &  & 0.15 &  \\
\hline
11.1 & 00:14:21.93 & -30:24:13.89 & \nodata & 0.4-2.8 & $2.82^{+0.18}_{-0.20}$ & 1.4-3.3 & 0.13 & 0.59 \\
11.2 & 00:14:23.34 & -30:24:05.23 &  &  &  &  & 0.52 &  \\
11.3 & 00:14:19.87 & -30:24:32.09 &  &  &  &  & 0.37 &  \\
11.4 & 00:14:22.69 & -30:24:23.55 &  &  &  &  & 0.28 &  \\
\hline
12.1 & 00:14:22.47 & -30:24:16.09 & \nodata & 1.4-3.1 & $4.51^{+0.62}_{-0.30}$ & 1.4-6.0 & 0.61 & 0.68 \\
12.2 & 00:14:22.38 & -30:24:11.72 &  &  &  &  & 0.47 &  \\
12.3 & 00:14:22.70 & -30:24:10.76 &  &  &  &  & 0.26 &  \\
12.4 & 00:14:19.07 & -30:24:35.83 &  &  &  &  & 0.44 &  \\
\hline
13.1 & 00:14:22.17 & -30:24:09.21 & \nodata & 0.6-2.6 & $1.44^{+0.06}_{-0.04}$ & 0.5-2.6 & 0.37 & 0.65 \\
13.2 & 00:14:22.51 & -30:24:07.79 &  &  &  &  & 0.41 &  \\
13.3 & 00:14:19.87 & -30:24:28.96 &  &  &  &  & 0.47 &  \\
\hline
14.1 & 00:14:21.54 & -30:23:40.69 & \nodata & 1.8-3.2 & $3.66^{+0.49}_{-0.43}$ & 1.8-5.0 & 0.35 & 0.58 \\
14.2 & 00:14:21.23 & -30:23:39.97 &  &  &  &  & 0.33 &  \\
\hline
16.1 & 00:14:13.57 & -30:22:32.91 & \nodata & 2.6-3.6\tablenotemark{4} & \nodata & \nodata & 0.12 & 0.50 \\
16.2 & 00:14:13.53 & -30:22:36.36 &  &  &  &  & 0.41 &  \\
16.3 & 00:14:13.10 & -30:22:45.51 &  &  &  &  & 0.10 &  \\
\hline
18.1 & 00:14:21.78 & -30:23:44.02 & \nodata & 1.5-5.4 & $3.56^{+0.84}_{-0.49}$ & 2.0-6.0 & 0.13 & 0.33 \\
18.2 & 00:14:21.21 & -30:23:44.29 &  &  &  &  & 0.09 &  
\enddata

\label{tab:a2744_arcs}
\end{deluxetable}

\begin{deluxetable}{ccccccccc}
\tablecaption{\MACSzerofour\ image constraints}
\tabletypesize{\tiny}
\tablehead{
    \colhead{Image} &
    \colhead{R.A.} &
    \colhead{Dec.} &
    \colhead{Spec $z$} &
    \colhead{Photo $z$}\tablenotemark{3}  &
    \colhead{Model $z$} &
    \colhead{$z$ prior} &
    \colhead{Image plane} &
    \colhead{Image plane}
    \\
    \colhead{system} &
    \colhead{} &
    \colhead{} &
    \colhead{} &
    \colhead{} &
    \colhead{} &
    \colhead{} &
    \colhead{individual rms (\arcsec)} &
    \colhead{system rms (\arcsec)}
}
\startdata
1.1 & 04:16:09.80 & -24:03:41.81 & 1.89\tablenotemark{1} & \nodata & \nodata & \nodata & 0.15 & 0.53 \\
1.2 & 04:16:10.43 & -24:03:48.56 &  &  &  &  & 0.42 &  \\
1.3 & 04:16:11.37 & -24:04:07.21 &  &  &  &  & 0.22 &  \\
\hline
2.1 & 04:16:09.86 & -24:03:42.48 & 1.89\tablenotemark{2} & \nodata & \nodata & \nodata & 0.29 & 0.53 \\
2.2 & 04:16:10.34 & -24:03:47.10 &  &  &  &  & 0.34 &  \\
2.3 & 04:16:11.39 & -24:04:07.74 &  &  &  &  & 0.21 &  \\
\hline
3.1 & 04:16:07.38 & -24:04:01.63 & 1.99\tablenotemark{2}  & \nodata & \nodata & \nodata & 0.73 & 0.74 \\
3.2 & 04:16:08.46 & -24:04:15.55 &  &  &  &  & 0.44 &  \\
3.3 & 04:16:10.03 & -24:04:32.61 &  &  &  &  & 0.42 &  \\
\hline
4.1 & 04:16:07.39 & -24:04:01.99 & 1.99\tablenotemark{2}  & \nodata & \nodata & \nodata & 0.72 & 0.73 \\
4.2 & 04:16:08.44 & -24:04:15.61 &  &  &  &  & 0.33 &  \\
4.3 & 04:16:10.04 & -24:04:32.98 &  &  &  &  & 0.48 &  \\
\hline
5.1 & 04:16:07.77 & -24:04:06.28 & \nodata & 2.1-2.8 & $1.79^{+0.11}_{-0.24}$ & 1.0-3.0 & 0.29 & 0.49 \\
5.2 & 04:16:07.84 & -24:04:07.15 &  &  &  &  & 0.23 &  \\
5.3 & 04:16:08.05 & -24:04:10.00 &  &  &  &  & 0.20 &  \\
\hline
6.1 & 04:16:09.61 & -24:03:42.62 & \nodata & 6-8 & $5.87^{+1.97}_{-1.39}$ & 4.0-8.0 & 0.01 & 0.15 \\
6.2 & 04:16:09.95 & -24:03:45.32 &  &  &  &  & 0.03 &  \\
\hline
7.1 & 04:16:09.55 & -24:03:47.11 & 2.09\tablenotemark{2}  & \nodata & \nodata & \nodata & 0.42 & 0.84 \\
7.2 & 04:16:09.76 & -24:03:48.91 &  &  &  &  & 0.95 &  \\
7.3 & 04:16:11.30 & -24:04:15.98 &  &  &  &  & 0.65 &  \\
\hline
8.1 & 04:16:08.78 & -24:03:58.03 & \nodata & 2.0-2.6 & $1.78^{+0.23}_{-0.21}$ & 1.2-2.6 & 0.08 & 0.28 \\
8.2 & 04:16:08.84 & -24:03:58.83 &  &  &  &  & 0.07 &  \\
\hline
9.1 & 04:16:06.49 & -24:04:42.87 & \nodata & 1.8-3.0 & $2.37^{+0.23}_{-0.28}$ & 1.0-3.5 & 0.13 & 0.36 \\
9.2 & 04:16:06.60 & -24:04:44.71 &  &  &  &  & 0.14 &  \\
\hline
10.1 & 04:16:06.25 & -24:04:38.13 & 2.30\tablenotemark{2} & \nodata & \nodata & \nodata & 0.70 & 0.78 \\
10.2 & 04:16:06.82 & -24:04:47.06 &  &  &  &  & 0.50 &  \\
\hline
12.1 & 04:16:09.25 & -24:04:25.92 & \nodata & 1.0-2.1 & $1.62^{+0.17}_{-0.10}$ & 1.0-2.3 & 0.06 & 0.25 \\
12.2 & 04:16:08.98 & -24:04:23.39 &  &  &  &  & 0.06 &  \\
\hline
13.1 & 04:16:06.62 & -24:04:21.60 & 3.22\tablenotemark{2}  & \nodata & \nodata & \nodata & 0.37 & 0.57 \\
13.2 & 04:16:07.72 & -24:04:30.37 &  &  &  &  & 0.35 &  \\
13.3 & 04:16:09.68 & -24:04:53.32 &  &  &  &  & 0.24 &  \\
\hline
14.1 & 04:16:06.30 & -24:04:27.60 & 2.05\tablenotemark{2}  & \nodata & \nodata & \nodata & 0.82 & 0.92 \\
14.2 & 04:16:07.45 & -24:04:44.22 &  &  &  &  & 0.29 &  \\
14.3 & 04:16:08.60 & -24:04:52.74 &  &  &  &  & 1.20 &  \\
\hline
16.1 & 04:16:05.78 & -24:04:51.20 & 1.96\tablenotemark{2}  & \nodata & \nodata & \nodata & 0.76 & 0.88 \\
16.2 & 04:16:06.80 & -24:05:04.34 &  &  &  &  & 0.65 &  \\
16.3 & 04:16:07.59 & -24:05:08.73 &  &  &  &  & 0.89 &  \\
\hline
17.1 & 04:16:07.16 & -24:05:10.91 & 2.21\tablenotemark{2}  & \nodata & \nodata & \nodata & 0.98 & 0.80 \\
17.2 & 04:16:06.86 & -24:05:09.45 &  &  &  &  & 0.44 &  \\
17.3 & 04:16:05.60 & -24:04:53.69 &  &  &  &  & 0.31 &  
\enddata

\tablenotetext{1}{\citet{Zitrin:2013lr}}
\tablenotetext{2}{\citet{Grillo:2014uq}, obtained from VLT program 186.A-0798 \citep{Balestra:2013uq}.}
\tablenotetext{3}{\citet{Jouvel:2014qy}, 95\% confidence levels on BPZ for entire image system from CLASH imaging.}
\label{tab:m0416_arcs}
\end{deluxetable}

\begin{deluxetable}{ccccccccc}
\tablecaption{\MACSzeroseven\ image constraints}
\tabletypesize{\tiny}
\tablehead{
    \colhead{Image} &
    \colhead{R.A.} &
    \colhead{Dec.} &
    \colhead{Spec $z$\tablenotemark{1}} &
    \colhead{Photo $z$\tablenotemark{2}} &
    \colhead{Model $z$} &
    \colhead{$z$ prior} &
    \colhead{Image plane} &
    \colhead{Image plane}
    \\
    \colhead{system} &
    \colhead{} &
    \colhead{} &
    \colhead{} &
    \colhead{} &
    \colhead{} &
    \colhead{} &
    \colhead{individual rms (\arcsec)} &
    \colhead{system rms (\arcsec)}
}
\startdata
1.1 & 07:17:34.88 & +37:44:28.22 & 2.90 & \nodata & \nodata & \nodata & 0.45 & 0.63 \\
1.2 & 07:17:34.52 & +37:44:24.33 &  &  &  &  & 0.59 &  \\
1.3 & 07:17:33.84 & +37:44:17.82 &  &  &  &  & 0.27 &  \\
1.4 & 07:17:32.24 & +37:44:12.97 &  &  &  &  & 0.29 &  \\
1.5 & 07:17:37.39 & +37:45:40.94 &  &  &  &  & 0.29 &  \\
\hline
3.1 & 07:17:35.65 & +37:44:29.39 & 1.80 & \nodata & \nodata & \nodata & 0.16 & 0.64 \\
3.2 & 07:17:34.67 & +37:44:21.01 &  &  &  &  & 0.62 &  \\
3.3 & 07:17:37.72 & +37:45:13.78 &  &  &  &  & 0.28 &  \\
\hline
4.1 & 07:17:31.41 & +37:45:00.42 & \nodata & 1.7-2.0 & $1.85^{+0.03}_{-0.04}$ & 1.7-2.0 & 0.10 & 0.37 \\
4.2 & 07:17:30.35 & +37:44:40.89 &  &  &  &  & 0.12 &  \\
4.3 & 07:17:33.86 & +37:45:47.84 &  &  &  &  & 0.17 &  \\
\hline
5.1 & 07:17:31.17 & +37:44:48.70 & \nodata & 4.4-4.8 & $4.02^{+0.20}_{-0.16}$ & 3.5-4.8 & 0.30 & 0.52 \\
5.2 & 07:17:30.70 & +37:44:34.09 &  &  &  &  & 0.32 &  \\
5.3 & 07:17:36.00 & +37:46:02.63 &  &  &  &  & 0.17 &  \\
\hline
6.1 & 07:17:27.44 & +37:45:25.53 & \nodata & 2.2-2.8 & $1.99^{+0.05}_{-0.05}$ & 1.8-2.8 & 0.39 & 0.57 \\
6.2 & 07:17:27.05 & +37:45:09.64 &  &  &  &  & 0.11 &  \\
6.3 & 07:17:29.73 & +37:46:10.94 &  &  &  &  & 0.39 &  \\
\hline
7.1 & 07:17:27.98 & +37:45:58.83 & \nodata & 1.2-3.0 & $1.80^{+0.11}_{-0.11}$ & 1.0-3.0 & 0.04 & 0.20 \\
7.2 & 07:17:27.61 & +37:45:50.85 &  &  &  &  & 0.04 &  \\
\hline
8.1 & 07:17:28.00 & +37:46:10.80 & \nodata & 2.7-3.6 & $2.23^{+0.07}_{-0.06}$ & 2.0-3.5 & 0.15 & 0.33 \\
8.2 & 07:17:26.90 & +37:45:47.29 &  &  &  &  & 0.06 &  \\
8.3 & 07:17:25.56 & +37:45:06.96 &  &  &  &  & 0.09 &  \\
\hline
12.1 & 07:17:32.44 & +37:45:06.63 & \nodata & 1.4-1.8 & $1.66^{+0.03}_{-0.02}$ & 1.4-1.8 & 0.49 & 0.55 \\
12.2 & 07:17:30.63 & +37:44:34.38 &  &  &  &  & 0.19 &  \\
12.3 & 07:17:33.89 & +37:45:38.24 &  &  &  &  & 0.02 &  \\
\hline
13.1 & 07:17:32.56 & +37:45:02.59 & 2.50 & \nodata & \nodata & \nodata & 0.27 & 0.64 \\
13.2 & 07:17:30.61 & +37:44:22.67 &  &  &  &  & 0.38 &  \\
13.3 & 07:17:35.09 & +37:45:47.96 &  &  &  &  & 0.54 &  \\
\hline
14.1 & 07:17:33.31 & +37:45:07.81 & 1.85 & \nodata & \nodata & \nodata & 0.67 & 0.73 \\
14.2 & 07:17:31.12 & +37:44:22.95 &  &  &  &  & 0.54 &  \\
14.3 & 07:17:35.08 & +37:45:37.51 &  &  &  &  & 0.37 &  \\
\hline
15.1 & 07:17:28.24 & +37:46:19.41 & 2.40 & \nodata & \nodata & \nodata & 0.30 & 0.48 \\
15.2 & 07:17:26.07 & +37:45:36.45 &  &  &  &  & 0.10 &  \\
15.3 & 07:17:25.57 & +37:45:16.69 &  &  &  &  & 0.25 &  \\
\hline
16.1 & 07:17:28.60 & +37:46:23.80 & \nodata & 4.0-4.7 & $3.04^{+0.11}_{-0.11}$ & 2.5-4.5 & 0.31 & 0.58 \\
16.2 & 07:17:26.06 & +37:45:34.34 &  &  &  &  & 0.41 &  \\
16.3 & 07:17:25.66 & +37:45:13.36 &  &  &  &  & 0.28 &  \\
\hline
17.1 & 07:17:28.65 & +37:46:18.71 & \nodata & 3.0-4.7 & $2.42^{+0.08}_{-0.05}$ & 2.0-3.5 & 0.38 & 0.85 \\
17.2 & 07:17:26.25 & +37:45:31.65 &  &  &  &  & 0.64 &  \\
17.3 & 07:17:25.98 & +37:45:12.96 &  &  &  &  & 1.00 &  \\
\hline
18.1 & 07:17:27.42 & +37:46:07.06 & \nodata & 1.6-3.7 & $1.76^{+0.18}_{-0.04}$ & 1.6-3.3 & 0.51 & 0.71 \\
18.2 & 07:17:26.69 & +37:45:51.61 &  &  &  &  & 0.51 &  
\enddata

\tablenotetext{1}{\citet{Limousin:2012fj}}
\tablenotetext{2}{\citet{Jouvel:2014qy}, 95\% confidence levels on BPZ for entire image system from CLASH imaging.}
\label{tab:m0717_arcs}
\end{deluxetable}

\begin{deluxetable}{cccccccccc}
\tablecaption{\MACSeleven\ image constraints}
\tabletypesize{\tiny}
\tablehead{
    \colhead{Image} &
    \colhead{R.A.} &
    \colhead{Dec.} &
    \colhead{Spec $z$\tablenotemark{1}} &
    \colhead{Photo $z$\tablenotemark{2}} &
    \colhead{Model $z$} &
    \colhead{$z$ prior} &
    \colhead{Image plane} &
    \colhead{Image plane}
    \\
    \colhead{system} &
    \colhead{} &
    \colhead{} &
    \colhead{} &
    \colhead{} &
    \colhead{} &
    \colhead{} &
    \colhead{individual rms (\arcsec)} &
    \colhead{system rms (\arcsec)}
}
\startdata
1.1 & 11:49:35.28 & +22:23:45.60 & 1.48 & \nodata & \nodata & \nodata & 0.97 & 0.83 \\
1.2 & 11:49:35.86 & +22:23:50.78 &  &  &  &  & 0.62 &  \\
1.3 & 11:49:36.82 & +22:24:08.78 &  &  &  &  & 0.31 &  \\
\hline
2.1 & 11:49:36.58 & +22:23:23.10 & 1.89 & \nodata & \nodata & \nodata & 0.37 & 0.66 \\
2.2 & 11:49:37.45 & +22:23:32.92 &  &  &  &  & 0.51 &  \\
2.3 & 11:49:37.58 & +22:23:34.39 &  &  &  &  & 0.41 &  \\
\hline
3.1 & 11:49:33.78 & +22:23:59.45 & 2.50 & \nodata & \nodata & \nodata & 0.26 & 0.49 \\
3.2 & 11:49:34.25 & +22:24:11.09 &  &  &  &  & 0.28 &  \\
3.3 & 11:49:36.31 & +22:24:25.88 &  &  &  &  & 0.15 &  \\
\hline
4.1 & 11:49:34.32 & +22:23:48.57 & \nodata & 2.7-3.1 & $2.83^{+0.16}_{-0.15}$ & 2.6-3.1 & 0.51 & 0.83 \\
4.2 & 11:49:34.65 & +22:24:02.65 &  &  &  &  & 0.87 &  \\
4.3 & 11:49:37.00 & +22:24:22.06 &  &  &  &  & 0.63 &  \\
\hline
5.1 & 11:49:35.94 & +22:23:35.02 & \nodata & 2.4-2.9 & $3.23^{+0.22}_{-0.30}$ & 2.4-3.5 & 0.35 & 0.63 \\
5.2 & 11:49:36.26 & +22:23:37.77 &  &  &  &  & 0.57 &  \\
5.3 & 11:49:37.90 & +22:24:12.79 &  &  &  &  & 0.19 &  \\
\hline
6.1 & 11:49:35.93 & +22:23:33.16 & \nodata & 0.1-3.3 & $3.13^{+0.16}_{-0.30}$ & 2.0-3.3 & 0.37 & 0.67 \\
6.2 & 11:49:36.44 & +22:23:37.89 &  &  &  &  & 0.67 &  \\
6.3 & 11:49:37.93 & +22:24:09.02 &  &  &  &  & 0.17 &  \\
\hline
7.1 & 11:49:35.75 & +22:23:28.82 & \nodata & 2.5-3.2 & $3.06^{+0.31}_{-0.27}$ & 2.5-3.5 & 1.09 & 0.95 \\
7.2 & 11:49:36.82 & +22:23:39.37 &  &  &  &  & 0.70 &  \\
7.3 & 11:49:37.82 & +22:24:04.47 &  &  &  &  & 0.88 &  \\
\hline
8.1 & 11:49:35.64 & +22:23:39.66 & \nodata & 1.1-2.8 & $3.31^{+0.17}_{-0.28}$ & 1.1-3.5 & 0.79 & 0.84 \\
8.2 & 11:49:35.95 & +22:23:42.16 &  &  &  &  & 0.90 &  \\
8.3 & 11:49:37.69 & +22:24:19.99 &  &  &  &  & 0.16 &  \\
\hline
9.1 & 11:49:37.24 & +22:25:34.44 & \nodata & 0.6-1.7 & $1.35^{+0.32}_{-0.35}$ & 0.6-1.7 & 0.19 & 0.53 \\
9.2 & 11:49:36.93 & +22:25:38.03 &  &  &  &  & 0.35 &  \\
9.3 & 11:49:36.78 & +22:25:38.02 &  &  &  &  & 0.27 &  \\
\hline
10.1 & 11:49:37.08 & +22:25:31.85 & \nodata & 1.0-2.2 & $1.61^{+0.56}_{-0.43}$ & 1.0-2.2 & 0.32 & 0.51 \\
10.2 & 11:49:36.87 & +22:25:32.29 &  &  &  &  & 0.28 &  \\
10.3 & 11:49:36.53 & +22:25:35.85 &  &  &  &  & 0.13 &  \\
\hline
13.1 & 11:49:36.89 & +22:23:52.03 & \nodata & 0.7-1.4 & $1.20^{+0.05}_{-0.02}$ & 0.7-1.4 & 0.15 & 0.60 \\
13.2 & 11:49:36.68 & +22:23:47.96 &  &  &  &  & 0.46 &  \\
13.3 & 11:49:36.01 & +22:23:37.89 &  &  &  &  & 0.40 &  \\
\hline
14.1 & 11:49:34.00 & +22:24:12.56 & \nodata & 0.7-4.0 & $2.57^{+0.21}_{-0.15}$ & 2.0-4.0 & 0.29 & 0.51 \\
14.2 & 11:49:33.80 & +22:24:09.53 &  &  &  &  & 0.22 &  
\enddata

\tablenotetext{1}{Spectroscopic redshifts reported by \citet{Smith:2009lr}.}
\tablenotetext{2}{\citet{Jouvel:2014qy}, 95\% confidence levels on BPZ for entire image system from CLASH imaging.}
\label{tab:m1149_arcs}
\end{deluxetable}

\begin{deluxetable}{cccccccccc}
\tablecaption{Abell S1063 image constraints}
\tabletypesize{\tiny}
\tablehead{
    \colhead{Image} &
    \colhead{R.A.} &
    \colhead{Dec.} &
    \colhead{Spec $z$} &
    \colhead{Photo $z$\tablenotemark{3}} &
    \colhead{Model $z$} &
    \colhead{$z$ prior} &
    \colhead{Image plane} &
    \colhead{Image plane}
    \\
    \colhead{system} &
    \colhead{} &
    \colhead{} &
    \colhead{} &
    \colhead{} &
    \colhead{} &
    \colhead{} &
    \colhead{individual rms (\arcsec)} &
    \colhead{system rms (\arcsec)}
}
\startdata
1.1 & 22:48:46.68 & -44:31:37.13 & 1.24\tablenotemark{1,2} & \nodata & \nodata & \nodata & 0.50 & 0.68 \\
1.2 & 22:48:47.01 & -44:31:44.22 &  &  &  &  & 0.44 &  \\
1.3 & 22:48:44.74 & -44:31:16.32 &  &  &  &  & 0.43 &  \\
\hline
2.1 & 22:48:46.25 & -44:31:52.28 & 1.26\tablenotemark{1,2} & \nodata & \nodata & \nodata & 0.16 & 0.69 \\
2.2 & 22:48:46.11 & -44:31:47.39 &  &  &  &  & 0.62 &  \\
2.3 & 22:48:43.16 & -44:31:17.62 &  &  &  &  & 0.53 &  \\
\hline
3.1 & 22:48:46.93 & -44:31:55.70 & \nodata & 1.8-2.3 & $2.08^{+0.11}_{-0.19}$ & 1.2-2.3 & 0.24 & 0.45 \\
3.2 & 22:48:46.54 & -44:31:43.43 &  &  &  &  & 0.16 &  \\
\hline
4.1 & 22:48:46.49 & -44:31:48.58 & \nodata & 0.9-1.8 & $1.25^{+0.04}_{-0.06}$ & 0.9-1.8 & 0.09 & 0.32 \\
4.2 & 22:48:46.40 & -44:31:45.91 &  &  &  &  & 0.10 &  \\
\hline
5.1 & 22:48:43.01 & -44:31:24.92 & 1.40\tablenotemark{1,2} & \nodata & \nodata & \nodata & 0.73 & 0.73 \\
5.2 & 22:48:45.08 & -44:31:38.32 &  &  &  &  & 0.56 &  \\
5.3 & 22:48:46.36 & -44:32:11.51 &  &  &  &  & 0.04 &  \\
\hline
6.1 & 22:48:41.82 & -44:31:41.99 & 1.43\tablenotemark{1,2} & \nodata & \nodata & \nodata & 0.50 & 0.93 \\
6.2 & 22:48:42.20 & -44:31:57.14 &  &  &  &  & 1.25 &  \\
6.3 & 22:48:45.23 & -44:32:24.00 &  &  &  &  & 0.66 &  \\
\hline
7.1 & 22:48:40.65 & -44:31:38.10 & \nodata & 1.8-2.8 & $1.92^{+0.05}_{-0.03}$ & 1.8-2.7 & 0.82 & 0.77 \\
7.2 & 22:48:41.82 & -44:32:13.60 &  &  &  &  & 0.60 &  \\
7.3 & 22:48:43.64 & -44:32:25.80 &  &  &  &  & 0.18 &  \\
\hline
8.1 & 22:48:40.31 & -44:31:34.32 & \nodata & 2.4-3.2 & $2.84^{+0.12}_{-0.07}$ & 2.4-3.1 & 0.63 & 0.65 \\
8.2 & 22:48:41.91 & -44:32:18.20 &  &  &  &  & 0.33 &  \\
8.3 & 22:48:43.39 & -44:32:27.17 &  &  &  &  & 0.18 &  \\
\hline
9.1 & 22:48:40.27 & -44:31:34.61 & \nodata & 2.4-3.1 & $2.87^{+0.08}_{-0.09}$ & 2.4-3.1 & 0.50 & 0.63 \\
9.2 & 22:48:41.95 & -44:32:19.00 &  &  &  &  & 0.36 &  \\
9.3 & 22:48:43.27 & -44:32:26.92 &  &  &  &  & 0.29 &  \\
\hline
11.1 & 22:48:42.01 & -44:32:27.71 & 3.12\tablenotemark{1} & \nodata & \nodata & \nodata & 0.20 & 0.67 \\
11.2 & 22:48:41.56 & -44:32:23.93 &  &  &  &  & 0.24 &  \\
11.3 & 22:48:39.74 & -44:31:46.31 &  &  &  &  & 0.72 &  \\
\hline
12.1 & 22:48:45.37 & -44:31:48.18 & 6.11\tablenotemark{4} & \nodata & \nodata & \nodata & 1.10 & 0.98 \\
12.2 & 22:48:43.45 & -44:32:04.63 &  &  &  &  & 1.10 &  \\
12.3 & 22:48:45.81 & -44:32:14.89 &  &  &  &  & 0.50 &  \\
12.4 & 22:48:41.11 & -44:31:11.32 &  &  &  &  & 0.98 &  \\
\hline
14.1 & 22:48:42.92 & -44:32:09.13 & \nodata & 3.1-3.6 & $3.22^{+0.16}_{-0.15}$ & 2.9-3.6 & 0.93 & 0.81 \\
14.2 & 22:48:44.98 & -44:32:19.28 &  &  &  &  & 0.31 &  \\
14.3 & 22:48:40.96 & -44:31:19.52 &  &  &  &  & 0.59 &  \\
\hline
15.1 & 22:48:46.01 & -44:31:49.87 & \nodata & 2.8-3.3 & $2.47^{+0.08}_{-0.11}$ & 2.0-3.1 & 1.38 & 1.17 \\
15.2 & 22:48:46.21 & -44:32:03.91 &  &  &  &  & 1.51 &  \\
15.3 & 22:48:42.22 & -44:31:10.74 &  &  &  &  & 1.21 &  \\
\hline
16.1 & 22:48:39.90 & -44:32:01.14 & \nodata & 3.0-3.5 & $3.12^{+0.10}_{-0.13}$ & 2.8-4.0 & 0.28 & 0.65 \\
16.2 & 22:48:40.03 & -44:32:05.75 &  &  &  &  & 0.25 &  \\
16.3 & 22:48:42.68 & -44:32:35.05 &  &  &  &  & 0.62 &  \\
\hline
17.1 & 22:48:44.60 & -44:32:19.86 & \nodata & 3.5-4.0 & $3.09^{+0.34}_{-0.18}$ & 2.5-3.9 & 0.21 & 0.48 \\
17.2 & 22:48:42.92 & -44:32:12.23 &  &  &  &  & 0.25 &  \\
\hline
18.1 & 22:48:41.32 & -44:32:11.83 & \nodata & 0.5-4.3 & $3.70^{+0.50}_{-0.29}$ & 0.5-4.5 & 0.31 & 0.57 \\
18.2 & 22:48:44.35 & -44:32:31.42 &  &  &  &  & 0.35 &  
\enddata

\tablenotetext{1}{This work.}
\tablenotetext{2}{\citet{Richard:2014gf}.}
\tablenotetext{3}{\citet{Jouvel:2014qy}, 95\% confidence levels on BPZ for entire image system from CLASH imaging.}
\tablenotetext{4}{\citet{Balestra:2013uq,Boone:2013lr}.}
\label{tab:as1063_arcs}
\end{deluxetable}

\begin{deluxetable}{ccccccccc}
\tablecaption{Abell 370 image constraints}
\tabletypesize{\tiny}
\tablehead{
    \colhead{Image} &
    \colhead{R.A.} &
    \colhead{Dec.} &
    \colhead{Spec $z$} &
    \colhead{Photo $z$\tablenotemark{3}} &
    \colhead{Model $z$} &
    \colhead{$z$ prior} &
    \colhead{Image plane} &
    \colhead{Image plane}
    \\
    \colhead{system} &
    \colhead{} &
    \colhead{} &
    \colhead{} &
    \colhead{} &
    \colhead{} &
    \colhead{} &
    \colhead{individual rms (\arcsec)} &
    \colhead{system rms (\arcsec)}
}
\startdata
1.1 & 02:39:52.09 & -01:34:37.28 & 0.81\tablenotemark{1} & \nodata & \nodata & \nodata & 1.37 & 1.15 \\
1.2 & 02:39:54.31 & -01:34:34.11 &  &  &  &  & 1.37 &  \\
1.3 & 02:39:52.48 & -01:34:36.20 &  &  &  &  & 1.25 &  \\
\hline
2.1 & 02:39:53.72 & -01:35:03.56 & 0.72\tablenotemark{1} & \nodata & \nodata & \nodata & 0.12 & 0.47 \\
2.2 & 02:39:53.03 & -01:35:06.65 &  &  &  &  & 0.23 &  \\
2.3 & 02:39:52.50 & -01:35:04.60 &  &  &  &  & 0.38 &  \\
2.4 & 02:39:52.65 & -01:35:05.36 &  &  &  &  & 0.15 &  \\
2.5 & 02:39:52.70 & -01:35:05.79 &  &  &  &  & 0.07 &  \\
\hline
3.1 & 02:39:51.75 & -01:34:01.10 & 1.42\tablenotemark{2} & \nodata & \nodata & \nodata & 0.23 & 0.59 \\
3.2 & 02:39:52.44 & -01:33:57.35 &  &  &  &  & 0.54 &  \\
3.3 & 02:39:54.54 & -01:34:02.25 &  &  &  &  & 0.14 &  \\
\hline
4.1 & 02:39:55.11 & -01:34:35.15 & 1.27\tablenotemark{2} & \nodata & \nodata & \nodata & 1.47 & 1.05 \\
4.2 & 02:39:52.98 & -01:34:34.94 &  &  &  &  & 0.37 &  \\
4.3 & 02:39:50.86 & -01:34:40.95 &  &  &  &  & 1.17 &  \\
\hline
5.1 & 02:39:53.63 & -01:35:21.05 & \nodata & 1.0-1.8 & $1.15^{+0.03}_{-0.03}$ & 1.0-1.8 & 1.79 & 1.17 \\
5.2 & 02:39:53.07 & -01:35:21.66 &  &  &  &  & 0.17 &  \\
5.3 & 02:39:52.51 & -01:35:20.91 &  &  &  &  & 1.54 &  \\
\hline
6.1 & 02:39:52.67 & -01:34:38.28 & 1.06\tablenotemark{2} & \nodata & \nodata & \nodata & 0.22 & 0.66 \\
6.2 & 02:39:51.48 & -01:34:42.10 &  &  &  &  & 0.62 &  \\
6.3 & 02:39:55.11 & -01:34:38.10 &  &  &  &  & 0.39 &  \\
\hline
7.1 & 02:39:52.74 & -01:34:49.88 & \nodata & 2.8-3.2 & $2.07^{+0.11}_{-0.06}$ & 0.9-3.2 & 0.74 & 0.90 \\
7.2 & 02:39:52.76 & -01:34:51.06 &  &  &  &  & 0.62 &  \\
7.3 & 02:39:52.51 & -01:35:08.60 &  &  &  &  & 0.87 &  \\
7.4 & 02:39:50.77 & -01:34:48.44 &  &  &  &  & 0.95 &  \\
7.5 & 02:39:56.11 & -01:34:41.09 &  &  &  &  & 0.80 &  \\
\hline
8.1 & 02:39:51.47 & -01:34:11.64 & \nodata & 2.8-3.4 & $2.26^{+0.07}_{-0.06}$ & 2.0-3.4 & 0.87 & 0.90 \\
8.2 & 02:39:50.85 & -01:34:25.65 &  &  &  &  & 0.65 &  \\
8.3 & 02:39:56.18 & -01:34:24.46 &  &  &  &  & 0.88 &  \\
\hline
9.1 & 02:39:50.98 & -01:34:40.84 & \nodata & 1.0-1.7 & $1.50^{+0.05}_{-0.03}$ & 1.0-1.7 & 0.25 & 0.46 \\
9.2 & 02:39:52.67 & -01:34:34.94 &  &  &  &  & 0.16 &  \\
9.3 & 02:39:55.68 & -01:34:35.98 &  &  &  &  & 0.23 &  
\enddata

\tablenotetext{1}{\citet{Richard:2010wd}.}
\tablenotetext{2}{\citet{Richard:2014gf}.}
\tablenotetext{3}{BPZ measured from \hst\ preliminary data reductions.}
\label{tab:a370_arcs}
\end{deluxetable}

\clearpage

\begin{deluxetable}{lccccccc}
\tablecaption{Abell 2744 model parameters}
\tabletypesize{\scriptsize}
\tablecolumns{8}
\tablehead{
    \colhead{Component} &
    \colhead{$\Delta$RA (\arcsec)} &
    \colhead{$\Delta$Dec (\arcsec)} &
    \colhead{$e$} &
    \colhead{$\theta$ (\degr)} &
    \colhead{$r_\mathrm{core}$ (kpc)} &
    \colhead{$r_\mathrm{cut}$ (kpc)} &
    \colhead{$\sigma\ (\mathrm{km\ s^{-1}})$}
}
\startdata
cluster halo \#1 (H1) & $1.25^{+1.2}_{-1.4}$ & $-5.45^{+1.9}_{-2.7}$ & [0] & \nodata& $57.7^{+15}_{-16}$ & [1500] & $549^{+76}_{-77}$ \\[5pt]
cluster halo \#2 (H2) & $19.7^{+1.4}_{-0.73}$ & $-18.3^{+0.98}_{-0.64}$ & $0.538^{+0.059}_{-0.041}$ & $33.7^{+7.5}_{-6.7}$ & $25.5^{+6.3}_{-4.9}$ & [1500] & $516^{+44}_{-43}$ \\[5pt]
cluster halo \#3 (H3) & $12.9^{+1.3}_{-3.8}$ & $47.9^{+8.5}_{-5.8}$ & [0] & \nodata & [20.0] & [1500] & $504^{+89}_{-58}$ \\[5pt]
cluster halo \#4 (H4) & [-102] & [84.7] & $0.12\ (<0.23)$ & $44.1^{+17}_{-21}$ & $88.6^{+14}_{-3.7}$ & [1500] & $885^{+82}_{-31}$ \\[5pt]
cluster halo \#5 (H5) & $-34.8^{+11}_{-15}$ & $133^{+17}_{-15}$ & [0] & \nodata & [150] & [1500] & $890^{+140}_{-220}$ \\[5pt]
cluster galaxy \#1 (G1) & [0] & [0] & [0.8] & $57.8^{+8.5}_{-6.2}$ & [1.35] & $457^{+150}_{-270}$ & $359^{+23}_{-18}$ \\[5pt]
cluster galaxy \#2 (G2) & [18.0] & [-20.1] & $0.551^{+0.23}_{-0.21}$ & $93.1^{+14}_{-13}$ & [0.159] & $363^{+110}_{-130}$ & $290^{+46}_{-33}$ \\[5pt]
\hline \\[-5pt]
$L^\star$ galaxy & \multicolumn{4}{c}{$m_\star=18.50,z=0.308$ (ACS F814W)} & [0.15] & [30] & [120]
\enddata

\tablecomments{Parameters for best fit-model and errors representing the 95\% confidence level of the parameter values from the MCMC chain. Values in brackets are not optimized, or fixed para2meters. $\Delta$RA and $\Delta$Dec are measured with respect to the galaxy at $\alpha$=00:14:20.70, $\delta$=-30:24:00.62, position angles are measured north of west, ellipticity is defined as $e=(a^2-b^2)/(a^2+b^2)$. The labels for the halos shown in Figure \ref{fig:crit_a2744} are given in parentheses. The cluster core is constrained by most of the lensing constraints and is composed of the cluster halos \#1 and \#2. The remaining halos lie outside of this region and are not well constrained.}
\label{tab:a2744_params}
\end{deluxetable}

\begin{deluxetable}{lccccccc}
\tablecaption{\MACSzerofour\ model parameters}
\tabletypesize{\scriptsize}
\tablecolumns{8}
\tablehead{
    \colhead{Component} &
    \colhead{$\Delta$RA (\arcsec)} &
    \colhead{$\Delta$Dec (\arcsec)} &
    \colhead{$e$} &
    \colhead{$\theta$ (\degr)} &
    \colhead{$r_\mathrm{core}$ (kpc)} &
    \colhead{$r_\mathrm{cut}$ (kpc)} &
    \colhead{$\sigma\ (\mathrm{km\ s^{-1}})$}
}
\startdata
cluster halo \#1 (H1) & $16.2^{+1.1}_{-0.9}$ & $16.6^{+2.0}_{-1.3}$ & $0.618^{+0.094}_{-0.056}$ & $-34.1^{+3.7}_{-2.7}$ & $91.3^{+13}_{-20}$ & [1500] & $938^{+69}_{-130}$ \\[5pt]
cluster halo \#2 (H2) & $-9.03^{+6.0}_{-6.9}$ & $-23.2^{+6.9}_{-8.7}$ & $0.731^{+0.07}_{-0.21}$ & $-41.7^{+5.3}_{-6.6}$ & $68.3^{+48}_{-22}$ & [1500] & $521^{+250}_{-78}$ \\[5pt]
foreground galaxy\tablenotemark{*} (F1) & [-20.7] & [-50.7] & [0.143] & [-44.4] & $93.4^{+14}_{-23}$ & [1500] & $774^{+75}_{-210}$ \\[5pt]
cluster galaxy \#1 (G1) & [-9.02] & [-21.0] & [0.0710] & [-44.7] & [0.152] & [30.4] & $320^{+55}_{-59}$ \\[5pt]
cluster galaxy \#2 (G2) & [-7.48] & [11.2] & [0] & \nodata & [0.5] & [250] & $58.0^{+23}_{-16}$ \\[5pt]
\hline \\[-5pt]
$L^\star$ galaxy & \multicolumn{4}{c}{$m_\star=19.33,z=0.396$ (ACS F775W)} & [0.15] & [30] & [120]
\enddata

\tablecomments{Parameters for best fit-model and errors representing the 95\% confidence level of the parameter values from the MCMC chain. Values in brackets are not optimized, or fixed parameters. $\Delta$RA and $\Delta$Dec are measured with respect to the galaxy at $\alpha$=4:16:08.331, $\delta$=-24:04:17.74, position angles are measured north of west, ellipticity is defined as $e=(a^2-b^2)/(a^2+b^2)$. The labels for the halos shown in Figure \ref{fig:crit_m0416} are given in parentheses.}
\tablenotetext{*}{Parameters intrinsic to galaxy ($r_\mathrm{core}$, $r_\mathrm{cut}$, and $\sigma$) derived by arbitrarily placing the galaxy at cluster redshift, $z=0.396$.}
\label{tab:m0416_params}
\end{deluxetable}

\begin{deluxetable}{lccccccc}
\tablecaption{\MACSzeroseven\ model parameters}
\tabletypesize{\scriptsize}
\tablecolumns{8}
\tablehead{
    \colhead{Component} &
    \colhead{$\Delta$RA (\arcsec)} &
    \colhead{$\Delta$Dec (\arcsec)} &
    \colhead{$e$} &
    \colhead{$\theta$ (\degr)} &
    \colhead{$r_\mathrm{core}$ (kpc)} &
    \colhead{$r_\mathrm{cut}$ (kpc)} &
    \colhead{$\sigma\ (\mathrm{km\ s^{-1}})$}
}
\startdata
cluster halo \#1 (H1) & $-5.70^{+0.81}_{-1.2}$ & $6.39^{+1.1}_{-1.3}$ & $0.314^{+0.088}_{-0.074}$ & $82.6^{+8.3}_{-7.3}$ & $48.1^{+17}_{-20}$ & [1500] & $832^{+67}_{-67}$ \\[5pt]
cluster halo \#2 (H2) & $-34.9^{+1.0}_{-0.4}$ & $-12.8^{+0.5}_{-1.5}$ & $0.869^{+0.039}_{-0.030}$ & $55.2^{+0.8}_{-2.2}$ & $28.3^{+9.8}_{-19}$ & [1500] & $694^{+33}_{-35}$ \\[5pt]
cluster halo \#3 (H3) & $-73.6^{+5.8}_{-2.6}$ & $39.3^{+0.88}_{-2.9}$ & $0.822^{+0.040}_{-0.040}$ & $10.1^{+2.0}_{-1.5}$ & $156^{+20}_{-38}$ & [1500] & $1080^{+55}_{-120}$ \\[5pt]
cluster halo \#4 (H4) & $-117^{+3.5}_{-4.8}$ & $72.3^{+2.4}_{-2.3}$ & $0.565^{+0.22}_{-0.14}$ & $9.64^{+5.8}_{-9.1}$ & $73.5^{+53}_{-9.4}$ & [1500] & $790^{+170}_{-31}$ \\[5pt]
foreground galaxy\tablenotemark{*} (F1) & [19.6] & [-21.8] & [0] & \nodata & $110^{+54}_{-33}$ & $336^{+43}_{-150}$ & $854^{+160}_{-79}$ \\[5pt]
cluster galaxy (G1) & [0.929] & [32.6] & [0.329] & [-20.0] & [0.315] & $18.1^{+35}_{-7.8}$ & $1010^{+310}_{-360}$ \\[5pt]
\hline \\[-5pt]
$L^\star$ galaxy & \multicolumn{4}{c}{$m_\star=20.66,z=0.545$ \citep[ACS F814W, ][]{Limousin:2012fj}} & [0.15] & [30] & [120]
\enddata

\tablecomments{Parameters for best fit-model and errors representing the 95\% confidence level of the parameter values from the MCMC chain. Values in brackets are not optimized, or fixed parameters. $\Delta$RA and $\Delta$Dec are measured with respect to the galaxy at $\alpha$=07:17:35.57, $\delta$=+37:44:44.80, position angles are measured north of west, ellipticity is defined as $e=(a^2-b^2)/(a^2+b^2)$. The labels for the halos shown in Figure \ref{fig:crit_m0717} are given in parentheses.}
\tablenotetext{*}{Parameters intrinsic to galaxy ($r_\mathrm{core}$, $r_\mathrm{cut}$, and $\sigma$) derived by arbitrarily placing the galaxy at cluster redshift, $z=0.545$.}
\label{tab:m0717_params}
\end{deluxetable}

\begin{deluxetable}{lccccccc}
\tablecaption{\MACSeleven\ model parameters}
\tabletypesize{\scriptsize}
\tablecolumns{8}
\tablehead{
    \colhead{Component} &
    \colhead{$\Delta$RA (\arcsec)} &
    \colhead{$\Delta$Dec (\arcsec)} &
    \colhead{$e$} &
    \colhead{$\theta$ (\degr)} &
    \colhead{$r_\mathrm{core}$ (kpc)} &
    \colhead{$r_\mathrm{cut}$ (kpc)} &
    \colhead{$\sigma\ (\mathrm{km\ s^{-1}})$}
}
\startdata
cluster halo \#1 (H1) & $6.79^{+2.8}_{-2.8}$ & $-5.14^{+2.1}_{-1.8}$ & $0.701^{+0.021}_{-0.097}$ & $29.8^{+1.3}_{-2.2}$ & $64.9^{+8.6}_{-10.}$ & [1500] & $812^{+55}_{-65}$ \\[5pt]
cluster halo \#2 (H2) & $-12.6^{+1.1}_{-1.9}$ & $26.3^{+3.7}_{-2.8}$ & [0] & \nodata & $107^{+31}_{-18}$ & [1500] & $919^{+130}_{-88}$ \\[5pt]
BCG (BCG) & [0] & [0] & [0.2] & [124] & [1] & [200] & $299^{+23}_{-58}$ \\[5pt]
cluster galaxy \#1 (G1) & [25.6] & [-32.2] & [0.205] & [47.0] & [0.233] & [40.2] & $544^{+32}_{-28}$ \\[5pt]
cluster galaxy \#2 (G2) & $16.9^{+0.39}_{-0.55}$ & $101^{+0.93}_{-1.1}$ & [0.800] & $-60.1^{+4.3}_{-6.9}$ & [0.261] & [300] & $371^{+57}_{-40}$ \\[5pt]
\hline \\[-5pt]
$L^\star$ galaxy & \multicolumn{4}{c}{$m_\star=20.3,z=0.543$  \citep[K band, ][]{Smith:2009lr}} & [0.15] & [30] & [120]
\enddata

\tablecomments{Parameters for best fit-model and errors representing the 95\% confidence level of the parameter values from the MCMC chain. Values in brackets are not optimized, or fixed parameters. $\Delta$RA and $\Delta$Dec are measured with respect to the galaxy at $\alpha$=11:49:35.695, $\delta$=+22:23:54.70, position angles are measured north of west, ellipticity is defined as $e=(a^2-b^2)/(a^2+b^2)$. The labels for the halos shown in Figure \ref{fig:crit_m1149} are given in parentheses.}
\label{tab:m1149_params}
\end{deluxetable}

\begin{deluxetable}{lccccccc}
\tablecaption{Abell S1063 model parameters}
\tabletypesize{\scriptsize}
\tablecolumns{8}
\tablehead{
    \colhead{Component} &
    \colhead{$\Delta$RA (\arcsec)} &
    \colhead{$\Delta$Dec (\arcsec)} &
    \colhead{$e$} &
    \colhead{$\theta$ (\degr)} &
    \colhead{$r_\mathrm{core}$ (kpc)} &
    \colhead{$r_\mathrm{cut}$ (kpc)} &
    \colhead{$\sigma\ (\mathrm{km\ s^{-1}})$}
}
\startdata
cluster halo \#1 (H1) & $-0.826^{+0.59}_{-0.32}$ & $0.0556^{+0.38}_{-0.35}$ & $0.573^{+0.025}_{-0.026}$ & $-36.2^{+0.56}_{-0.52}$ & $84.9^{+8.7}_{-8.5}$ & [1500] & $1190^{+24}_{-29}$ \\[5pt]
cluster halo \#2 (H2) & $386^{+170}_{-67}$ & $212^{+89}_{-22}$ & [0] & \nodata & [50.0] & [1500] & $1820^{+650}_{-260}$ \\[5pt]
cluster halo \#3 (H3) & $12.9^{+31}_{-9.9}$ & $-111^{+24}_{-87}$ & [0] & \nodata & [50.0] & [1500] & $592^{+390}_{-180}$ \\[5pt]
BCG (BCG) & [0] & [0] & [0.269] & [-37.7] & [0.208] & [41.5] & $356^{+77}_{-76}$ \\[5pt]
cluster galaxy \#1 (G1) & [31.6] & [17.6] & [0.246] & [-86.3] & [0.107] & [21.4] & $115^{+68}_{-99}$ \\[5pt]
cluster galaxy \#2 (G2) & [-29.4] & [-34.7] & [0.635] & [89.2] & [0.0580] & [11.6] & $85.8^{+67}_{-81}$ \\[5pt]
cluster galaxy \#3 (G3) & [-42.6] & [-14.1] & [0.250] & [4.90] & [0.0410] & [8.23] & $74.2^{+25}_{-7.5}$ \\[5pt]
\hline \\[-5pt]
$L^\star$ galaxy & \multicolumn{4}{c}{$m_\star=18.82,z=0.348$ (ACS F814W)} & [0.15] & [30] & [120]
\enddata

\tablecomments{Parameters for best fit-model and errors representing the 95\% confidence level of the parameter values from the MCMC chain. Values in brackets are not optimized, or fixed parameters. $\Delta$RA and $\Delta$Dec are measured with respect to the galaxy at $\alpha$=22:48:43.970, $\delta$=-44:31:51.22, position angles are measured north of west, ellipticity is defined as $e=(a^2-b^2)/(a^2+b^2)$. The labels for the halos shown in Figure \ref{fig:crit_as1063} are given in parentheses.}
\label{tab:as1063_params}
\end{deluxetable}

\begin{deluxetable}{lccccccc}
\tablecaption{Abell 370 model parameters}
\tabletypesize{\scriptsize}
\tablecolumns{8}
\tablehead{
    \colhead{Component} &
    \colhead{$\Delta$RA (\arcsec)} &
    \colhead{$\Delta$Dec (\arcsec)} &
    \colhead{$e$} &
    \colhead{$\theta$ (\degr)} &
    \colhead{$r_\mathrm{core}$ (kpc)} &
    \colhead{$r_\mathrm{cut}$ (kpc)} &
    \colhead{$\sigma\ (\mathrm{km\ s^{-1}})$}
}
\startdata
cluster halo \#1 (H1) & $0.143^{+0.56}_{-0.28}$ & $34.9^{+2.4}_{-0.8}$ & $0.0913^{+0.023}_{-0.062}$ & $89.4^{+10.}_{-3.9}$ & $94.7^{+3.7}_{-15}$ & [1500] & $1040^{+45}_{-120}$ \\[5pt]
cluster halo \#2 (H2) & $-2.46^{+0.27}_{-0.30}$ & $1.96^{+2.4}_{-1.3}$ & $0.473^{+0.024}_{-0.027}$ & $80.8^{+0.99}_{-0.74}$ & $88.2^{+8.7}_{-5.7}$ & [1500] & $969^{+100}_{-46}$ \\[5pt]
BCG (BCG) & [0] & [0] & [0.373] & [-83.8] & [0.196] & [39.2] & $405^{+25}_{-26}$ \\[5pt]
cluster galaxy (G1) & [-7.95] & [-9.80] & [0.8] & $121^{+31}_{-19}$ & [0.0790] & [15.8] & $114^{+18}_{-15}$ \\[5pt]
\hline \\[-5pt]
$L^\star$ galaxy & \multicolumn{4}{c}{$m_\star=19.04,z=0.375$ (ACS F814W)} & [0.15] & [30] & [120]
\enddata

\tablecomments{Parameters for best fit-model and errors representing the 95\% confidence level of the parameter values from the MCMC chain. Values in brackets are not optimized, or fixed parameters. $\Delta$RA and $\Delta$Dec are measured with respect to the galaxy at $\alpha$=2:39:53.125, $\delta$=-1:34:56.420, position angles are measured north of west, ellipticity is defined as $e=(a^2-b^2)/(a^2+b^2)$. The labels for the halos shown in Figure \ref{fig:crit_a370} are given in parentheses.}
\label{tab:a370_params}
\end{deluxetable}

\end{document}